\documentclass[final,3p,times]{elsarticle}

\makeatletter
\def\ps@pprintTitle{%
  \let\@oddhead\@empty
  \let\@evenhead\@empty
  \let\@oddfoot\@empty
  \let\@evenfoot\@oddfoot
}
\makeatother
\usepackage{amssymb}
\usepackage{bm,morefloats}
\usepackage{dcolumn}
\usepackage[T2A,T1]{fontenc} % load the desired encodings
\usepackage{fonttable}
\usepackage{graphicx}
\usepackage{balance}
\usepackage[T1]{fontenc}
		\usepackage{cellspace}
\usepackage{latexsym,color}
\usepackage{centernot}
\usepackage{mathtools}
\usepackage{amssymb}

\usepackage{amsmath,amssymb,amsthm,mathrsfs,amsfonts,dsfont}

\usepackage{color}
\usepackage{bm}		% Bold maths symbols, including upright Greek
\usepackage{rotating}
\usepackage[dvipsnames]{xcolor}
\definecolor{LinkBlue}{RGB}{6,69,173}
\definecolor{DarkBlue}{RGB}{11,0,128}
\definecolor{red}{rgb}{1,0.,0.}
\usepackage[colorlinks=true,linkcolor=LinkBlue,urlcolor=magenta,
	 citecolor=green,hyperfootnotes=true,urlcolor=red]{hyperref}

\usepackage{enumitem}
\count\footins = 1000
\usepackage{amsmath}
\begin{document}

\begin{frontmatter}



\title{On the  red--shift  emission from  the  black hole horizons replicas } %% Article title


\author{D. Pugliese\&Z. Stuchl\'{\i}k} %% Author name


\affiliation{organization={Research Centre for Theoretical Physics and Astrophysics, Institute of Physics,Silesian University in Opava},%Department and Organization
            addressline={Bezru\v{c}ovo n\'{a}m\v{e}st\'{i} 13},
            city={Opava},
            postcode={74601},
            country={Czech Republic}}

%% Abstract
\begin{abstract}
We  examine   the   red-shift emission  from special orbits (\emph{horizons replicas}) where photons  have the same angular velocity,  in magnitude, as those of the  Kerr black hole (\textbf{BH}) Killing horizons.
We focus on two particular  contexts of special  significance for  their observational implications.
A set  of these  orbits is  located in the
   \textbf{BH} photons shell. Therefore the  analysis also connects the red-shifting or  blue-shifting  from replicas (defined through the \textbf{BH} light surfaces) to the    \textbf{BH}  shadow boundaries.
   We then concentrate on the  equatorial, general relativistic, axially symmetric accretion disks, orbiting around the central attractor. In the analysis we adopted  in particular   the Polish doughnut models of geometrically thick toroids,  to  investigate the  signals frequency--shifting   from the horizons  replica  as related to  the disks surface, inner edge and accretion flows.  The findings  map  the red-shifting and blue-shifting regions of  the signals in  dependence to  the  photons impact parameter, view-angle, angular velocity of the emitter and \textbf{BH} spin.  Co--rotating and counter--rotating replica  could be identified  through  the  red--shifting  or  blue--shifting of the emitted  signals from the mapped regions.
\end{abstract}



%% Keywords
\begin{keyword}


Lense--Thirring--Black holes-- Accretion disks--Accretion; Hydrodynamics --Galaxies: actives
\end{keyword}

\end{frontmatter}



\def\bea{\begin{eqnarray}}
\def\eea{\end{eqnarray}}
\newcommand{\tb}[1]{\textbf{#1}}
\newcommand{\actaa}{Acta Astronomica}
\newcommand{\laa}{\mathcal{L}}
\newcommand{\ba}{\mathcal{B}}
\newcommand{\Sie}{\mathcal{S}}
\newcommand{\Mie}{\mathcal{M}}
\newcommand{\La}{\mathcal{L}}
\newcommand{\Em}{\mathcal{E}}
\count\footins = 1000
\newcommand{\aap}{A\&A}
\newcommand{\mnras}{MNRAS}
\newcommand{\apss}{Ap\&SS}
\newcommand{\apjs}{APJS}
\newcommand{\apjj}{ApJ}
\newcommand{\apj}{ApJ}
\newcommand{\prdd}{Phys. Rev. \textbf{D}}

\newcommand{\pp}{\textbf{()}}
\def\be{\begin{equation}}
\def\ee{\end{equation}}
\newcommand{\mso}{\mathrm{mso}}
\newcommand{\mbo}{\mathrm{mbo}}
\newcommand{\Da}{{\mbox{\scriptsize  \textbf{\textsf{D}}}}}

\newcommand{\rtb}[1]{\textcolor[rgb]{1.00,0.00,0.00}{\tb{#1}}}
\newcommand{\gtb}[1]{\textcolor[rgb]{0.17,0.72,0.40}{\tb{#1}}}
\newcommand{\ptb}[1]{\textcolor[rgb]{0.07,0.04,0.95}{\tb{#1}}}
\newcommand{\btb}[1]{\textcolor[rgb]{0.00,0.00,1.00}{\textbf{#1}}}
\newcommand{\otb}[1]{\textcolor[rgb]{1.00,0.50,0.25}{\tb{#1}}}
\newcommand{\non}[1]{{\LARGE{\not}}{#1}}

\newcommand{\cc}{\mathrm{C}}

\newcommand{\il}{~}
\newcommand{\la}{\mathcal{A}}
  \newcommand{\Qa}{\mathcal{Q}}
\newcommand{\Sa}{\mathcal{\mathbf{S}}}
\newcommand{\Ta}{\mathbf{T}}
\newcommand{\Ca}{\mathcal{\mathbf{C}}}

%%%%%%%%%%%%%%%%%%%
\section{Introduction}

The   observation of the closest  proximity of the black hole (\textbf{BH})  horizon (\textbf{BH} shadow) was made accessible with the study  in  the 2019  of the \textbf{BH} located at the center of the giant elliptical galaxy
\textbf{M87},  by the \textbf{E}vent \textbf{H}orizon \textbf{T}elescope (\textbf{EHT}) Collaboration\footnote{https://eventhorizontelescope.org}--(\cite{EHT1}) .
This analysis opened up a new way to the observational astronomy  of  several optical phenomena,  connected with the  accretion physics in the vicinity of the innermost circular orbit and the  \textbf{BH}  horizon.
The study   of the  (Doppler and gravitationally)  red--shifted signals    emitted from the surface of an accretion disk  and its  inner edge, is an important aspect of  the spectral observational analyses that can provide insights into the disk's morphological and kinematic characteristics, as well as focus on  the deeper aspects of its background space-time structure in the accretion regions. These areas are   approximately within the marginally stable orbit and near the rotation axis of the central spinning attractor.  The  region, for this class of singularities, also comprises  a spacetime ergoregion,  which plays a  crucial role  in the energy extraction processes from the  \textbf{BH},    consequently to the  interaction with the surrounding environment\footnote{However this region is also involved  in   the quantum effects (as  Hawking semiclassical  radiation) related to  characteristic thermal emission, unrelated to the interaction of the singularity and its  horizons with the surrounding  matter and field environments--\cite{Hawking71,Hawking74,Hawking75,Penrose69} and \cite{Daly:2008zk,Daly}.
The (intrinsic and geometrical) \textbf{BH}   radiation emission is  regulated by the temperature defined by its surface gravity. These  processes arise from the vacuum fluctuation
 in the regions  in the proximity of (any)  \textbf{BH} horizon (and they  could be interpreted as
a derivation of the  "Unruh effect" in  general relativistic background).}.
Red--shifted or blue-shifted emissions    combine the relativistic and kinematic effects acting on  the emitting particles, and the deformation effects, on the particles motion and photons paths, induced by the central attractor.
In \cite{tobesubmitted}  the red--shift of signals      emitted from    accretion   disks  orbiting  in the spacetime of a Kerr \textbf{BH} was discussed. In more general terms,  conditions for  signals blue--shifiting  and   red--shifting  were   provided   according to different conditions on the emitting radius and the photon  impact parameter.  Emission  from the outer ergoregion was investigated and   the maximum and minimum values  of the red-shift  function were mapped, according to the model parameters.  Frame dragging effects on the emission red-shift  were examined considering, in particular, emission  from co-rotating disks orbiting entirely or in part  in the  spacetime ergoregion.
 More generally,
red-shift of signals, emitted from  the surface of  the  aggregates of axially-symmetric     co--rotating and  counter--rotating orbiting disks, was mapped  discussing  the red-shift    properties  from the emitting particle located on  the tori cusps, the  centers,  the outer edges, and  the accretion flows from the disk   into  central \textbf{BH}, and for particles located on  the "accretion throat" surface.

In this work, we study the red--shift function,  which  expresses  the emissions red--shift, by focusing on particular orbits, called \emph{horizons replica}, associated with the Killing horizons definition in a Kerr \textbf{BH} spacetime.
These orbits  can be very close to the  \textbf{BH}  rotation axis  and    poles,  providing also a pattern for  the analysis in this special  region of  the spinning attractor  spacetime\footnote{
A significant aspect of the horizons replica, in fact,  is that the  inner horizons  replicas  are confined to observation,for sufficiently large view angles, close to the \textbf{BH} equatorial plane, being detectable only close to the \textbf{BH} rotational axis and the poles.}--\cite{2021EPJC...81..258P}.

Replicas are photons circular orbits where  the photon orbital frequency coincide with the horizon angular frequency.
The  horizon replica  concept has been then extended also  to include photons  counter--rotating with respect to the central attractor--\cite{Pugliese:2023liv,Pugliese:2022xry,Pugliese:2021ivl,Pugliese:2020azr,Pugliese:2019age,Pugliese:2021aeb,2024NuPhB100816700P,2021EPJC...81..258P}.
In this analysis we consider both co--rotating and counter--rotating replicas.
The set of replica orbits, of a given \textbf{BH} horizon, generates an object defined  in all Kerr spacetimes (including naked singularities) known as metric Killing bundle, which also plays a role in naked  singularity physics and provides an alternative,  bundle-based definition of the Killing horizons.
The Kerr \textbf{BHs} (inner and outer) horizons  emerge as the envelope surface of the all curves  representing,  in a given plane, called \emph{extended plane},   the  bundles in all Kerr  spacetimes.
A bundle is then defined by the  frequency of the horizon of the corresponding tangent curve in the extended plane, which is the bundle characteristic frequency.

In this work, we will limit our investigation  mostly to the study of Kerr \textbf{BH} spacetimes,  analyzing the properties of the red--shift function for photons on replicas in the extended plane. This will produce a  map  allowing the  identification of  the   red--shifted (or also blue--shifted) signals, as related   to these special orbits.  We  will also consider  accretion disk models, by investigating  the particles emitting from  the disk surface, accretion flow or jet-emission. This will connect
 our  study to two phenomenological  aspects in   the region under consideration. 1. We shall  relate   photons from the  horizons  replicas to the  \textbf{BH}  shadow boundary, at certain view-angles,  considering those replicas, and the associated red--shift function, present   in the \textbf{BH}  photons shell--\cite{2024NuPhB100816700P}.
2.   The investigation will focus also  on   the  replicas  which can be directly connected   to the accretion disk  physics  (also through the definition of light surfaces, constraining, for example, a spinning attractor magnetosphere). Specifically replicas will be connected   to the accretion disks  inner edge, which we study here in the extended plane, using a axially symmetric and geometrically thick Polish Doughnut  disk model--see for example \cite{abrafra}.
To implement this analysis we  will re--phrase      the  main quantities defining the tori morphology of  the axially symmetric, equatorial, general relativistic disks, in the extended plane.

Below  we define   the notion of horizons replica in the Kerr spacetimes  formalizing   more precisely these concepts.
The Kerr horizons are Killing horizons (a Killing horizon is a hypersurface where a Killing vector of the metric becomes null) with respect to the spacetime  Killing vector  $\mathfrak{L}^{a}$  (generator or Killing horizons) which is defined as
$\mathfrak{L}^{a}\partial_{a}=\partial_t+\omega \partial_\phi$, in terms of the Killing fields of  "time translations"  $\xi_{t}=\partial_{t}  $ and   the rotational Killing field $\xi_{\phi}=\partial_{\phi} $ for  axial symmetry of  the Kerr geometry
(the vector $\xi_{t}$ becomes    spacelike in the ergoregion).
(In the  static and spherically symmetric limit of the Kerr geometry, the (Schwarzchild  spacetime), the Killing  vector
$\xi_t$   (now generator of  the event horizon of the Schwarzchild spacetime) is hypersurface-orthogonal.)
Here
$\mathfrak{L}^{a}$  is a null vector at the  spacetime horizons, where  at the horizons $r_\pm$, the  quantities  $\omega=\omega_H^\pm$  are known as frequencies  or the angular velocities of  the horizons.
 The vector  $\mathfrak{L}^a$ is also normal to the horizon, and  it satisfies the condition   $\mathfrak{L}^a\nabla_a\mathfrak{L}^b=\kappa \mathfrak{L}^b$.
Then,
$\nabla_b (\mathfrak{L}_N)$ is normal  to the Killing horizon where
$\mathfrak{L}_N\equiv(\mathfrak{L}_b\mathfrak{L}^b)$.
Quantity $\mathfrak{L}_N$ is constant and zero on the Killing horizon by definition. The \textbf{BH} surface gravity therefore can be defined,  in particular, in terms of the
null generators of the horizon. Here   $\kappa $  (which  can be different on each  generator) is constant on each null   generator of the horizon.
With the horizons replica, we consider the \emph{null} vector  $\mathfrak{L}^{a}$ in \emph{all} points of the Kerr geometries (including naked singularities spacetimes).  The set of geometries and \emph{null} orbits  with the same null frequency  $\omega$ (which is, therefore, also an horizon frequency)  is a  curve on  the  extended plane, where the curves of  all the \textbf{BH} horizons is the envelope surface of all the bundles curves.
(Note these quantities  inherit some of the properties of the Killing vector $\mathfrak{L}$, in particular they are conformal invariants of the metric.)
Hence, in particular,   solutions of the condition $\mathfrak{L}_N=0$, which are special light surfaces,  in a fixed spacetime, determine replicas of  that \textbf{BH} spacetime inner or outer horizon--\cite{Pugliese:2023liv,Pugliese:2022xry,Pugliese:2021ivl,Pugliese:2020azr,Pugliese:2019age,Pugliese:2021aeb,2024NuPhB100816700P,2021EPJC...81..258P}.
 These light surfaces  are solutions  $r_s(\omega,a,\theta)\equiv r:{\mathfrak{L}_N}=0$, for fixed  spin $a$ and  polar angle\footnote{(the \textbf{BH} equatorial plane is at $\theta=\pi/2$)} $\theta$.
That is,  each  replica is also a point of a spacetime  light surface.
It is worth noting that  notion of  (co-rotating)  light--surfaces,  defined as regions
		where the speed of a particle moving purely toroidally equals the light velocity, is  widely applied  in astrophysics as   ``velocity of light" surfaces. They  constrain the analysis of  numerous  phenomena, from jet emission to the  analysis of the \textbf{BHs}  and Pulsars magnetosphere\footnote{Similarly to the stationary observers,  the  \textbf{BH} magnetosphere, rigidly rotating with angular velocity $\omega$,  is divided into
		regions of sub-luminal and super-luminal rotation, depending on the
		sign of  $\mathfrak{L}_N\equiv \mathfrak{L}\cdot\mathfrak{L}$. The separating surfaces, defined via $\mathfrak{L}_N=0$, are  light surfaces
		\cite{Komissarov,Uz05}.
	The outer light surface in the Kerr spacetime  corresponds to the  Pulsar light cylinder only
		asymptotically ($r=1/\omega \sqrt{\sigma}$). (Note that   $\omega< \omega_H^+$ (outer horizon frequency) constraints the  energy extraction from the \textbf{BH}
		via the Blandford--Znajek mechanism\cite{BZ77,Z77}.)} --see for example \cite{Punsly,Camenzind,Mahlmann18,KMCK}.
		Light surfaces can, therefore,  be defined  as the surfaces separating sub--luminal and super--luminal  (co-rotating) motion.

\medskip

More in details the   plan of this article is as follows:
The Kerr spacetime metric is introduced in Sec.\il(\ref{Sec:quaconsta}).
In Sec.\il(\ref{Sec:carter-equa}) we discuss  spacetime symmetries and constants of motion. We introduce the concept of stationary and static observers, light surfaces, horizons replica and Killing bundles.
In this section  we define also  the  inversion surfaces, constituted by points  where there is   $\dot{\phi}=0$ that is:  flow, having an initial counter--rotating component, due to the Lense--Thirring effect of the Kerr central attractor, reverses the rotation orientation (the toroidal component of the velocity in the proper frame),  along its trajectory.
The inversion points are similarly defined for photon paths. Therefore, the  flow  is characterized by the presence inversion points, defined by the condition  $u^{\phi}=\Omega=0$, on the   axial component of the  velocity, $u^{\phi}$,  and the  relativistic angular velocity, $\Omega$ (related to the distant static observer).
Finally we provide the   Carter geodesics equations of motion.
The Kerr \textbf{BH} photons shell will be  introduced in Sec.\il(\ref{Sec:shadows}), studying  the shell  inner and outer boundaries  and the internal regions  filled with  co-rotating  and counter-rotating photon  orbits and, for the counter--rotating photons we define  the corresponding  inversion surface in the photons shell.
In  Section\il(\ref{Appendix:INNer-Edge}) we introduce the tori models and their constraints  considered in this analysis.
Sec.\il(\ref{Sec:redshift-replicas-First}) constitutes the second part of this  work, with  the investigation of the
red-shift  function $g_{red}$ properties in relation to the horizons replicas.
In Sec.\il(\ref{Sec:redhsift}) we define the red-shift  function $g_{red}$.
Red-shift function from  the horizons replicas will be  addressed in Sec.\il(\ref{Sec:from-de-replicas}).
As the  limiting angular velocities $\omega_\pm$ of the stationary observers  define the collections of replicas, and  constrain the $g_{red}$ definition,  we will  investigate, in   Sec.\il(\ref{Sec:limitng-velocitues}), the properties  of  these limiting functions.
In particular,  in Sec.\il(\ref{Sec:photon-shells}), the extreme points of the limiting frequencies $\omega_\pm$  are studied  in  relation with the \textbf{BH} photons shell boundaries.
The red-shift function on the  horizons replica,  defined on the  photons shell boundaries, is discussed in Sec.\il(\ref{Sec:redffhidft-shellsreplicas}).
Co--rotating and counter--rotating horizons  replicas are characterized  in more details in Sec.\il(\ref{Sec:co--rotating-counter-rpliAs}),  and in Sec.\il(\ref{Sec:absidered-extended-plane}) we  discuss more generally   the properties of the red-shift function in relation to the horizons replicas.
 Discussions and conclusions follow in  Sec.\il(\ref{Sec:Conclusions}).

%%%%%%%%%%%\inp
\section{The Kerr \textbf{BH}  geometry }\label{Sec:quaconsta}
The   Kerr  spacetime line element   reads\footnote{In the Boyer-Lindquist (BL)  coordinates
\( \{t,r,\theta ,\phi \}\) and we adopt the
geometrical  units $c=1=G$ and  the $(-,+,+,+)$ signature, Latin indices run in $\{0,1,2,3\}$.  The radius $r$ has unit of
mass $[M]$, and the angular momentum  units of $[M]^2$, the velocities  $[u^t]=[u^r]=1$
and $[u^{\phi}]=[u^{\theta}]=[M]^{-1}$ with $[u^{\phi}/u^{t}]=[M]^{-1}$,
$[u_{\phi}/u_{t}]=[M]$ and an angular momentum per
unit of mass $[L]/[M]=[M]$.},
%r
%
\bea \label{alai}&& ds^2=-\left(1-\frac{2Mr}{\Sigma}\right)dt^2+\frac{\Sigma}{\Delta}dr^2+\Sigma
d\theta^2+\left[(r^2+a^2)+\frac{2M r a^2}{\Sigma}\sin^2\theta\right]\sin^2\theta
d\phi^2
-\frac{4rMa}{\Sigma} \sin^2\theta  dt d\phi,
\\&&\label{Eq:delta}
\mbox{where}\quad
\Delta\equiv a^2+r^2-2 rM\quad\mbox{and}\quad \Sigma\equiv a^2 (1-\sin^2\theta)+r^2.
\eea
 Parameter  $a=J/M\geq0$ is the metric spin, where  total angular momentum is   $J$  and  the  gravitational mass parameter is $M$ (the {ADM  and Komar}  mass).
  Here there is   $r\in[0,+\infty)$, $t\in [0,+\infty)$, $\theta\in[0,\pi]$ and $\phi\in[0,2\pi]$.
 The Kerr naked singularities (\textbf{NSs}) have  $a>M$.
  A Kerr \textbf{BH} is defined  by the condition $a\in[0,M]$. The non-rotating   case $a=0$ is the   Schwarzschild \textbf{BH} solution. The extreme Kerr \textbf{BH}  has dimensionless spin $a/M=1$.
 The \textbf{BH}  outer and inner  horizons are
\bea
 \quad r_{\pm}\equiv M\pm\sqrt{M^2-a^2}, \quad \mbox{where}\quad
   r_-\leq r_+.
\eea
The ergoregion is $[r_\epsilon^-,r_{\epsilon^+}]$, the  outer ergoregion of the spacetime is  $]r_+,r_{\epsilon}^+
]$, where the  outer and inner  ergosurfaces are the radii    $r_{\epsilon}^\pm$ where
\bea\label{Eq:sigma-erg}
r_{\epsilon}^{\pm}\equiv M\pm \sqrt{M^2- a^2 (1-\sigma)}\quad\mbox{with}\quad \sigma\equiv \sin^2\theta\in [0,1].
%&&\nonumber
\eea
(Unless otherwise specified,  in the following,  with  ergoregion and ergosurface we will always refer to the  outer ergoregion and the  outer ergosurface.)
 In  the equatorial plane, where $\sigma=1$,  there is
  $r_{\epsilon}^+=2M$. The equatorial plane, $\sigma=1$,  is the symmetry plane for the metric, and the constant $r$
orbits on this plane are circular.
%and
Where appropriate,  to easy the reading of   complex expressions, we will  use, in the following,
dimensionless variables with $M=1$ (where $r\rightarrow r/M$  and $a\rightarrow a/M$).
\subsection{Symmetries and constants of motion}\label{Sec:carter-equa}
We introduce $p^a=dx^a/d \tau\equiv u^a\equiv\{ \dot{t},\dot{r},\dot{\theta},\dot{\phi}\}$  for   the geodesic  tangent four-vector, where $\tau$ is an affine parameter, normalized so  that $p^ap_a=-\mu^2$,
and  $\mu$ is the rest mass  of the test particle where  a null geodesics has $\mu=0$ \footnote{For the seek of simplicity,    notation $\dot{q}$ or $u^a$ for  photons and particles  has been adopted. The context should avoid any  misunderstanding.}.

\medskip

\emph{Static  observers}, having   four-velocity   $\dot{\theta}=\dot{r}=\dot{\phi}=0$,
cannot exist inside the ergoregion. Whereas, trajectories   $\dot{r}\geq0$, including photons  crossing the outer ergosurface  and escaping outside
in the region $r\geq r_{\epsilon}^+$, are possible.

\medskip

\emph{Stationary observers} have
     four-velocity  $u^a$  which is
linear combination of the two Killing vectors, $\xi_{\phi}$ and $\xi_{t}$ that is the Kerr geometry  rotational  Killing field   $\xi_{\phi}\equiv \partial_{\phi}$,
  and  the Killing field  $\xi_{t}\equiv \partial_{t}$
representing the stationarity of the  background.
Therefore, for the stationary observers  there is
$
u^a= \gamma (\xi_{t}^a+\omega \xi_{\phi}^a$),  where
 $\gamma$ is a normalization factor and $d\phi/{dt}={u^{\phi}}/{u^t}\equiv\omega$.
The  quantity $\omega$  is the orbital frequency (angular velocity) of the stationary observers,  bounded in the range $\omega\in]\omega_-,\omega_+[$, where  the  limiting frequencies,  $\omega_{\pm}$,  are photon orbital angular velocities.
Surfaces $(r,\theta): \omega_\pm(r,\theta)=\omega$ define the stationary observers \emph{light surfaces}.
The  angular velocities  $\omega_H^{\pm}$ of the \textbf{BH} Killing horizons are
\bea\label{Eq:omegahmo}
\omega_H^+\equiv \omega_{\mp}(r_+)=\frac{r_-}{2a},\quad\mbox{and}\quad   \omega_H^-\equiv \omega_{\mp}(r_-)=\frac{r_+}{2a}
\eea
where $\omega_H^\pm$ are the outer and inner \textbf{BH} horizons angular velocity respectively.
At fixed $\omega= \omega_H^\pm$, all the (photon) circular orbits defined by the conditions $(r\neq r_\pm,\theta): \omega_\pm=\omega= \omega_H^\pm$, define special orbits of the spacetime where the relativistic angular frequency is coincident with a \textbf{BH} horizons angular frequency, these orbits are called \emph{horizons replicas}--\cite{2021EPJC...81..258P,2024NuPhB100816700P}.
The collection of all orbits at fixed $\theta$ and $\omega$, in the plane $(a,r)$,
constitutes a metric \emph{Killing bundle (or metric bundle)}, and  $\omega$ is the metric bundle angular velocity.  We will consider more closely the notion of metric Killing bundles and replica orbits in Sec.\il(\ref{Sec:from-de-replicas}) and, in particular, in Sec.\il(\ref{Sec:co--rotating-counter-rpliAs}).
Quantity $\omega_H^{\pm}$, expressing the \textbf{BH} rigid rotation, regulates  the \textbf{BH} thermodynamic laws\footnote{For example
the variation of the   \textbf{BH} irreducible  mass
$
\delta M_{irr}\geq 0$  constrained  by $ (\delta M-\delta J \omega^+_H)\geq 0 $ for a variation of mass and momentum.}.
 The null vector fields  $\mathfrak{L}_H^{\pm}=\xi^t +\omega_H^{\pm} \xi^{\phi}$ in fact define the
horizons of the Kerr \textbf{BH}  as   Killing horizons\footnote{The Kerr  horizons can,  therefore, be interpreted as  {null} hypersurfaces generated by the flow of a Killing vector,
whose {null} generators coincide with the orbits of an
one-parameter group of isometries, i.e., in general,    there exists a Killing field $\mathfrak{L}$, which is normal to the null surface.} (being generators of Killing event  horizons)\footnote{For $a=0$, the horizon  of the  Schwarzschild \textbf{BH}  is a Killing horizon with respect to the
Killing field
$\xi^t$, consequently   the
event, apparent, and Killing horizons   coincide.}.

\medskip

Constants of geodesic  motions, associated to the spacetime Killing field $(\xi_t,\xi_\phi)$ are
quantities $(\Em, \La)$ defined as
\bea&&\label{Eq:EmLdef}
\Em=-(g_{t\phi} \dot{\phi}+g_{tt} \dot{t}),\quad \La=g_{\phi\phi} \dot{\phi}+g_{t\phi} \dot{t}, %
\eea
 representing  the  energy of the test particle
 related to the static observers at infinity and the axial component of the angular momentum  of a test    particle following
timelike geodesics, respectively.
There is
\bea
 u^\phi=-\frac{\mathcal{E} g_{t\phi}+g_{\phi\phi} \La}{g_{t\phi}^2-g_{\phi\phi} g_{tt}};\quad u^t=\frac{\mathcal{E} g_{\phi\phi}+g_{t\phi} \La}{g_{t\phi}^2-g_{\phi\phi} g_{tt}}, \mbox{and}\quad \Omega=\frac{u^\phi}{u^t}=\frac{\dot{\phi}}{\dot{t}}=-\frac{\mathcal{E} g_{t\phi}+g_{tt} \La}{\mathcal{E} g_{\phi\phi}+g_{t\phi} \La},
 \eea
where  $\Omega$ is the relativistic angular velocity.
We  introduce also  the specific  angular momentum $\ell$ (called also {impact parameter})
 \bea&&\label{Eq:flo-adding}
\ell\equiv\frac{\La}{\Em}=-\frac{g_{\phi\phi}u^\phi  +g_{\phi t} u^t }{g_{tt} u^t +g_{\phi t} u^\phi}.
\eea
The relativistic angular velocity $\Omega(\ell)$  can be expressed as
\bea\label{Eq:omegadefinl}
&&
\Omega(\ell)=-\frac{g_{t\phi}+g_{tt} \ell }{g_{\phi\phi}+g_{t\phi} \ell }.
\eea
With  $a>0$,    counter-rotation  (co-rotation) is defined by $\ell a<0$ ($\ell a>0$).

\medskip

An \emph{inversion surface}  is a  closed   surface,   defined by condition $\Omega=0$ or, equivalently,   $\dot{\phi}=0$.
This surface, defined by the points  where $\dot{\phi}=0$,
  in centered on the  Kerr \textbf{BH} and  is located out of the   outer ergosurface--\citep{2022MNRAS.512.5895P,2024EPJC...84.1082P}.
In general, within the condition $\dot{\phi}=0$,  there is
 $\ell=\ell_\Ta\equiv \left.-\frac{g_{t\phi}}{g_{tt}}\right|_\Ta$, and the inversion surface is parameterized with the constant of motion $\ell$, depending on $\ell$ only, and is defined, in the Kerr \textbf{BH} spacetimes,  for counter--rotating motion ($a\ell<0$).
(Quantities $\mathrm{Q}$, related to the   inversion surfaces, will be indicated with the notation $\mathrm{Q}_\Ta$.)

\begin{figure}
\centering
\includegraphics[width=8cm]{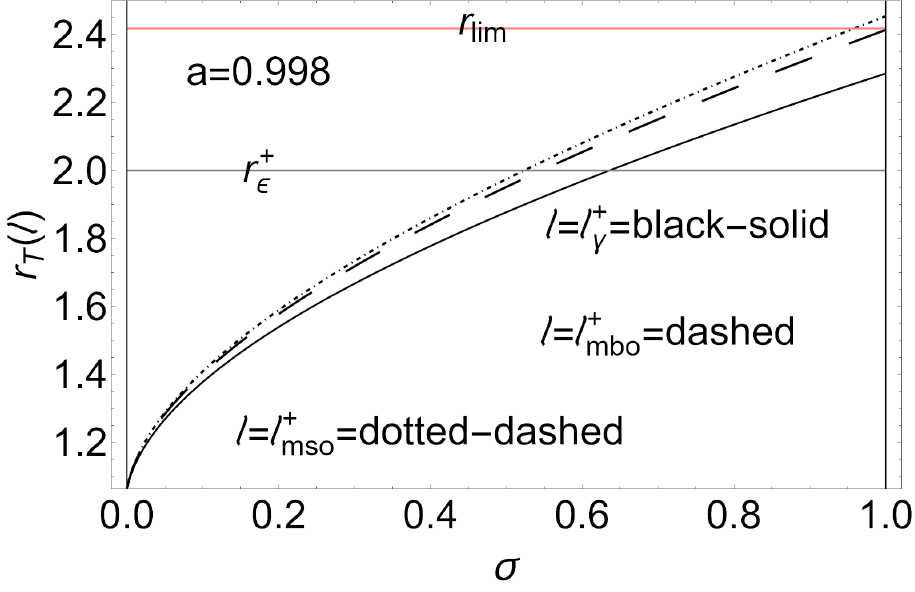}
\includegraphics[width=8cm]{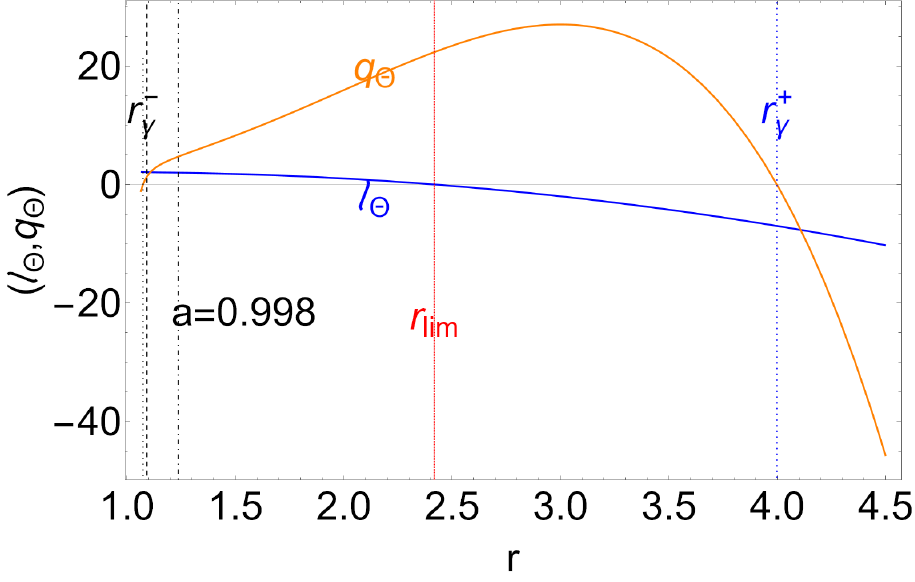}
\includegraphics[width=8cm]{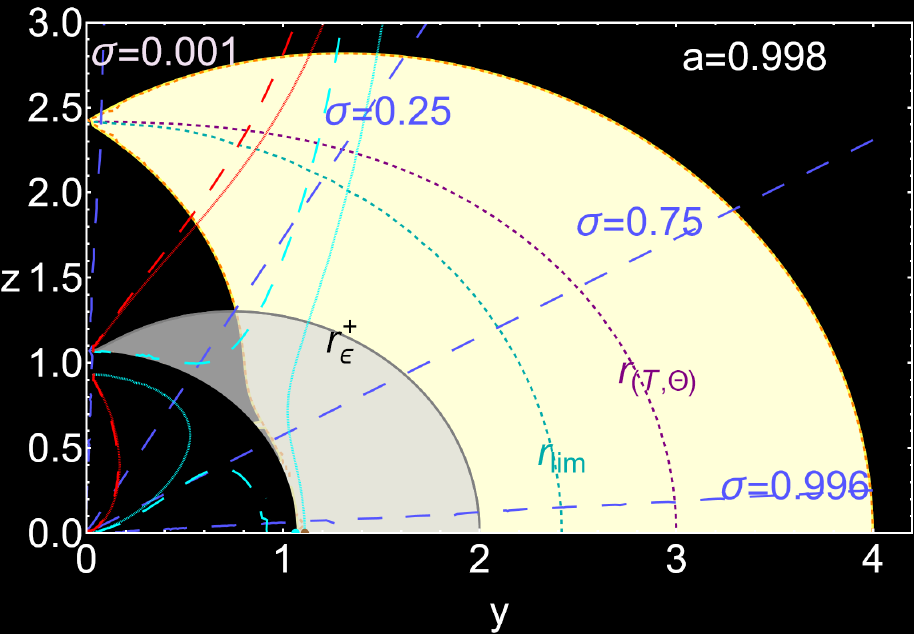}
\includegraphics[width=8cm]{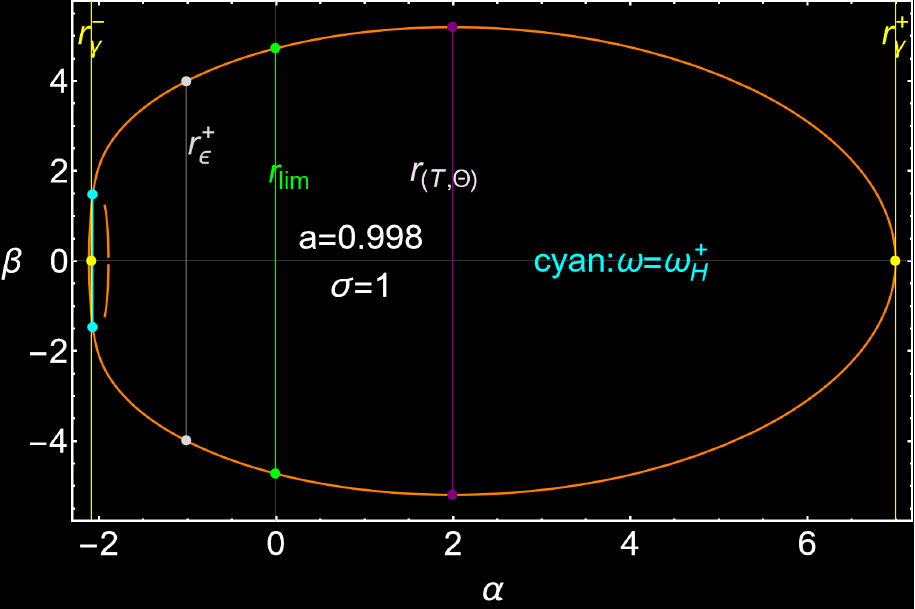}
\caption{Upper left panel: inversion radius $r_\Ta$ of Sec.\il(\ref{Sec:carter-equa}) as function of the angle $\sigma$ for \textbf{BH} spin $a=0.998$ and different specific angular momentum $\ell=\ell^\pm$ (calculated on the notable  geodesic radii) as signed on the panels, for co--rotating $(-)$ and counter--rotating $(+)$ particles. Radius $r_{\lim}$  of Eq.\il(\ref{Eq:rlimite}) is also shown. Upper right panel: photon paths parameters $(q_\Theta,\ell_\Theta)$ of Eqs\il(\ref{Eq:lqsoluzione}) defining the photons shell as function of the radius $r$. Radii $r_\gamma^\pm$ are the counter--rotating and co--rotating circular photon orbits. Bottom left  panel: photons shell (yellow region) of the \textbf{BH} (inner black region) with spin $a=0.998$. Gray region is the outer ergoregion, radius $r_\epsilon^+$ is the outer ergosurface. Radii $r_{\lim}$ where $\ell_p=0$ and $r_{(\Ta,\Theta)}$, where $\dot{\phi}=0$, are also shown. Dashed purple lines are angles $\sigma=$constant. The panel shows   the horizon replica surfaces, i.e. the surfaces $\omega_\pm=\pm\omega_H^\pm$, where $\omega_\pm$ are the stationary observer limiting angular velocity, and $\omega_H^\pm$ are the outer and inner horizon angular velocity respectively.  Red is for counter--rotating replicas,
cyan for co--rotating replicas,
dashed lines  for the outer horizon replicas,
solid lines for inner horizon replica. Bottom right panel:  \textbf{BH} shadow boundaries  in the celestial plane $(\alpha,\beta)$ for the view angle $\sigma=1$ (the equatorial plane)  relative to the right panel. For example, the gray points  correspond  to the outer ergosurface, green points to $\alpha=0$ (corresponding to the $r_{\lim}$), yellow points to the outer and inner boundary of the photons shell, corresponding to the counter--rotating and co--rotating  photon circular orbits $r_\gamma^\pm$, respectively. }\label{Fig:Plotelemtempor}
\end{figure}

The presence of an inversion surface can be interpreted as an effect of frame dragging on the motion (of photon and particles) with a counter-rotating component.
 Free particles arriving, for example, towards the  \textbf{BH} with $\ell<0$ and $\dot{\phi}<0$,  enter the ergoregion with (constant  of motion)  $\ell<0$ and with $\dot{\phi}>0$,
and there must be an  inversion point a $r=r_\Ta>r_\epsilon^+(a,\sigma,r): \dot{\phi}_\Ta=0$.
 The inversion surface has   radius $r_\Ta(\ell,\sigma,a):\ell=\ell_\Ta $.
  (In Figs\il(\ref{Fig:Plotelemtempor})--upper left panel we show the inversion radius $r_\Ta$  as function of the angle $\sigma$, for \textbf{BH} spin $a=0.998$ and different specific angular momentum $\ell<0$.)
An additional  constant of motion of the Kerr geometry (related to a further Killing field defined by the principal null vectors)   is the Carter constant of motion  $\Qa$, regulating the non--equatorial motion,  where
the complete geodesic equations  of motion are as follows:
\bea&&\label{Eq:eqCarter-full}
 \dot{t}=\frac{1}{\Sigma}\left[\frac{P \left(a^2+r^2\right)}{\Delta}-{a \left[a \Em\sigma-\La\right]}\right],\quad
\dot{r}=\pm \frac{\sqrt{R}}{\Sigma};\quad \dot{\theta}=\pm \frac{\sqrt{\Theta}}{\Sigma},\quad\dot{\phi}=\frac{1}{\Sigma}\left[\frac{a P}{\Delta}-\left[{a \Em-\frac{\La}{\sigma}}\right]\right];
\eea
and there is
\bea\label{Eq:eich}
& P\equiv \Em \left(a^2+r^2\right)-a \La,\quad & R\equiv P^2-\Delta \left[(\La-a \Em)^2+\mu^2 r^2+\Qa\right],\quad \Theta\equiv \Qa-
(\cos\theta)^2 \left[a^2 \left(\mu^2-\Em^2\right)+\left(\frac{\La^2}{\sigma}\right)\right],
\eea
 (see \cite{Carter}).
\subsection{\textbf{BH} photons shell}\label{Sec:shadows}
In this section  we introduce the \textbf{BH}  photons shell and notion of \textbf{BH} shadow  and shadow boundary.

 At fixed  spin $a$ and view angle  $\sigma$, the   shadow boundary  appears as a closed curve in  the   celestial coordinates  \footnote{The  celestial coordinates are
$
\alpha=-{\ell}/{\sqrt{\sigma}},\quad
\beta=\pm\sqrt{q-q_c},\quad\mbox{where}
\quad
 q_{c}\equiv{(1-\sigma ) \left(\ell^2-\ell_c^2 \right)}/{\sigma },\quad \mbox{and}\quad \ell_c^2\equiv a^2 \sigma
$-- here $q\equiv \Qa/\Em^2$ see \cite{Bardeen1973}.} %, read % in
 $(\alpha,\beta)$--(see Figs\il(\ref{Fig:Plotelemtempor})--bottom right panel).
The points of the shadow boundary are unstable (geodesic) photon orbits  solution of
\bea\label{Eq:radial-condition}
R=\partial_r R=0,\quad\mbox{with}\quad \partial_r^2R>0.
\eea
These bound null geodesics   are contained in a shell, \emph{photons shell}, of (unstable) bound photon orbits
surrounding the  central \textbf{BH}
(see for further details  also \cite{2024EPJP..139..531P,2024ApJS..275...45P,2025JHEAp..4700368P}).
%%%
%%

\begin{figure}
\centering
\includegraphics[width=8cm]{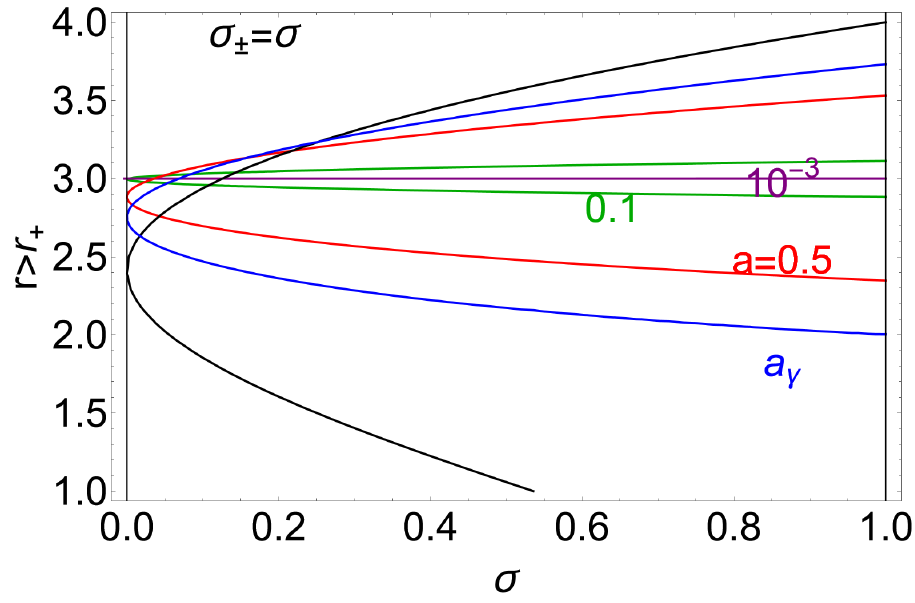}
\includegraphics[width=8cm]{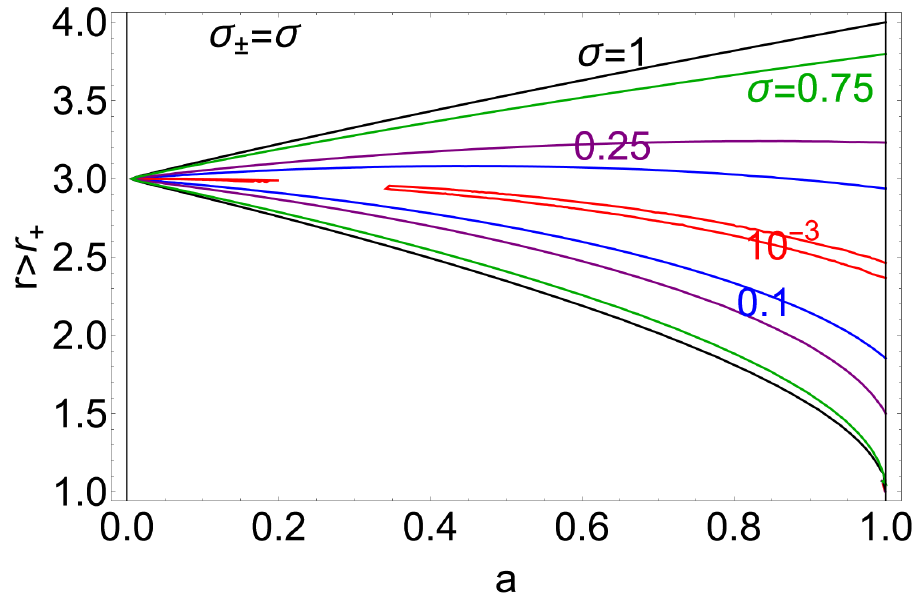}
\caption{Photons shell boundaries  angles $\sigma=\sigma_\pm$  of Eq.\il(\ref{Eq:sigmapm}) in the plane $(r,\sigma)$ for different \textbf{BH} spins (left panel), and in the plane $(r,a)$ for different $\sigma$ (right panel) as shown on the panel. Spin $a_\gamma=1/\sqrt{2}$.}\label{Fig:PlottuMinaD}
\end{figure}

More specifically, let us introduce   the constant of motion $q$, defined  from the Carter constant as
$
q\equiv {\Qa}/{\Em^2}$.
Solving $R =
 0$  and $R' =
  0$ of  Eq.\il(\ref{Eq:radial-condition})  we obtain the functions
  \bea\label{Eq:lqsoluzione}
\ell=\ell_\Theta\equiv\frac {a^2 (r + 1) + (r - 3) r^2} {a(1- r)},
\quad\mbox{and}\quad
q=q_\Theta\equiv - \frac {r^3\left[(r - 3)^2 r - 4 a^2 \right]} {a^2 (r - 1)^2}%\in[0,27]
\eea
Photons parameters $(q_\Theta,\ell_\Theta)$,  defining the photons shell as functions of the radius $r$, are shown in Figs\il(\ref{Fig:Plotelemtempor})--upper right panel, as function of the radius $r$, for the \textbf{BH} spin $a=0.998$.
%
%%%
Using  $q_\Theta$ and $\ell_\Theta$,
from equation
$ \Theta=
 0$, we obtain  the  photons shell  angles
\bea\label{Eq:sigmapm}&&
\sigma_\pm\equiv %1-\cos^2\theta_\pm=
\frac{r^2 \left(a^2+r^2-3\right)}{a^2 (r-1)^2}-2 \sqrt{\frac{r^2 \left[a^2+(r-2) r\right] \left[a^2+r^2 (2 r-3)\right]}{a^4 (r-1)^4}}+\frac{1}{(r-1)^2},
%\\
%&&\mbox{w
\eea
 defining the photons shell boundaries.
The boundaries $\sigma_\pm$ are shown as functions of  the angle $\sigma$ and spin $a$ in
Figs\il(\ref{Fig:PlottuMinaD}).
 It is  clear  that   $ \sigma_\pm=1$, on the co--rotating and counter--rotating  photon circular orbits  $r_\gamma^\mp$:
 \bea
r_\gamma^-\equiv 2 \left[\sin \left(\frac{1}{3} \sin ^{-1}\left(1-2 a^2\right)\right)+1\right],
\quad r_\gamma^+\equiv4 \cos ^2\left[\frac{1}{6} \cos ^{-1}\left(2 a^2-1\right)\right],
 \eea
respectively.
The photons shell is split in two   regions of co--rotating and counter--rotating photon  orbits, bounded by a radius  $r_{\lim}\in]r_\gamma^-,r_\gamma^+[$, located out of the outer  ergoregion where
\bea\label{Eq:rlimite}
&&
r_{\lim}\equiv -\frac{a^2-3}{\sqrt[3]{3} \sqrt[3]{S_r}}+\frac{\sqrt[3]{S_r}}{3^{2/3}}+1,\quad\mbox{with}\quad
S_r\equiv 9(1- a^2)+\sqrt{3} \sqrt{a^2 \left(a^4+18 a^2-27\right)}
\eea
where there is
$\ell_{\Theta}(r_{\lim})=0$.
Radius $r_{\lim}$  is shown in Figs\il(\ref{Fig:Plotelemtempor}) in the photons shell of  a \textbf{BH} with spin $a=0.998$.
In general the \textbf{BH}   photons shell is in the ergoregion in the co--rotating range $]r_{\gamma}^-,r_{\lim}]$, for $a\geq a_\gamma=1/\sqrt{2}$ where,  in this spacetime there is $a_\gamma: r_\gamma^-=r_\epsilon^+$, on the equatorial plane.
The counter--rotating region of the photons shell is split    an  inversion surface,  where for the photon impact parameter there is $\ell_\Ta=\ell_\Theta$,  and defined by the radius,
\bea&&\label{Eq:rthetat-definition}
r_{(\Ta,\Theta)}\equiv-\frac{\sqrt[3]{2} S_t}{3 \sqrt[3]{S_{\text{ttt}}}}+\frac{\sqrt[3]{S_{\text{ttt}}}}{3 \sqrt[3]{2}}+1,\quad \mbox{with} \\&&\nonumber  S_{\text{ttt}}\equiv S_{\text{tt}}+\sqrt{4 S_t^3+S_{\text{tt}}^2},
\quad S_{\text{tt}}\equiv 54 (a^2 \sigma - a^2+1),\quad \mbox{and}\quad S_{\text{t}}\equiv 3 (a^2- a^2 \sigma -3),
\eea
 where   there is
  $r_{(\Ta,\Theta)}\in]r_{\lim},r_\gamma^+]$.
Radius $r_{(\Ta,\Theta)}$ is shown  in Fig.\il(\ref{Fig:Plotelemtempor})  in the  photons shell (yellow region) of the \textbf{BH} with spin $a=0.998$ and on the shadow boundary for $\sigma=1$.  All the photons  with impact parameter $\ell<0$,  orbit  with $\dot{\phi}>0$ in the region  $r<r_{(\Ta,\Theta)}$.
(In the following we sometime use the subscript $p$ in  $\ell_p$  (or $l_p$) to indicate the  photon impact parameter  distinguished from a fluid particle parameter $\ell$.)

%%%%%%%%%%%%%%%\i

\section{Constraints of  accretion disk models}\label{Appendix:INNer-Edge}
In this Section we discuss constraints on accretion disk models as determined by the spacetime geodesic structure.
In our analysis we   relate some aspects of the red-shifed (or blue--shifted) signals emitted from  replicas, to the accretion physics,  under the  conditions in which the signal is  emitted by a replica coincident with the inner edge radius  of the disk, for example, or the center of a very thin disk orbiting the central \textbf{BH}.
We will  consider the  radii
$\{r_\gamma^\pm,r_{mbo}^{\pm},r_{mso}^{\pm}\}$ of the background  geodesic structure,    for co-rotating $(-)$ and counter-rotating $(+)$ motion.
Radii  $r_{mso}^{\pm}$ are the marginal  circular stable orbits,   and radii $r_{mbo}^{\pm}$ are marginal bounded circular orbits.
The inner edge of a general relativistic, equatorial, axially symmetric   accretion disk  is located on the equatorial plane,   in the orbital range  $[r_{mbo}^\pm, r_{mso}^\pm]$,   for  counter--rotating and co-rotating fluids respectively\footnote{The inner edge of a small disk and  of thin, ring--like, disks  are  close to the marginally stable orbit   $r=r_{mso}^\pm$.}
We  will focus on   disks and tori, modelled by  axially symmetric, general relativistic, equatorial geometrically thin disks or  geometrically thick accretion disks, orbiting the equatorial plane of a central   Kerr \textbf{BH}.
   Polish Doughnut (\textbf{PD})   is a  particularly remarkable example of tori models,  strongly constrained by the spacetime structure, and   featuring a barotropic, a general relativistic hydro-dynamic model of  a geometrically thick, opaque,  stationary and axially--symmetric  accretion disk--(see for example \cite{abrafra}).
Because of the symmetries,
the   continuity equation for the   fluid
is  identically satisfied, and the  fluid dynamics  is  governed by the Euler equation only. Furthermore,  it is often practical   to parametrize the   tori  by   fluid specific angular momentum $\ell$ constant  and conserved  in the disk. This assumption  affects the force balance in the disk, and  the
 toroids are regulated by  the     effective potential function only,
encoding the background   geometry (through  the parameter $a$), and the centrifugal effects, through the fluid specific angular momentum $\ell$\footnote{These effects are based on the Boyer  condition  of the  equilibrium configurations  of rotating rigid bodies  in general relativity  \cite{Boyer} and the application of the von Zeipel conditions. In these models  the surfaces of constant  relativistic velocity $\Omega$ and of constant specific angular momentum $\ell$ coincide. Therefore,  we   use the rotation law $\ell=\ell(\Omega)$   independently by  the equation of state\cite{M.A.Abramowicz,Chakrabarti0,Chakrabarti}.}. Tori are mostly determined by the properties of the pressure gradients on their equatorial plane.
 Within these conditions, toroids are   equi-pressure (and equi--potential) surfaces, and could be   proto-jets (open  surfaces with a cusp on equatorial plane  with matter funnels along the \textbf{BH} rotational axis) or  quiescent or  cusped tori.
Therefore,   minima of the effective potential as function of the radius on the equatorial plane  are the maxima of  the pressure in the torus (torus center $r_{center}$). Whereas, the minimum of pressure are the maximum points of the effective potential as function of the radius on the equatorial plane,  and are instability points correspondent to the    surfaces cusps
$r_{\times}$, for closed cusped tori or  cusps  $r_j$ for proto-jets--see \cite{2024EPJP..139..531P,
2024ApJS..275...45P,
2025JHEAp..4700368P}. The parameter $\ell$ is   determined by the  Keplerian specific angular momentum
distribution $\ell^\pm(r;a)\equiv \La/\Em$   on the equatorial plane,    that is
\bea\label{Eq:definlpm}
\ell^{\mp}(r)\equiv\left.\frac{\La}{\Em}\right|_{\sigma=1}\equiv  \frac{a^3+a r (3 r-4)\mp\sqrt{r^3 \left[a^2+(r-2) r\right]^2}}{a^2-(r-2)^2 r}. %\frac
\eea
Note, for  $a>0$ there is $\ell^->0$  for co-rotating tori and, viceversa, there is   $\ell^+<0$, with $a>0$, for counter-rotating tori.
On the other hand, for  $a<0$ there is $\ell^->0$  for counter-rotating tori, and  $\ell^+<0$  for co-rotating tori.
For counter--rotating and co- rotating tori, the disks cusps and centers are solutions of $\ell^\pm(r;a)=\ell=$constant  for $r=\{r_{\times}^\pm,r_{center}^\pm\}$,   and each toroid is parameterized by a constant  parameter   $\ell=\ell(r_{center})$, (there is $\ell=\ell(r_{center})=\ell(r_\times)$ or $\ell=\ell(r_{center})=\ell(r_j)$).
Therefore, solving   $\ell^\pm(r;a)=\ell=$constant  we obtain
\bea
&&r_{\times}^\pm=\frac{\mathfrak{C}_\star-\mathfrak{G}_\ast }{12}\quad\mbox{and}\quad r_{center}^\pm=\frac{\mathfrak{C}_\star+\mathfrak{G}_\ast }{12},
\eea
 \mbox{where}
 \bea
&&\nonumber
 \mathfrak{C}_\star\equiv \sqrt{9  \ell^4+\frac{48 \sqrt[3]{2}
\mathfrak{V}_{\triangleright}}{
\mathfrak{B}_{\oplus}}+6\ 2^{2/3}
\mathfrak{B}_{\oplus}-48  \mathfrak{F}_{\diamond}},\quad
\mathfrak{B}_{\oplus}\equiv \sqrt[3]{\mathfrak{I}_{\oslash}+\sqrt{\mathfrak{I}_{\oslash}^2-4 (4 \mathfrak{V}_{\triangleright})^3}},
\\\nonumber
&&\mathfrak{G}_\ast \equiv\sqrt{6} \sqrt{\frac{3 \sqrt{3} \left(\ell^6+32 \mathfrak{M}_{\ominus}-8 \mathfrak{F}_{\diamond} \ell^2\right)}{\mathfrak{C}_\star/\sqrt{3}}-16 \mathfrak{F}_{\diamond}+3  \ell^4-\frac{8 \sqrt[3]{2} \mathfrak{V}_{\triangleright}}{
\mathfrak{B}_{\oplus}}-2^{2/3}
\mathfrak{B}_{\oplus}};
\\&&\nonumber
\mathfrak{M}_{\ominus}\equiv  \mathfrak{F}_{\diamond}+ a \ell-\ell^2,\quad  \mathfrak{V}_{\triangleright}\equiv\left(\mathfrak{F}_{\diamond}+ a^2-\ell^2\right)^2,\quad \mathfrak{F}_{\diamond}\equiv a^2-3 a  \ell+2 \ell^2;
\\\nonumber
&&\mathfrak{I}_{\oslash}\equiv-(a- \ell)^2 \left[128 a^4-320 a^3  \ell-9 a^2 \left(3  \ell^4-8 \ell^2+48\right)+8 a  \ell \left(13 \ell^2+108\right)+16 \ell^2 \left(\ell^2-27\right)\right].
\eea
There is  $r_{center}^\pm>r_{mso}^\pm$, and $r_{inner}^\pm\in]r_{\gamma}^\pm,r_{mso}^\pm[$, where
$r_{inner}^\pm\in\{r_j^\pm,r_\times^\pm\}$.
It is also convenient to introduce the radii  $r_{\mathrm{[mbo]}}^{\pm}$ and $r_{\mathrm{[\gamma]}}^{\pm}$, solutions of
\bea&&\label{Eq:def-nota-ell}
r_{\mathrm{[mbo]}}^{\pm}:\;\ell^{\pm}(r_{\mathrm{mbo}}^{\pm})=
 \ell^{\pm}(r_{\mathrm{[mbo]}}^{\pm})\equiv \mathbf{\ell_{\mathrm{mbo}}^{\pm}},\quad\mbox{and}\quad
  r_{[\gamma]}^{\pm}: \ell^{\pm}(r_{\gamma}^{\pm})=
  \ell^{\pm}(r_{[\gamma]}^{\pm})\equiv \ell_{\gamma}^{\pm} %\quad  r_{(\mathcal{M})}^{\pm}:\ell^{\pm}(r_\Mie^{\pm})\equiv \ell_\Mie^{\pm}=\ell^{\pm}(r_\Mie^{\pm}),
\eea
respectively,
{where}   there is
\bea
 r_{\gamma}^{\pm}<r_{\mathrm{mbo}}^{\pm}<r_{\mathrm{mso}}^{\pm}<
 r_{\mathrm{[mbo]}}^{\pm}<
  r_{[\gamma]}^{\pm}.
  \eea
(In general  the notation $\mathrm{Q}_\bullet$  is for any quantity evaluated at radius $r_\bullet$.)
For  $\mp \ell^{\pm}\in[\mp \ell_{mso}^{\pm},\mp\ell_{mbo}^{\pm}[$ there are quiescent (i.e. not cusped) and cusped tori.
The cusp is   $r^{\pm}_{\times}\in]r^{\pm}_{mbo},r^{\pm}_{mso}]$.
The  torus  center is  $r^{\pm}_{center}\in]r^{\pm}_{mso},r^{\pm}_{[mbo]}]$.  For
 $\mp\ell^{\pm}\in[\mp \ell_{mbo}^{\pm},\mp\ell_{\gamma}^{\pm}[ $ there are quiescent  tori  and proto-jets with cusp  in
 $r_{j}^{\pm}\in]r_{\gamma}^{\pm},r_{mbo}^{\pm}]$  and
 the  torus center  is in   $r_{center}^{\pm}\in]r_{[mbo]}^{\pm},r_{[\gamma]}^{\pm}]$.
Finally, for $\mp \ell^{\pm}\geq\mp\ell_{\gamma}^{\pm}$ there are only quiescent  tori with
 center in   $r^{\pm}_{center}>r_{[\gamma]}^{\pm}$.
Some co--rotating configurations are partially or completely  included in the ergoregion, or in the \textbf{BH}  photons shell.
For large \textbf{BH} spins,  the photons shell is included  in the \textbf{BH} ergoregion, and the material swallowed by the \textbf{BH}  must have $\dot{\phi}>0$ (and could have  $\ell_p<0$).
 A counter--rotating disk orbits  always out of the photons shell.
 The co--rotating tori can orbit in the   \textbf{BH} photons shell (and ergoregion) and can be entirely embedded in the {photons} shell  for sufficiently high spins.
%
% %
The photons shell  can always host   co--rotating proto-jets cusps and the  inner region,  $]r_{inner}^-,r_{center}^-[$,  of the co--rotating quiescent  tori with   $\ell^-\in[\ell_{mbo}^-,\ell_{\gamma}^-]$--\cite{2024EPJP..139..531P,2024ApJS..275...45P,2025JHEAp..4700368P}.

%%%%%%%%%%%%%%%%%%%%%%%%%%%%
%

%
%%%%%%%%%%%%%%%%%%%%%%%%%%%
\section{Red-shift function and horizons replicas}\label{Sec:redshift-replicas-First}
In this Section we start the analysis of the signals frequency shifts  in relation to the  horizons  replicas.
In Sec.\il(\ref{Sec:redhsift}) we introduce the red-shift  function $g_{red}$.
Red-shift function from  the horizons replicas will be  addressed in Sec.\il(\ref{Sec:from-de-replicas}).
We will  investigate the properties of the  limiting angular velocities of the stationary observers in  Sec.\il(\ref{Sec:limitng-velocitues}), which define the metric bundles and  constrain the $g_{red}$ definition.
The extreme points of the limiting frequencies $\omega_\pm$  are addressed in Sec.\il(\ref{Sec:photon-shells}), in  relation with the \textbf{BH} photons shell boundaries.
The red-shift function on the  horizons replica  defined on the  photons shell boundaries is discussed in Sec.\il(\ref{Sec:redffhidft-shellsreplicas}).
Co--rotating and counter--rotating horizons  replicas are characterized  in more details in Sec.\il(\ref{Sec:co--rotating-counter-rpliAs}).
We conclude this section in Sec.\il(\ref{Sec:absidered-extended-plane}), by discussing  the properties of the red-shift function in the extended plane.
\subsection{Red-shift  function $g_{red}$}\label{Sec:redhsift}
\begin{figure*}
\centering
\includegraphics[width=8cm]{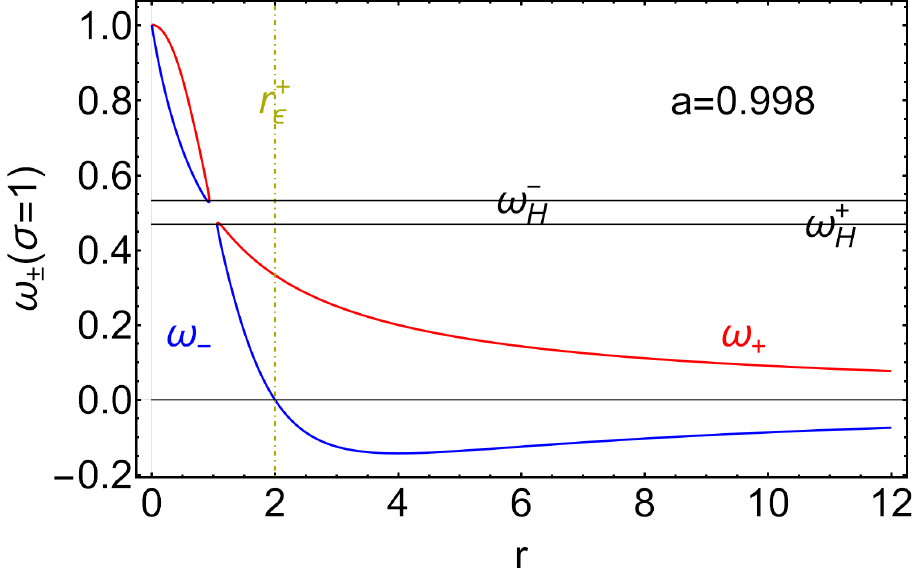}
\includegraphics[width=8cm]{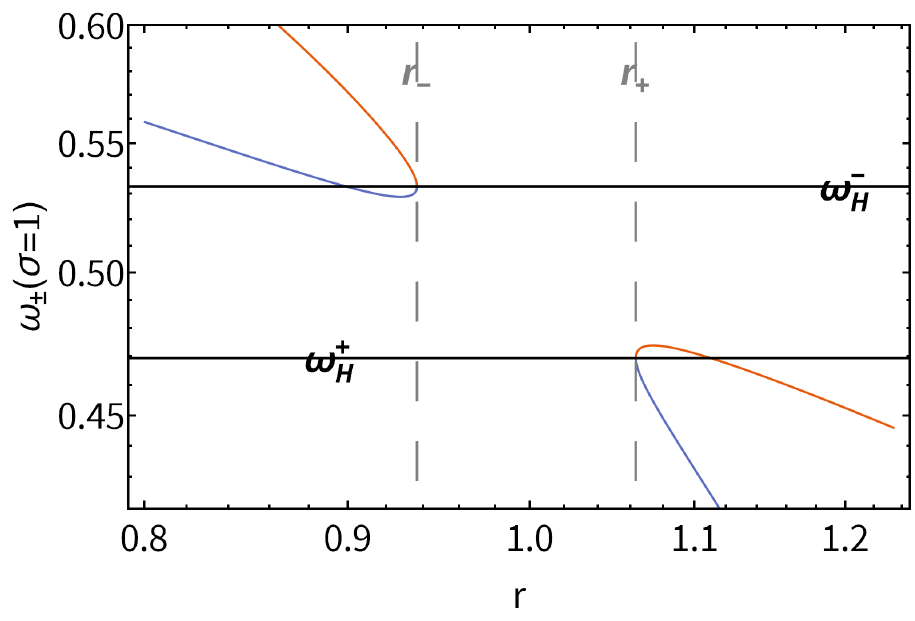}
        \caption{Limiting stationary angular frequencies $\omega_\pm $ on the equatorial plane $\sigma=1$ of the     Kerr \textbf{BH} geometry with  dimensionless spin $a=0.998$, as functions of the radius $r$. Radii $r_\pm$ are the outer and inner \textbf{BH} horizons respectively. Radius $r_\epsilon^+=2$ is the outer ergosurface.  $\omega_H^\pm$ are the outer and inner horizons frequencies respectively. Right panel is a close--up view of the left panel. Counter-rotating  outer (orange surfaces) and inner (green surfaces)  horizons  replicas  in   Kerr \textbf{BH} geometry with  dimensionless spin $a=0.998$. Cross of the horizontal lines $\omega=\omega_H^\pm$ with curves  $\omega_\pm(r)$ are the outer and inner horizons and their replicas.  }\label{Fig:Plotlocalwesit}
\end{figure*}

Let us consider a  circularly   orbiting particle. There is $u^r=0$ and $u^\theta=0$.
 We use the test particle  normalization condition  $g_{ab}u^a u^b=-1$,  on the four-velocity,  the constant of motions  $(\mathcal{E},\La)$,    and  the definition of relativistic angular velocity, $\Omega=u^\phi/u^t$   in   Sec.\il(\ref{Sec:carter-equa}).
Therefore, for  the particle constant of motions  $(\mathcal{E},\La)$,  there is
\bea&&
\mathcal{E}=-\frac{g_{t\phi} \Omega +g_{tt}}{\sqrt{-g_{\phi\phi} \Omega ^2-2 g_{t\phi} \Omega -g_{tt}}}\quad \mbox{and}\quad \La=\frac{g_{\phi\phi} \Omega +g_{t\phi}}{\sqrt{-g_{\phi\phi} \Omega ^2-2 g_{t\phi} \Omega -g_{tt}}},
\eea
in terms of the relativistic angular velocity,   where
\bea
\label{Eq:succeff-uo}
\dot{t}=\frac{1}{\sqrt{-g_{\phi\phi} \Omega^2-2 g_{t\phi} \Omega -g_{tt}}}.
\eea
%
%su
We define the \emph{red-shift function} $g_{red}$ as  the  energy (frequency) ratio
\bea\label{Eq:gredequa}
g_{red}\equiv \frac{\left.k_a u^a\right|_{o}}{\left.k_b u^b\right|_{e}},
\eea
where the subscript denotes the observer  $(o)$ and emitter  $(e)$ positions respectively\footnote{We will sometimes keep the subscript relative to the emitter  or observer to avoid confusion.}.
According to Eq.\il(\ref{Eq:gredequa}), a signal is blue--shifted if $g_{red}>1$ and red-shifted if \footnote{Note, in literature  the red-shift  function is also defined  as the emitted energy to observed energy, i.e. as $g=g_{red}^{-1}=\Em/\Em_o$,  with $g\in [0,1]$ and $g_{red}\in[0,1]$. In this case,  the emission (signal) is  red-shifted if $\Em/\Em_o> 1$. This for example occurs for receding side of the disks  orbiting  \textbf{BH}. Viceversa, the emission is  blue-shifted, i.e.  $\Em/\Em_o< 1$,  for the approaching
side of the disk with respect to the observer at infinity.  On the other hand, the red-shift $z$, differently from the red-shift function (or also red-shift  factor)  $g$ would be defined as   $z\equiv g-1$. (See also, for example, \cite{Bambi,Chen}).} $g_{red}<1$.
In details, in Eq.\il(\ref{Eq:gredequa}) the four-velocity of the observer at infinity is   $ u_o^a=(1,0,0,0) $.
Whereas there is  $
   k^a=(k^t,k^r,k^\theta,k^\phi)$, for the  four--momentum of the  photon, constrained by the normalization condition $k_a k^a=0$.
   Note 	here, considering a circular orbiting particle  in  Eq.\il(\ref{Eq:gredequa}), only components  $k^t$ and $k^\phi$ are relevant.  In fact    there is  $ u_e^a= (u_e^t,0,0, u_e^\phi) $, where $ u_e^t=\dot{t}$ in Eq.\il(\ref{Eq:succeff-uo}), is the  four--velocity of the  particles of gas.
   We introduce  the constant of (photon) motion (photon impact parameter) $ l_p\equiv -k_\phi/k_t$ (analogue  to the definition of the   gas specific/test particle specific angular momentum $ \ell=-u_\phi/u_t$).
 While  quantity
$\Omega $ is the (in general not conserved) angular frequency of the emitter, the quantity $g_{red}$ in Eq.\il(\ref{Eq:gredequa}) is  expressed in terms   of the photon constant of motion $l_p$.
In \cite{tobesubmitted}  the red-shift  function  has been studied for  signals emitted from    the orbiting disks  and,
 in more general term,   function $g_{red}$   was investigated by considering the conditions for   $g_{red}\geq 0$ and $g_{red}\gtrless1$ (the
conditions for  blue--shifted and   red--shifted signals).  Emission  from the outer ergoregion,  and the maximum and minimum values  of the red-shift  function $g_{red}$  were discussed   according to different conditions on the emitting radius and the photon  impact parameter $l_p$.    Here,  we note that
function $g_{red}$  reads explicitly
\bea\label{Eq:stat-usa-wash}
g_{red}=\frac{\sqrt{-g_{\phi\phi} \Omega ^2-2 g_{t\phi} \Omega -g_{tt}}}{1-l_p \Omega}.%=
\eea
Therefore, the sign of  $g_{red}={1}/{(1-l_p\Omega)\dot{t}}$, defined by the condition  $(-g_{\phi\phi} \Omega ^2-2 g_{t\phi} \Omega -g_{tt})\geq 0$, is  determined by  the  fluid relativistic angular velocity  $\Omega$ relative to  the photon impact parameter $l_p$, that is by the  quantity $(1-l_p\Omega(\ell))$.
As discussed in Sec.\il(\ref{Sec:carter-equa}), sign of $\Omega(\ell)$ can change along a (free test particle or photon) path. In  the Kerr \textbf{BH}   geometry, for  $\ell>0$ there is $\Omega>0$,  while for  $\ell<0$ there is $\Omega>0$ for $r<r_\Ta(\ell)$.
However , at fixed $(r,\ell)$, the relativistic angular velocity $\Omega$  is fixed as well.
Therefore, for $\Omega=0$, that is for  $r=r_\Ta$,  there is  $g_{red}=\sqrt{ -g_{tt}}$.
On the other hand, for $ l_p=0$, there is $g_{red}={1}/{\dot{t}}$.
We note that  in the \textbf{BH} photons shell,  such condition would correspond  to the orbit $r_{\lim}$.
Finally   red-shift function is null, $g_{red}=0$, correspondent to  $ \sqrt{(-g_{ab} u^a u^b)/ (u^t)^2}=\sqrt{-g_{\phi\phi}\Omega^2-2 g_{t\phi} \Omega -g_{tt}}=0$, for $\Omega(a,\sigma,r)=\omega_{\pm}(a,\sigma,r)$, coincident with the limiting stationary observer frequencies  $\omega_\pm$ (on the limiting stationary observers orbits), that is the light surfaces with frequencies $\omega_\pm$. Therefore,  in particular  this condition occurs on the \textbf{BH} horizons  $r=r_\pm$   where $\Omega(a,\sigma,r)=\omega_H^{\pm}$.
Explicitly, the limiting angular velocities of the stationary observers are:
\bea\label{Eq:limiting-stationary}
\omega_{\mp}(a,r,\sigma)\equiv \frac{-g_{t\phi}\mp\sqrt{g_{t\phi}^2-g_{\phi\phi} g_{tt}}}{g_{\phi\phi}},
\eea
--see Figs\il(\ref{Fig:Plotlocalwesit}).
The \textbf{BH} horizons angular frequencies $\omega_H^\pm$ of Eqs\il(\ref{Eq:omegahmo}) are  the frequencies $\omega_{\mp}$ evaluated on $r_\pm$.
In  the following  analysis we will consider  the
 red-shift function $g_{red}$     for emission from the \textbf{BH} photons shell boundaries $\omega=\sigma_\pm$ and from replica orbits.
\subsection{Red-shift function from  horizons replicas}\label{Sec:from-de-replicas}
\begin{figure*}
\centering
\includegraphics[width=4.75cm]{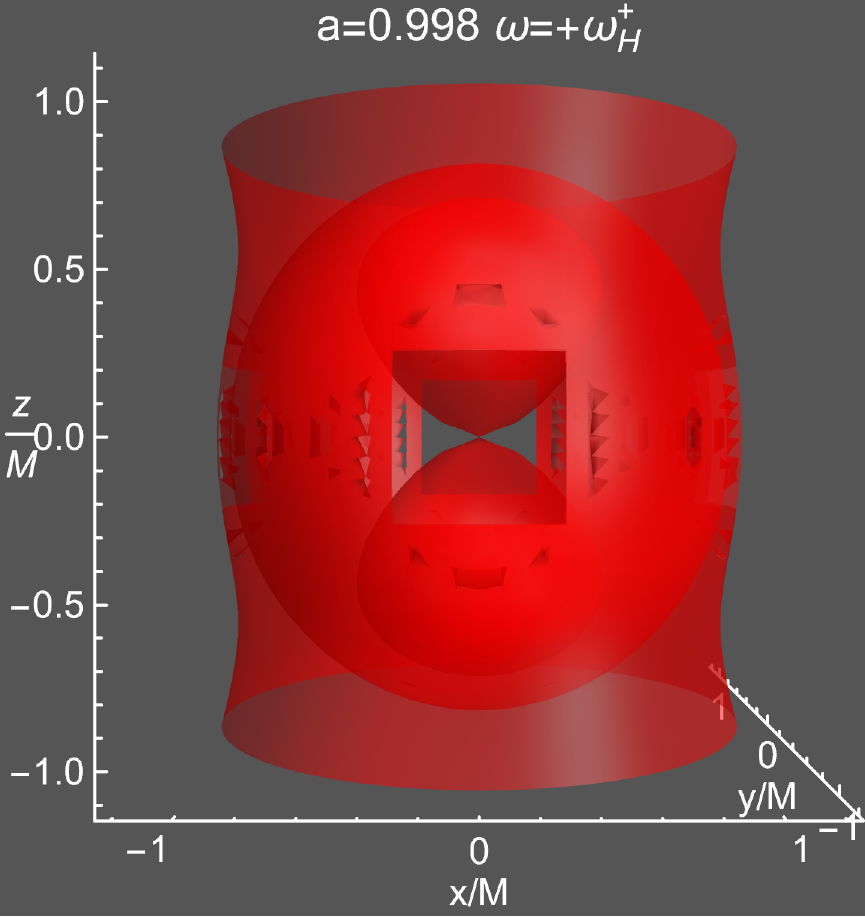
}\includegraphics[width=4.88cm]{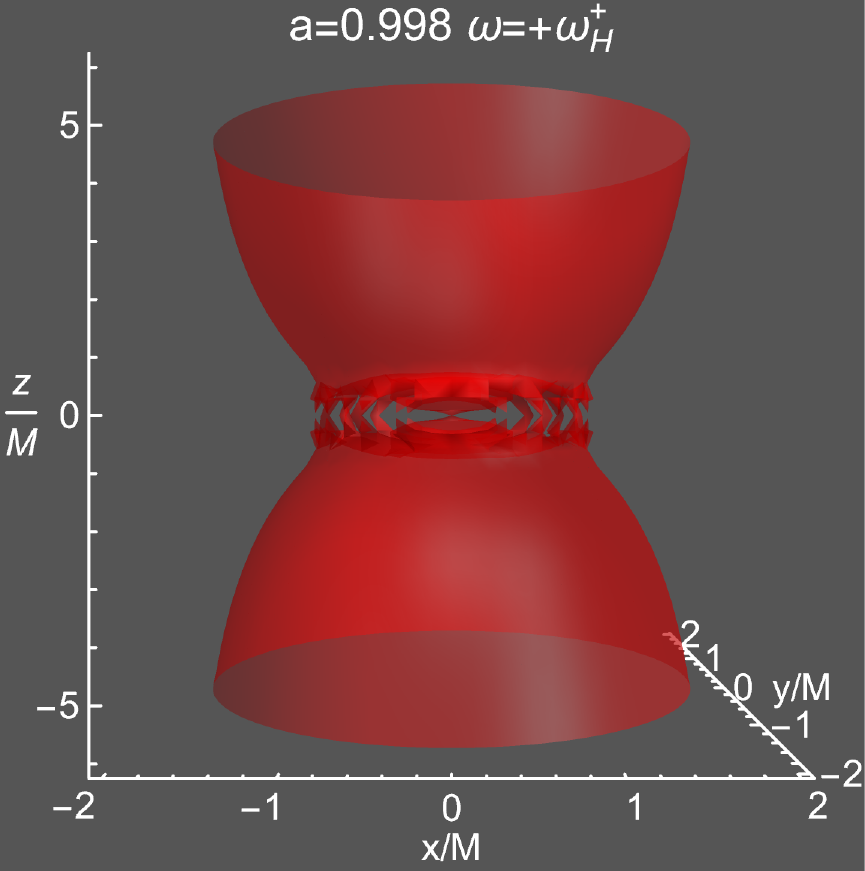}\includegraphics[width=4.75cm]{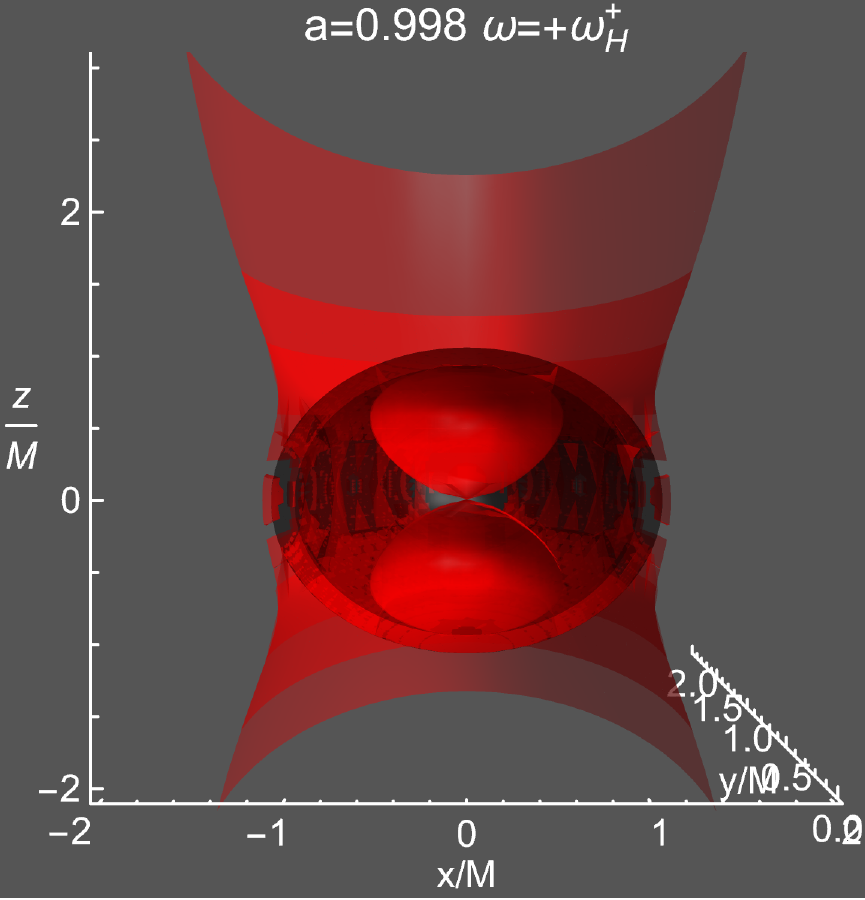}\\\includegraphics[width=4.75cm]{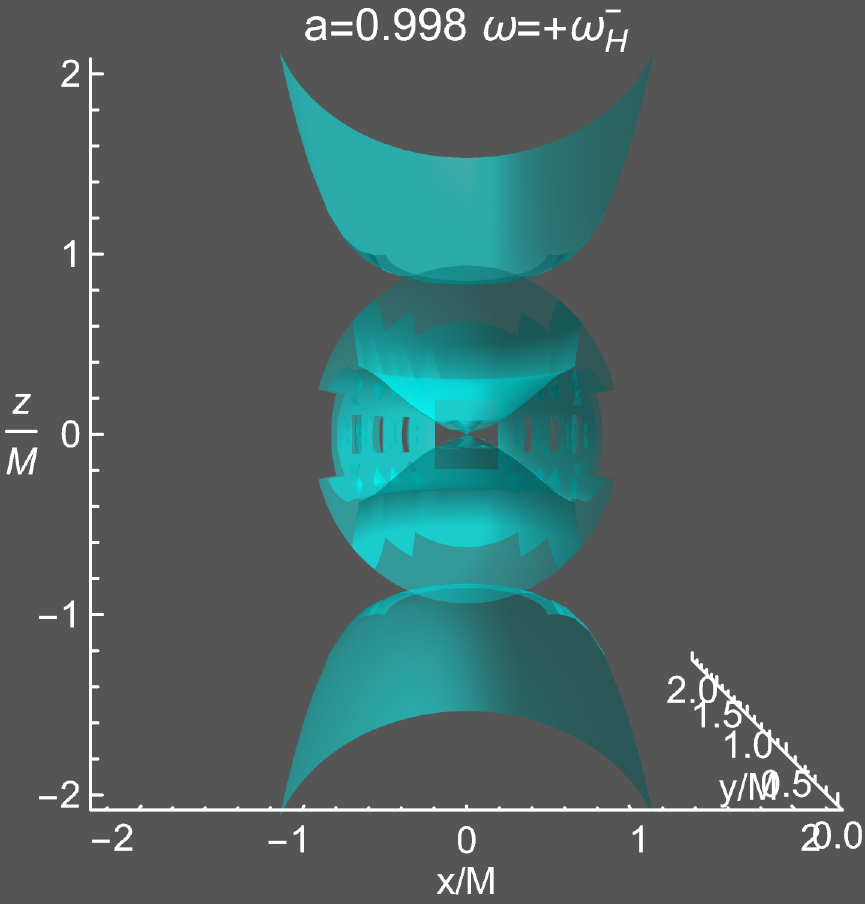}\includegraphics[width=4.75cm]{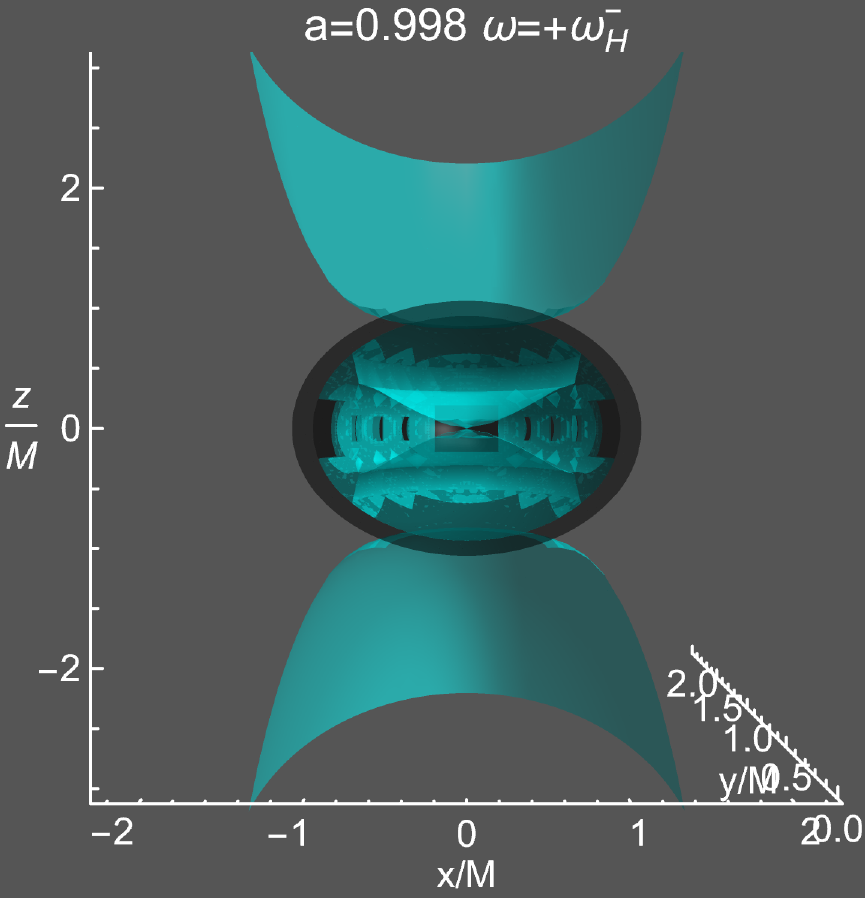}\includegraphics[width=5cm]{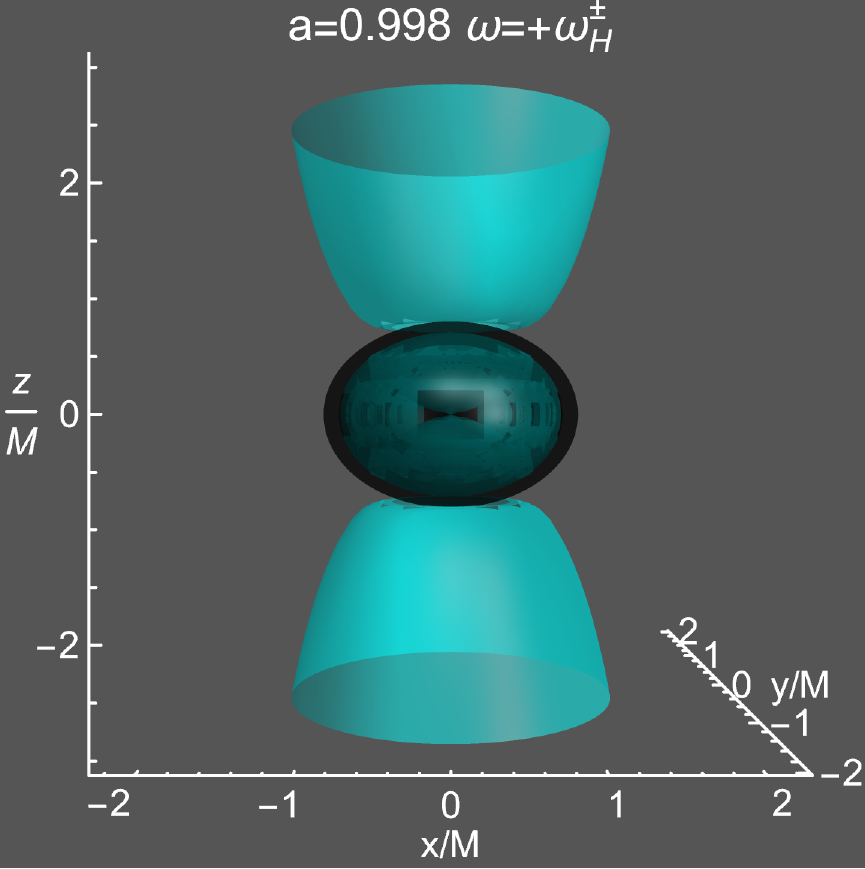}
        \caption{Co-rotating  outer (red surfaces--upper panels) and inner (cyan surfaces--bottom panels)  horizons  replicas  in the    Kerr \textbf{BH} geometry with  dimensionless spin $a=0.998$.    $\omega_H^\pm$ are   the outer and inner \textbf{BH}  horizons frequencies, respectively.  The inner and outer horizons spheres are shown as  gray spheres.
        	(In some panels, the inner and outer horizon spheres can be seen as  embedded in  replicas.) In the plots $\{z=r \cos\theta, y=r \sin\theta \sin\phi, x=r \sin\theta\cos\phi\}$. Left, middle and right panels show different views of the replica surfaces.}\label{Fig:PlotBri0999m}
\end{figure*}

\begin{figure*}
\centering
\includegraphics[width=4.75cm]{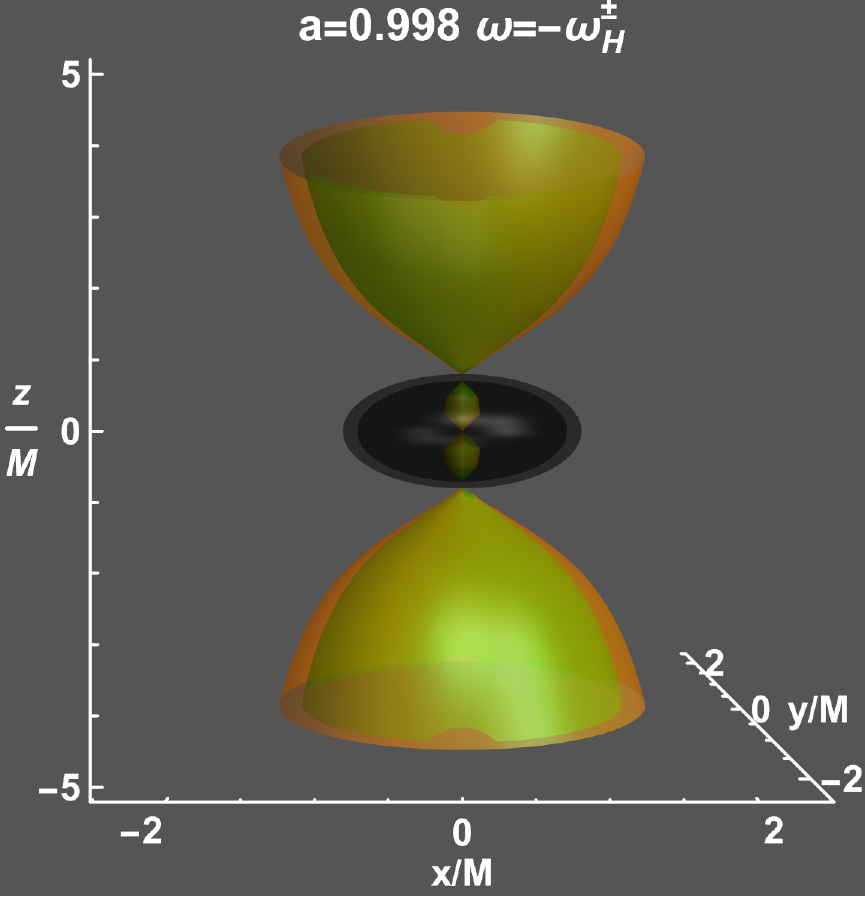}
        \caption{Counter-rotating  outer (orange surfaces) and inner (green surfaces)  horizons  replicas  in  the  Kerr \textbf{BH} geometry with  dimensionless spin $a=0.998$. For further details see  Figs\il(\ref{Fig:PlotBri0999m})--caption.}\label{Fig:PlotBri0999mm}
\end{figure*}

In this section we consider the red-shift  function in relation to the "horizons replica",  for the emission    from orbits located on  the  co--rotating or counter--rotating replicas,  considering the case when $\Omega=\pm\omega_H^\pm$.
As introduced in Sec.\il(\ref{Sec:carter-equa}), horizons replicas  are defined as  circular orbits of photons (at any angle $\sigma$) that have the same angular frequency as the \textbf{BH} inner or outer horizons $\omega_H^\pm$. Horizons  replica were also extended to consider counter--rotating orbits, where $\omega=-\omega_H^\pm$. These conditions are  satisfied  in  special sets of  points $(r,\sigma)$, in selected spacetimes.  The conditions and properties of replicas for fixed spacetime have been discussed  extensively and in details in several analysis--\cite{2021EPJC...81..258P,2024NuPhB100816700P}.
 In Figs\il(\ref{Fig:PlotBri0999m}) we show  the \emph{co-rotating}  outer  and inner horizons  replicas  in   Kerr \textbf{BH} geometry with  dimensionless spin $a=0.998$.  In Figs\il(\ref{Fig:PlotBri0999mm})
are the  \emph{counter-rotating } outer  and  inner horizons  replicas.

For convenience we  review here    main properties of the replica.
 The collection of all replicas,  of a specific horizon frequency, at fixed $\sigma$, in \emph{all} Kerr  space-times,  including naked singularities,  forms a closed curve,  called  \emph{Killing metric bundle} (or \emph{bundle}), in plane $(a,r)$,   the \emph{extended plane}.
The curve $a_\pm=\sqrt{r(2-r)}$ of all horizons in the $(a,r)$ plane,  emerges as the envelope surface of all metric bundles, at fixed $\sigma$.
 It should be noted that  for a particle with  $\Omega=\omega_H^\pm$  (and $\Omega=-\omega_H^\pm$) , there is, \emph{in general}, $g_{red}\neq 0$,  while the red-shift function is zero, in particular, on the  horizons and, more generally, on the solution  $\omega_\pm: g_{ab}u^au^b= g_{tt} \omega+2g_{r\phi}\omega+g_{\phi\phi}\omega^2=0$,  which define the stationary observers light surfaces, having (photons) angular velocities $\omega=\omega_\pm$.
 On the horizons, where  $\Omega(r_\pm)=\omega_\pm(\pm)=\omega_H^\pm$,  there is  $\left.g_{red}\right|_{r_\pm}=0$.
 Therefore, according to the definitions of horizons replicas  (for photons), there  are particular light--like orbits  at $r>r_+$ (the horizons replicas) and at some $\sigma$, where  there is $\omega_\pm=\omega_H^\pm$, and
 therefore  it is $g_{red}= 0$.
\subsubsection{The limiting angular velocities of the stationary observers}\label{Sec:limitng-velocitues}
\begin{figure}
\centering
\includegraphics[width=8cm]{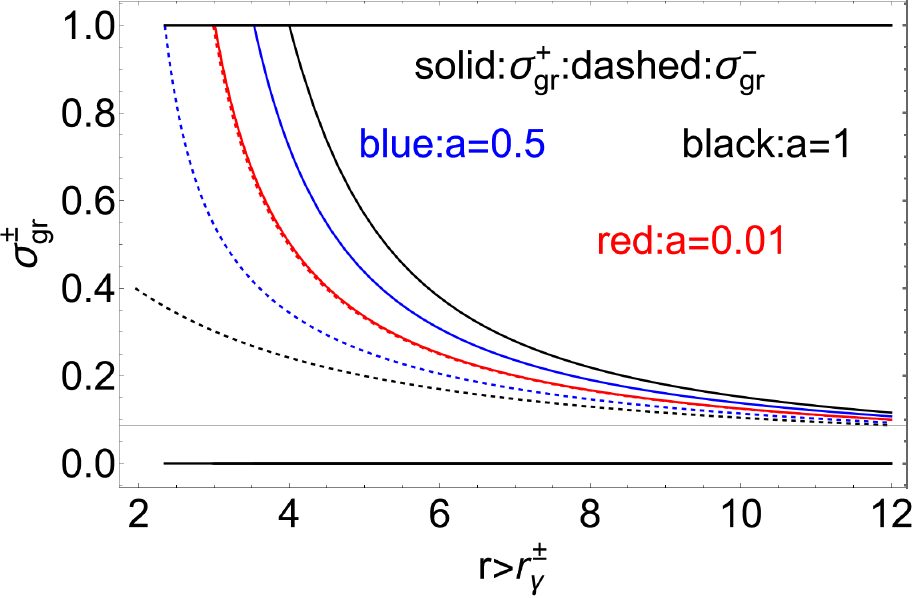}
\includegraphics[width=8cm]{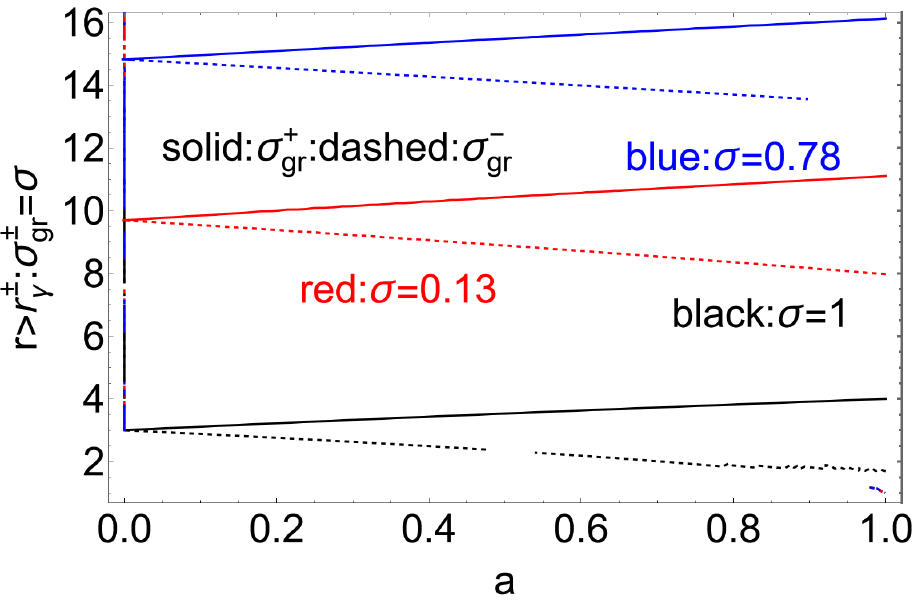}
\caption{Left panel: planes $\sigma_{gr}^\mp$:  $\ell^\pm\in[1/\omega_-,1/\omega_+]$ for  $r>r_\gamma^\pm$ defined in Eq.\il(\ref{Eq:sigmacrdef}), plotted as function of $r$, and different \textbf{BH} spins, signed on the panels. Right panel:  curves  $\sigma_{gr}^\mp=\sigma=$constant in the plane $(r,a)$  for different $\sigma$ signed on the panel.}\label{Fig:PlotDiret9}
\end{figure}

Let us consider the stationary observer limiting angular frequencies $\omega_\pm$  of Eqs\il(\ref{Eq:limiting-stationary}).
There is $\omega_+(r,a,\sigma)>0$ for all $\{a,\sigma,r\}$, while  there is $\omega_-(r,a,\sigma)=0$ on the outer ergosurface $r=r_\epsilon^+(a,\sigma)$.
Therefore,  there is  $ \omega_-(r,a,\sigma)\lessgtr 0$, for $r\gtrless r_\epsilon^+(r,a,\sigma)$, respectively--(see Figs\il(\ref{Fig:Plotlocalwesit})).
It  is
 $\omega_{\pm}:(\dot{t})^{-1}=0$,    therefore there is
 $g_{red}=0$, for $\Omega=\omega_{\pm}$. While for the fluid particles there is    $\Omega\in]\omega_{+},\omega_{-}[$.
(In Figs\il(\ref{Fig:Plotlocalwesit}), limiting stationary angular frequencies $\omega_\pm $ are shown on the equatorial plane $\sigma=1$ of the     Kerr \textbf{BH} geometry with  dimensionless spin $a=0.998$, as functions of the radius $r$. Crosses of the horizontal lines $\omega=\omega_H^\pm$ with curves  $\omega_\pm(r)$ are the outer and inner horizons and their replicas. )
 From the  conditions $\Omega(\ell)=\omega_{\mp}$, (see Eq.\il(\ref{Eq:omegadefinl})  and Eqs\il(\ref{Eq:flo-adding})), we find  the limiting specific angular momentum $\ell_{(\mp)}$:
\bea\label{Eq:defoniitn}
\ell_{(\mp)}\equiv -\frac{g_{\phi\phi} \omega_{\mp} +g_{t\phi}}{g_{t\phi} \omega_{\mp} +g_{tt}}=\frac{1}{ \omega_{\mp}}=\frac{-g_{t\phi}\pm\sqrt{g_{t\phi}^2-g_{\phi\phi} g_{tt}}}{g_{tt}}.
\eea
for  the fluid particles,   using
 Eq.\il(\ref{Eq:defoniitn}),
there is $\ell^\pm(r)\in[1/\omega_-,1/\omega_+]$ or,  equivalently, $\Omega(\ell^\pm(r))\in[\omega_-,\omega_+]$),  for radii   $r>r_\gamma^\pm$ and  angle $\sigma>\sigma_{gr}^\pm$, where
 \bea\label{Eq:sigmacrdef}
\sigma_{gr}^\mp\equiv \frac{\mathbf{C}_\xi-\sqrt{\mathbf{C}_o\mp \mathbf{C}_\mu}\mp2 \sqrt{\mathbf{C}_c}}{2},
\eea
 for counter--rotating and co--rotating fluids  respectively,
with
\bea&&\nonumber
\mathbf{C}_\xi\equiv \frac{2 a^6-a^4 r \left(r^2-14 r+16\right)+a^2 r^2 \left(r^4-9 r^3+34 r^2-56 r+32\right)+(r-2)^3 r^5}{a^2 \left[a^2-(r-2)^2 r\right]^2};\\\nonumber
&&\mathbf{C}_c\equiv \frac{r^3 \left[a^4+3 a^2 (r-2) r+2 (r-2)^2 r^2\right]^2}{a^2 \left[a^2-(r-2)^2 r\right]^4};
\\\nonumber
&&
\mathbf{C}_\mu\equiv \frac{4 r^5\hat{\mathbf{C}}_\tau \left[a^2+(r-2) r\right]^2}{a^4 \sqrt{\mathbf{C}_c} \left[a^2-(r-2)^2 r\right]^6},\quad\mbox{and}\quad
\mathbf{C}_o \equiv 2 \mathbf{C}_\xi^2-\hat{\mathbf{C}}_\iota-\hat{\mathbf{C}}_b,
\eea
{where}
\bea
&&\nonumber
\hat{\mathbf{C}}_\iota\equiv \frac{6 a^8+2 a^6 r \left(r^2+18 r-24\right)+a^4 r^2 \left(r^4-4 r^3+70 r^2-144 r+96\right)+2 a^2 r^5 \left(r^3-5 r^2+16 r-20\right)+(r-2)^2 r^8}{a^4 \left[a^2-(r-2)^2 r\right]^2};
\\&&\nonumber
\hat{\mathbf{C}}_b\equiv \frac{2 \left[a^6+a^4 r \left(r^2+6 r-8\right)+a^2 r^2 \left(2 r^3+5 r^2-24 r+16\right)+(r-2)^2 r^5\right]}{a^2 \left(a^2-(r-2)^2 r\right)^2};		
\\\nonumber
&&
\hat{\mathbf{C}}_\tau\equiv a^8 (3 r-4)-a^6 r \left(r^3-23 r^2+66 r-48\right)-a^4 (r-2)^2 r^2 \left(5 r^2-46 r+48\right)-
\\\nonumber
&&\nonumber \qquad \qquad 4 a^2 (r-2)^3 r^3 \left(r^2-7 r+8\right)+4 (r-2)^5 r^5.
\eea
As clear from  Fig.\il(\ref{Fig:PlotDiret9}), the limiting angle  $\sigma_{gr}^\pm$ decreases with $r>r_\gamma^\pm$.  Angle  $\sigma_{gr}^+$ increases with the spin,  while   $\sigma_{gr}^-$ decreases  with the spin .
\subsubsection{Extreme points of the limiting frequencies $\omega_\pm$ and relation with the \textbf{BH} photons shell}\label{Sec:photon-shells}
\begin{figure}
\centering
\includegraphics[width=8cm]{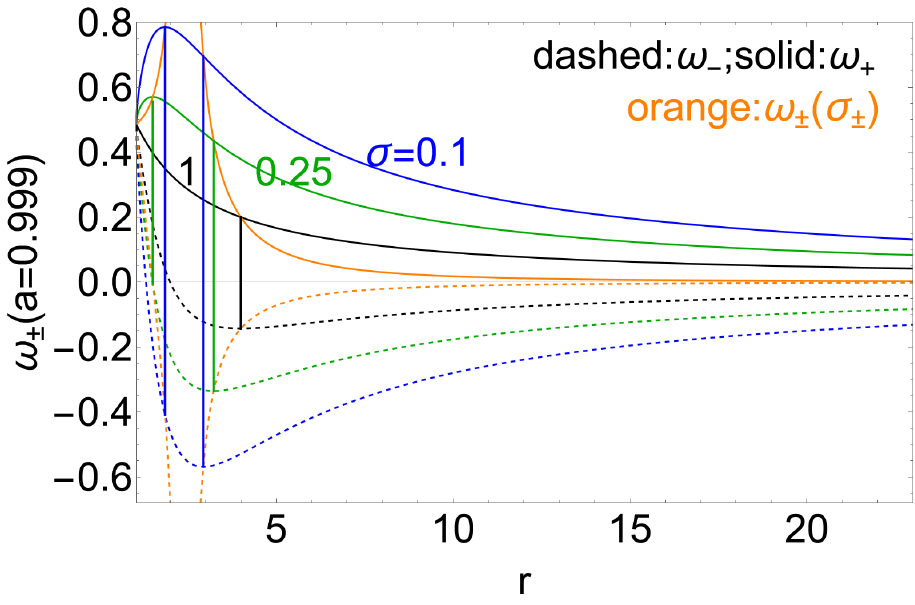}
\includegraphics[width=8cm]{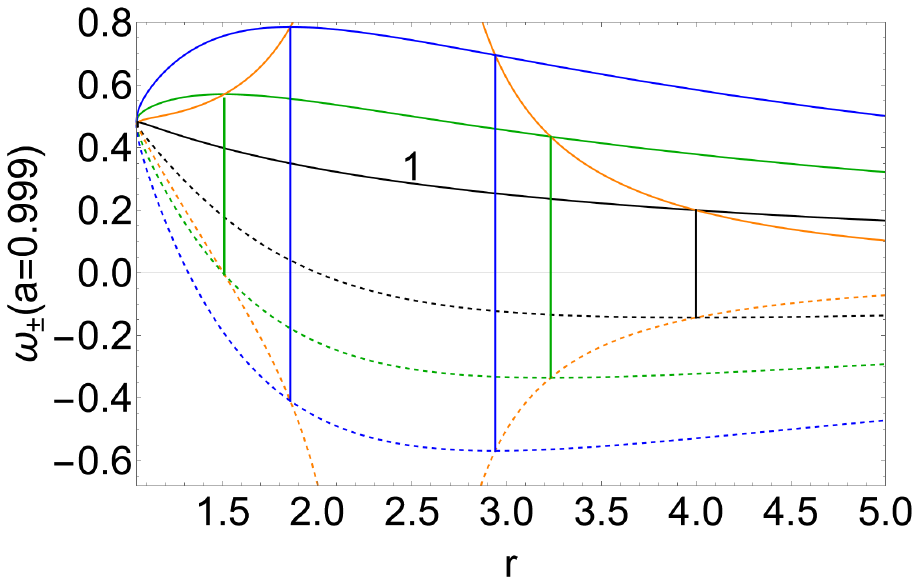}
\includegraphics[width=8cm]{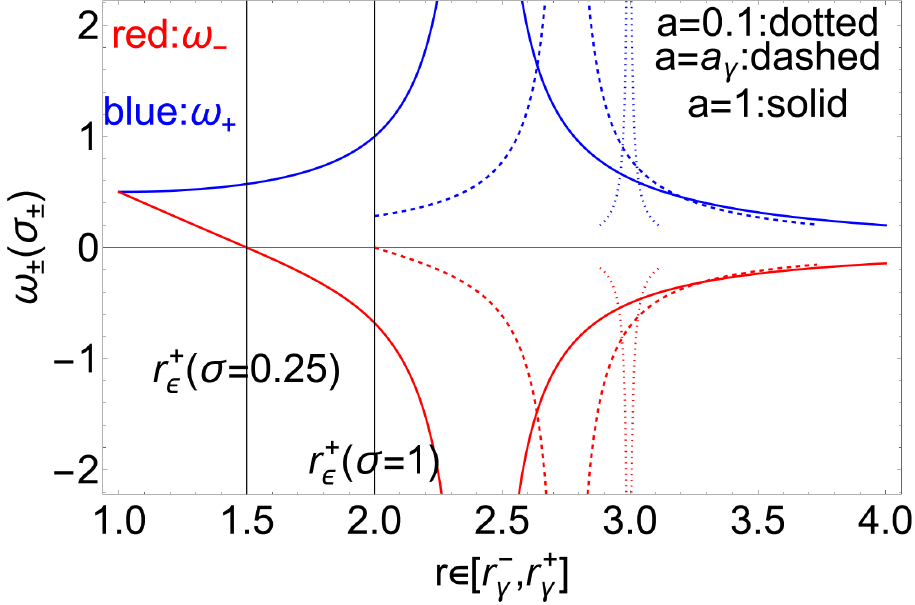}
\caption{
Upper left panel:  stationary observer limiting angular velocity  $\omega_\pm$ are plotted as function of $r$ for different angles $\sigma$ signed on the panel and for  \textbf{BH} spin $a=0.999$. Orange curve is the maximum/minimum points $\omega_\pm(\sigma=\sigma_\pm)$, where $\sigma_\pm$ are the photons shell boundary. Upper right panel is a close up  view of the left panel.  Note, for fixed $(a,\sigma)$, the crossing points  $r:\omega_\pm=\omega_\pm(\sigma_\pm)$ coincide for $\omega_\pm$ (see vertical lines). Bottom panel: The maximum/minimum points $\omega_\pm(\sigma=\sigma_\pm)$ are plotted  as functions of $r$ for different spins signed on the panel. $r_\gamma^\pm$ are the counter--rotating/co--rotating photon circular orbits. Spin $a_\gamma\equiv 1/\sqrt{2}$.  $r_\epsilon^+$ is the outer ergosurface, where there is $\omega_-=0$ on the plane $\sigma_\pm$ written on the panel
}\label{Fig:Plotturenesa}
\end{figure}

\begin{figure}
\centering
\includegraphics[width=8cm]{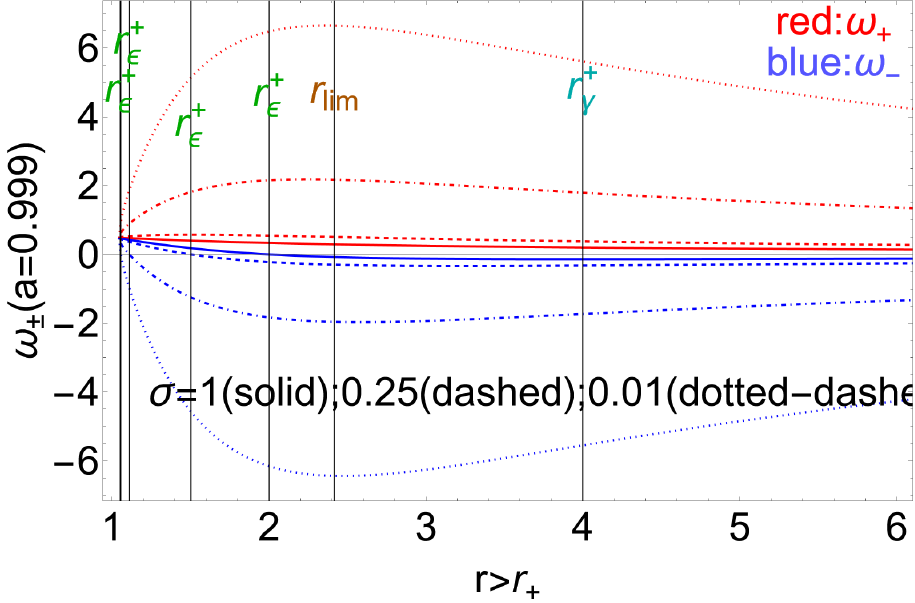}
\includegraphics[width=8cm]{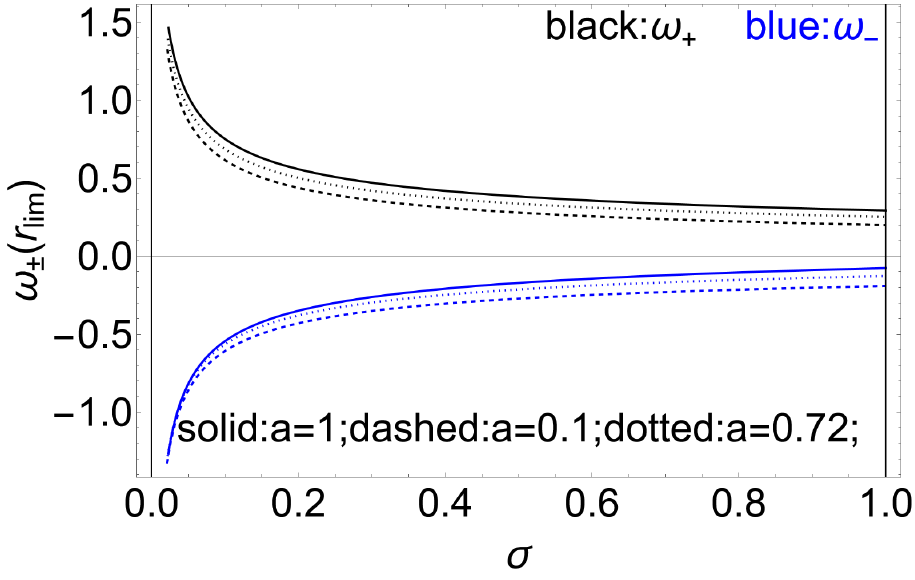}
\caption{Right panel: limiting (stationary observers) angular velocity $\omega_\pm$  evaluated on the radius $r_{\lim}:\ell_q=0$ as function of the angle $\sigma$ for different \textbf{BH}  spins as signed on the panel. Left panel:  for spin $a=0.999$, angular velocities  $\omega_\pm$ are plotted as function of $r$ for different angles $\sigma$ signed on the panel. $r_\epsilon^+$ are the ergosurface radius at each angle (where $\omega_-=0$). $r_\gamma^+$ is the counter--rotating circular photon orbit. See Eqs\il(\ref{Eq:estremeoemgapojnts}). The photons shell  boundaries $\sigma_\pm$ are the extreme points of $\omega_\pm$ as functions of the radius $r$.}\label{Fig:PlottuMinaE}
\end{figure}

\begin{figure}
\centering
\includegraphics[width=8cm]{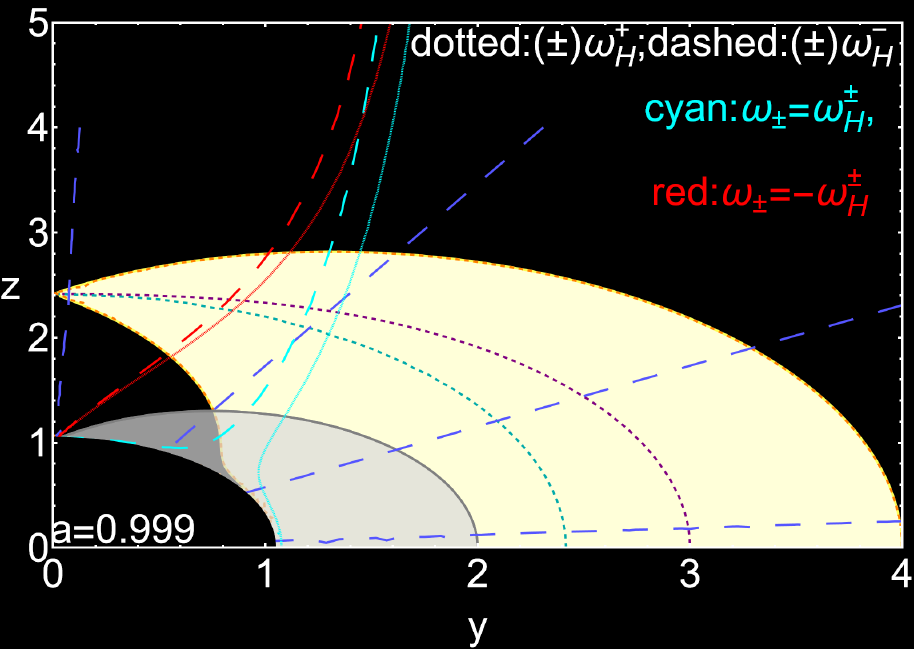}
\includegraphics[width=8cm]{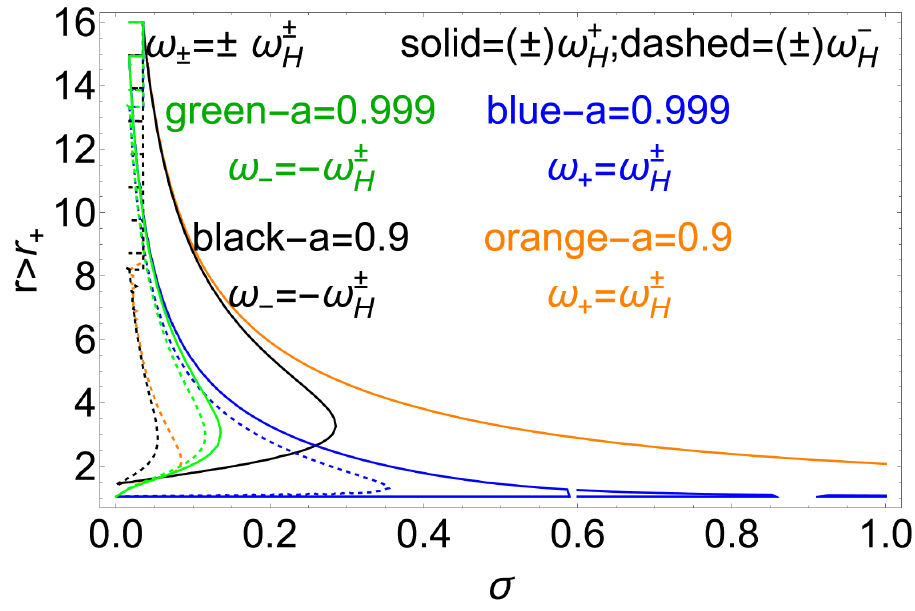}
\caption{Left panel: photons shell as in  Fig\il(\ref{Fig:Plotelemtempor})-bottom panel  for the \textbf{BH} with spin $a=0.999$, with the horizon replica surfaces, i.e. the surfaces $\omega_\pm=\pm\omega_H^\pm$, where $\omega_\pm$ are the stationary observer limiting angular velocity and $\omega_H^\pm$ are the outer and inner horizon angular velocity respectively. Crossing of the surfaces with the photons shell boundary are the extreme points of $\omega_\pm$ for the radius $r$. Right panel: horizon replica in the plane $(r,\sigma)$ for different spins as signed on the panel. Note, in the planes  the counter--rotating horizons replicas are also shown.}\label{Fig:PlotturenesaA}
\end{figure}

\begin{figure}
\centering
\includegraphics[width=8cm]{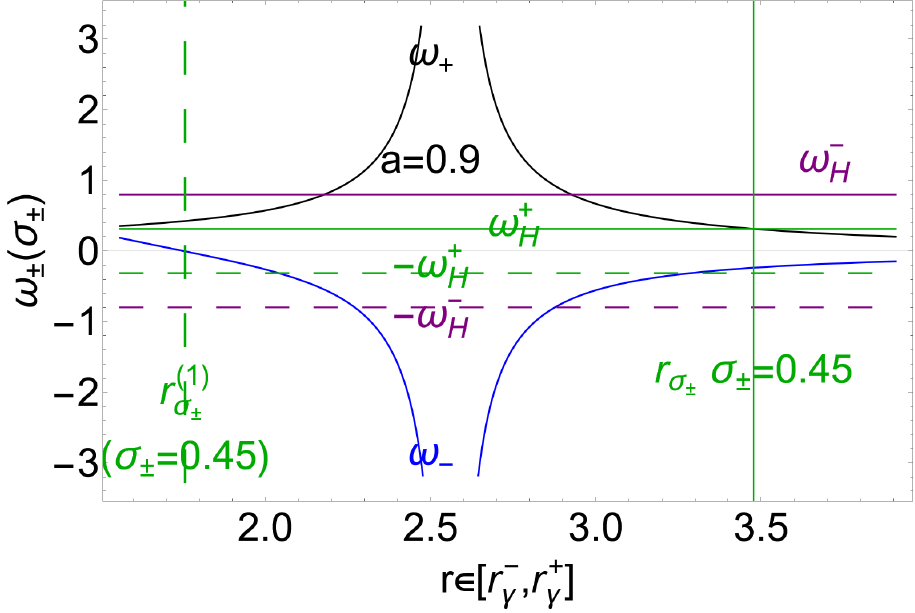}
\includegraphics[width=8cm]{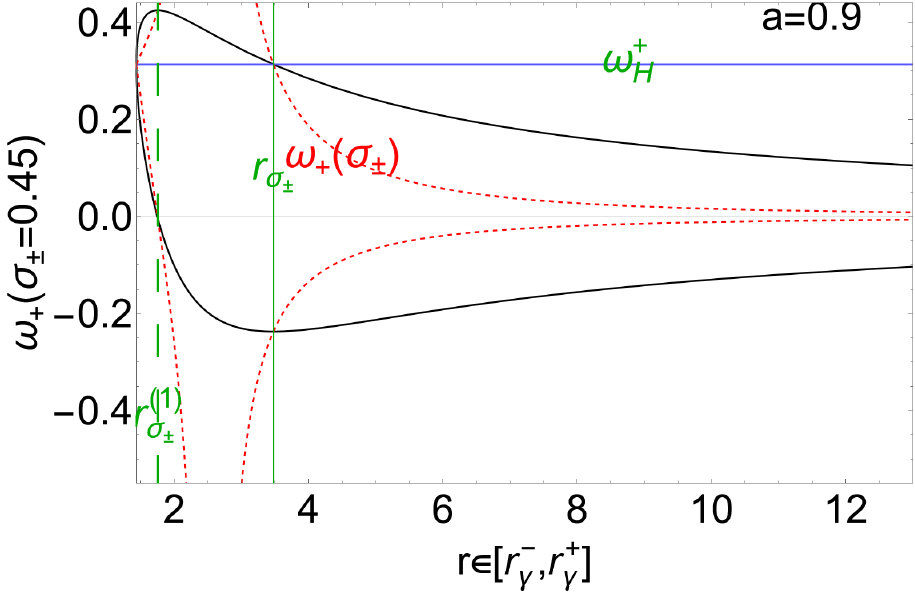}
\includegraphics[width=8cm]{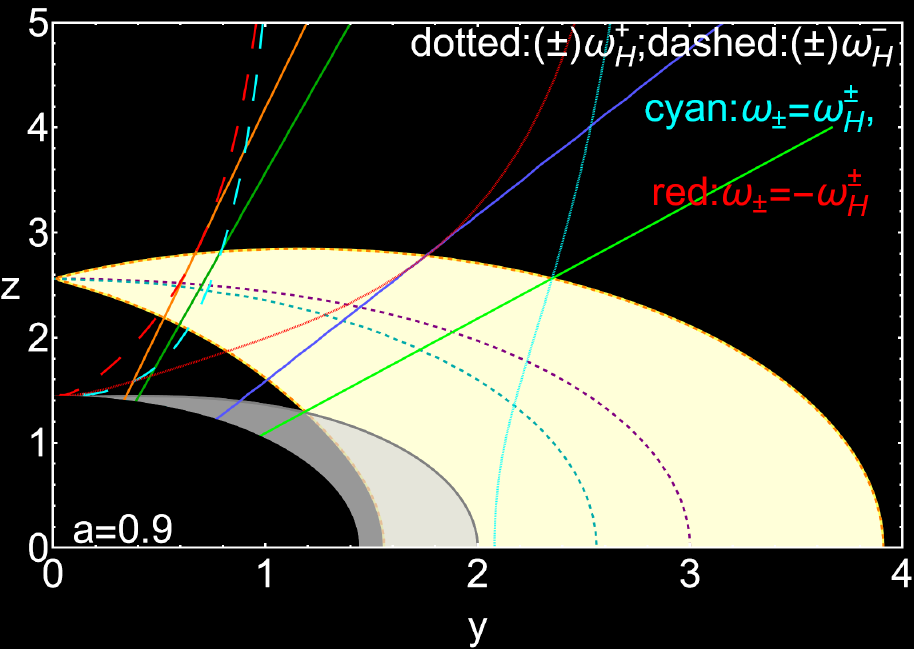}
\caption{Horizons replicas $\omega_H^\pm=\omega_\pm$ for  \textbf{BH} spin $a=0.9$ on the \textbf{BH} photons shell $\sigma=\sigma_\pm$ (extreme points of the functions $\omega_\pm(r)$). Upper left panel:  maximum/minimum points $\omega_\pm(\sigma=\sigma_\pm)$ are shown as function of the  radius $r$, $\omega_\pm$ are the limiting stationary observers angular velocities, $\omega_H^\pm$ are the outer and inner horizon angular velocity respectively.  Solid vertical  green line is the point $(\sigma_\pm,r_{\sigma_\pm})$ where $\omega_H^+=\omega_+(\sigma_\pm)$ (outer horizon replica on the photons shell boundary).  Green dashed line is the radius $r_{\sigma_\pm}^{(1)}$, where is the maximum of $\omega_+$ for $\sigma=0.45$.  This situation is illustrated in details in the upper right panel. See also Figs\il(\ref{Fig:Plotturenesa}).
Bottom panel: photons shell for the \textbf{BH} with spin $a=0.9$. For further details see also Figs\il(\ref{Fig:PlotturenesaA}).  Horizons replicas surfaces $\mp\omega_H^\pm=\omega_\pm$ are show. Diagonal lines are the angles of crossing of the replicas with the photons shell boundaries.}\label{Fig:Plotelemtemporeplicas}
\end{figure}

Considering  $\omega_\pm$ as function of the radius $r$,
there is
 \bea\label{Eq:estremeoemgapojnts}
\partial_r \omega_+=0, \quad \mbox{for}\quad
 \sigma=\sigma_\pm,  \quad \mbox{on}\quad   r\in[r_\gamma^-, r_{\lim}],
\eea
and
 \bea\label{Eq:estremeoemgapojnts1}
\partial_r \omega_-=0, \quad \mbox{for}\quad
 \sigma=\sigma_\pm,  \quad \mbox{on}\quad  r\in[ r_{\lim},r_\gamma^+].
\eea
Therefore  the \textbf{BH}   photons shell boundary  provides  the extreme points of the  frequencies  $\omega_\pm$ in  the radial range $r\in[1,4]$ (see Figs\il(\ref{Fig:PlottuMinaE})).
(Note that the region $[r_\gamma^-,r_{\lim}]$ is filled with spherical co--rotating $(\ell>0)$ photon orbits, while region  $[r_{\lim},r_\gamma^+]$ is filled with spherical counter--rotating $(\ell<0)$ photon orbits  and, for $r=r_{\lim}$, there is $\ell=0$.)
For  $a\in[0,1]$, there is  $r_{\sigma_\pm}\in[r_\gamma^-,r_\gamma^+]$ while, for $a>0$, there is  $r_{\sigma_\pm}<r_\gamma^+$, where we introduced the radii  $r_{\sigma_\pm}(a,\sigma):\sigma_\pm(a,r)=\sigma$, shown in Figs\il(\ref{Fig:PlottuMinaD}).
For fixed spin and angle,  the surfaces defined by $\omega_\pm=$constant cross the photons shell in general in two points, one being an extreme point of the frequencies---see Figs\il(\ref{Fig:PlotturenesaA}) and Figs\il(\ref{Fig:PlotDiret9}).
In Figs\il(\ref{Fig:Plotturenesa})-upper  panels the stationary observer limiting angular velocity  $\omega_\pm$ are plotted  for  \textbf{BH} spin $a=0.999$,  as functions of $r$, for different angles $\sigma$.  We show also the  curves of maximum/minimum points $\omega_\pm(\sigma=\sigma_\pm)$. For fixed  spin and angle,  the crossing points  $r:\omega_\pm=\omega_\pm(\sigma_\pm)$, for  $\omega_\pm$,  respectively, coincide.
In Figs\il(\ref{Fig:Plotturenesa})-bottom panel  the extremes  $\omega_\pm(\sigma=\sigma_\pm)$ are plotted  as functions of $r\in[r_{\gamma}^-,r_\gamma^+]$, for different spins.  It is evident that, for  the  outer ergosurface   $r=r_\epsilon^+$, there is $\omega_-=0$.
Figs\il(\ref{Fig:PlottuMinaE})-left panel shows the    angular velocities  $\omega_\pm$ as functions of the radius  $r$, for  the  \textbf{BH} spin $a=0.999$, and  different values of the angle $\sigma$. On  the outer ergosurface   there is $\omega_-=0$.  The photons shell  boundaries  $\sigma_\pm$ are the extreme points of $\omega_\pm$ as functions of the radius $r$. We have shown also the  radius $r_{\lim}$ separating  the outer counter--rotating photons in the photons shell, and having a role for the extreme of the angular velocities $\omega_\pm$, according to Eqs\il(\ref{Eq:estremeoemgapojnts}). The extreme points radii  of   $\omega_-$ are in the range $[r_{\lim},r_\gamma^+]$.
The limiting   angular velocity $\omega_\pm$ are shown in
Figs\il(\ref{Fig:PlottuMinaE})-right panel,  evaluated on the radius $r=r_{\lim}$,  as function of the angle $\sigma$, for different \textbf{BH}  spins. The quantities $\omega_\pm(r_{\lim})$ increase with the spin, and decrease, in magnitude, with the  angle $\sigma$.
This could  be seen also in Figs\il(\ref{Fig:PlotturenesaA})-right panel, where $\omega_\pm=$constant  are shown
crossing the outer photons shell.
Note, solutions $\omega=\omega_\pm$ can be solved for the functions $\sigma_\omega^\pm$:
\bea\label{Eq:sigmaomega}
&&\sigma_\omega^\pm\equiv \frac{\Gamma_{(a)}\pm \Gamma_{(b)}}{2 a^2  \omega ^2\Delta},\quad\mbox{where}
\\
&&\nonumber
\Gamma_{(b)}\equiv \left[a-\omega  \left(a^2+r^2\right)\right] \sqrt{ \Gamma_{(a)}+2 a \omega \Delta},\quad\mbox{and}\quad \Gamma_{(a)}\equiv\omega ^2 \left(a^2+r^2\right)^2+a^2-4 a r \omega,
\eea
(see Figs\il(\ref{Fig:PlotDiret9})).
In Figs\il(\ref{Fig:PlotturenesaA})--left panel we show  the horizon replica curves  as the circular orbits $(r,\sigma):\omega_\pm=\pm\omega_H^\pm$.
We have considered  the \textbf{BH} and \textbf{BH} photons shell in Fig\il(\ref{Fig:Plotelemtempor})-bottom panel, showing the replicas curves   crossing with the photons shell boundaries,   being  the  extreme points of  the  relativistic velocity  $\omega_\pm$, for the radius $r$. These are  separated  by the limiting radius $r_{\lim}$ (therefore with co--rotating orbits at $r<r_{\lim}$, and counter--rotating orbits in the region  $r>r_{\lim}$).
In Figs\il(\ref{Fig:Plotelemtemporeplicas}) we have repeated   the analysis  of  Figs\il(\ref{Fig:Plotturenesa}) for the horizons replicas $\omega_H^\pm=\omega_\pm$, and the  relative counter--rotating horizons replicas, for  \textbf{BH} spin $a=0.9$, focusing  on the  role of the  \textbf{BH} photons shell boundaries, $\sigma=\sigma_\pm$, as the extreme points of the functions $\pm\omega_\pm(r)$.
 In Figs\il(\ref{Fig:Plotelemtemporeplicas})--upper left panel, in particular,  we show the  extreme  relativistic angular velocity  $\omega_\pm(\sigma=\sigma_\pm)$, as  function of the  radius $r$,  in relation with the horizons $\omega_H^\pm$ (horizontal lines), where solid vertical   green lines  define  the  coordinates where the outer horizon replica crosses the photons shell boundary, as  point   $(\sigma_\pm,r_{\sigma_\pm})$, where $\omega_H^+=\omega_+(\sigma_\pm)$.  On  radius  $r_{\sigma_\pm}^{(1)}$,  there  is the maximum of $\omega_+$, for $\sigma=0.45$.  The case for angle $\sigma=0.45$ is   illustrated in details in the Figs\il(\ref{Fig:Plotelemtemporeplicas})--upper right panel,  and in
Figs\il(\ref{Fig:Plotelemtemporeplicas})--bottom panel,  where horizons replicas surfaces, $\mp\omega_H^\pm=\omega_\pm$, are shown for   the  photons shell, for the \textbf{BH} with spin $a=0.9$.
\subsubsection{The red-shift function  and the photons shell horizons replicas}\label{Sec:redffhidft-shellsreplicas}
\begin{figure}
\centering
\includegraphics[width=8cm]{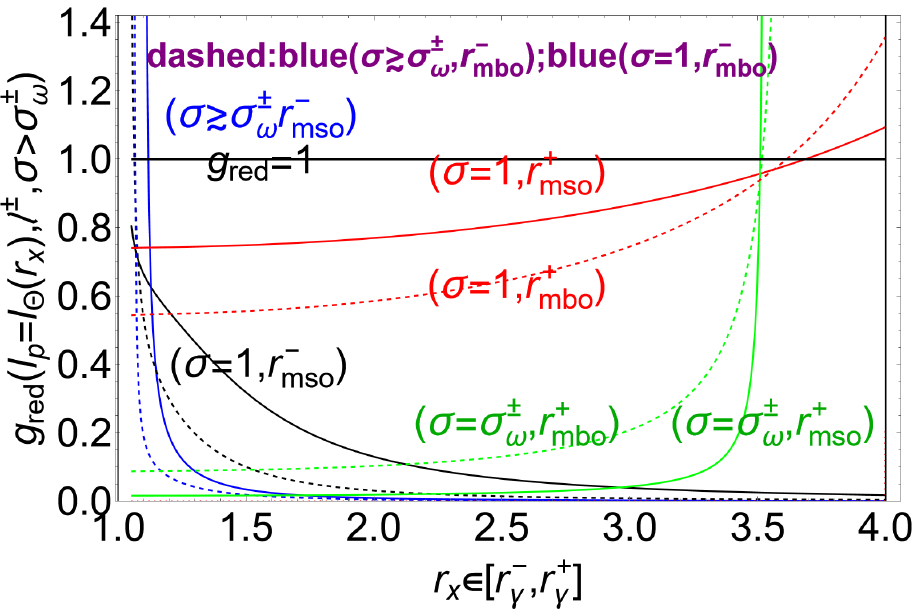}
\caption{Red-shift function $g_{red}\left(l_p=l_{\Theta } (r_\times),\ell^\pm,\sigma >\sigma _{\omega }^{\pm }\right)$ evaluated for the photon impact parameter $l_p=\ell_\Theta(r)$, where $\ell_\Theta(r_\times)$ is the photons shell photon impact parameter in Eq.\il(\ref{Eq:lqsoluzione}), as function of the radius $r_\times\in[r_\gamma^-,r_\gamma^+]$, included in the photons shell, where $r_\gamma^\pm$ are the counter--rotating and co--rotating photon circular orbits. There is the gas angular  momentum $\ell=\ell^\pm(r)$ evaluated on the marginally stable orbit $r_{mso}^\pm$ and marginally bounded orbits $r_{mso}^\pm$, for different values of the angle $\sigma$, where $\sigma_\omega^\pm$ is defined in Eq.\il(\ref{Eq:sigmaomega}). That is here
$l_p=\ell_\Theta=\ell_\Theta(r_\times)$
$
g_{red}(l_p,\ell^\pm(r),r,\sigma)$
where $r\neq r_\times$ is fixed on the panel.}\label{Fig:PlottSta}
\end{figure}

In Figs\il(\ref{Fig:PlottSta})  the red-shift function $g_{red}$ is evaluated  for $\ell=\ell^\pm$  and $\sigma >\sigma _{\omega }^{\pm}$, and  for the photons shell photon impact parameter $l_p=\ell_\Theta(r_\times)$ in Eq.\il(\ref{Eq:lqsoluzione}). The quantity $g_{red}$  is then showed  as function of the radius $r_\times\in[r_\gamma^-, r_\gamma^+]$,  in the \textbf{BH}  photons shell. The gas angular  momentum $\ell=\ell^\pm(r)$ is evaluated on the marginally stable orbit $r_{mso}^\pm$ and marginally bounded orbits $r_{mso}^\pm$. Different values of the angle $\sigma>\sigma_\omega^\pm$ are considered. Angle $\sigma_\omega^\pm$ defines  the constraints   in Eq.\il(\ref{Eq:sigmaomega}).  Therefore, there is
$l_p=\ell_\Theta=\ell_\Theta(r_\times)$, while it
$
g_{red}(l_p,\ell^\pm(r),r,\sigma)$, for  fixed  $r\neq r_\times$.
 The signals are prevalently red-shifted close to the boundaries  $r=r_\gamma^\pm$. The red-shift function  increases (smaller red-shift effects on the signal), in general,  with the angle, and decreases  with the  distance from the central attractor,  for the counter--rotating fluids. Viceversa, for co--rotating fluids,  the red-shift function increases  with the  angle, and decreases with distance from the attractor. %\btb
\subsubsection{Co--rotating and counter--rotating horizons  replicas}\label{Sec:co--rotating-counter-rpliAs}
In this section we consider more closely the  notion of the co--rotating and counter--rotating \textbf{BH}  horizons replicas, in relation to the properties of the orbiting disks and the \textbf{BH} photons shell.

\begin{figure}
\centering
\includegraphics[width=8cm]{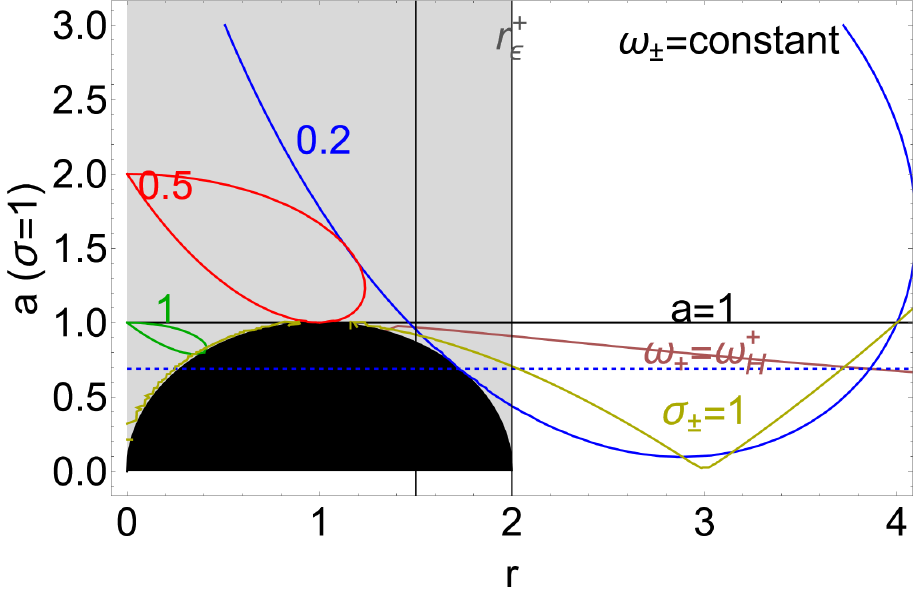}
\includegraphics[width=8cm]{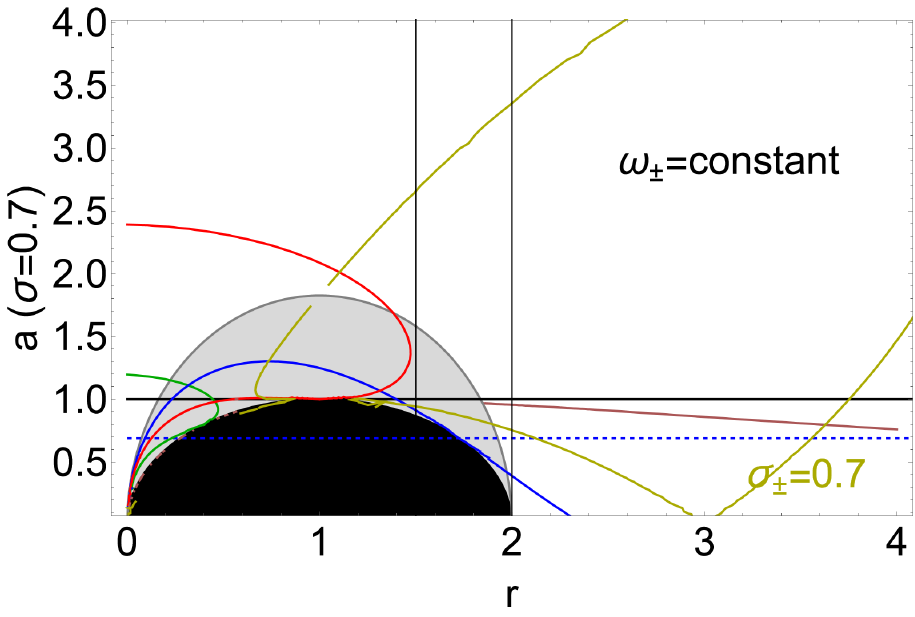}
\includegraphics[width=8cm]{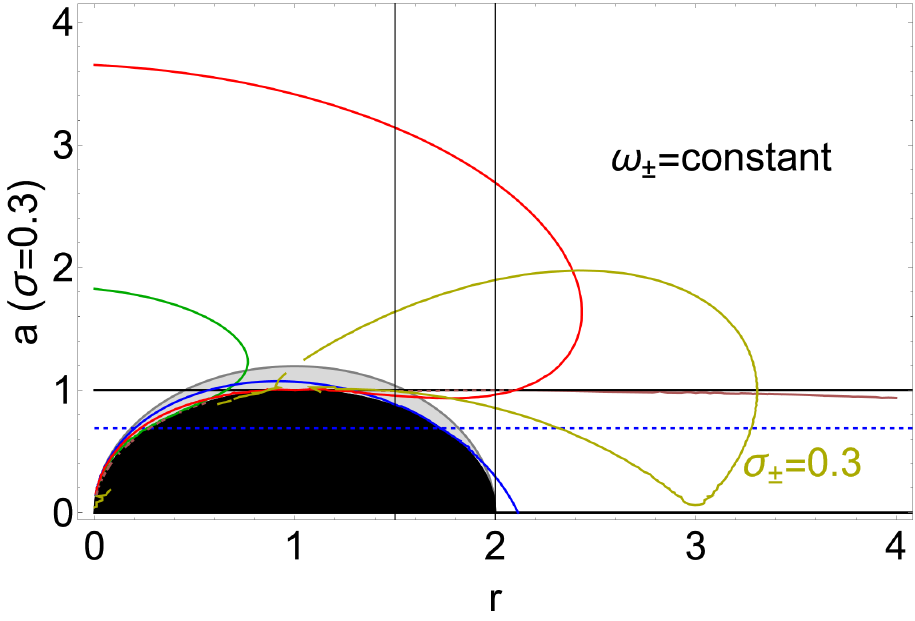}
\includegraphics[width=8cm]{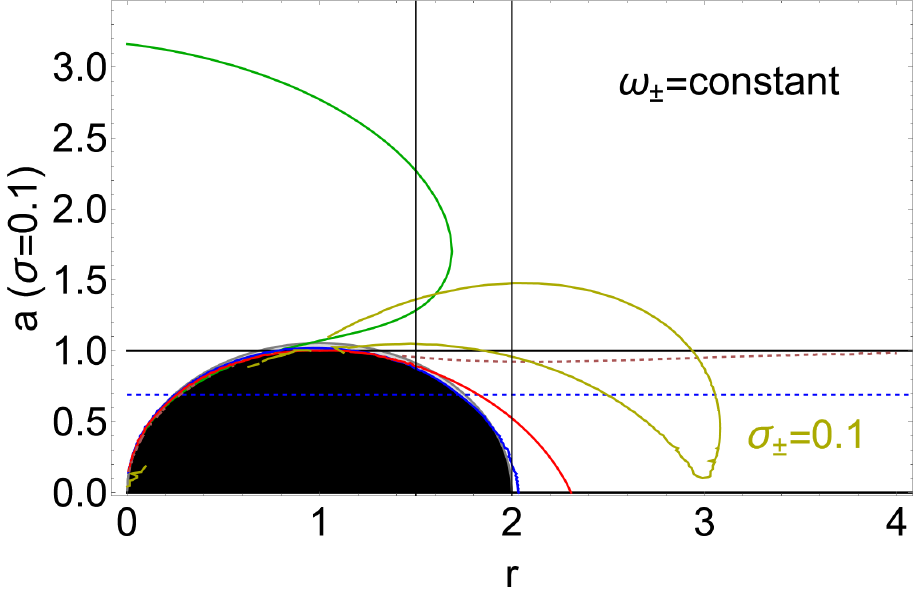}
\caption{Metric Killing bundles, horizons replicas and photons shell. Black region is the \textbf{BH} region in the (extended) plane $(a,r)$. The boundary of the  region  for $r\in[1,2]$ ($r\in[0,1]$) comprises  the outer (inner) horizons of all the  Kerr \textbf{BH} (there is $a\in[0,1]$).  For $a>1$ is the \textbf{NS} region. Gray region is the outer and inner   ergoregion. Colored curves  are the Metric Killing bundles (the curves $\omega_\pm=\omega$). Each curve is tangent to the an horizon point, the metric bundle angular velocity $\omega_\pm$ is the angular velocity of the horizon identified by the tangent point. Therefore each point of the metric bundle curve is an horizon replica. Yellow curve is the photons shell, $\sigma_\pm=\sigma$ in the   extended plane. Each panel is a for a fixed angle $\sigma$. Pink curve is the curves of all the outer horizons replica. As  example we show the outer horizon replica for $\omega_H^+=0.2$ (the corresponding bundle is the blue curve)  corresponding  \textbf{BH} with $a=0.689655$, and the replica in this spacetime distinguished by the crossing  blue curve with dotted blue curve (the curve $a=0.689655$).}\label{Fig:PlotturenesaAsecc1}
\end{figure}

In Figs\il(\ref{Fig:PlotturenesaAsecc1}) we show the metric Killing bundles, the horizons replicas,  the spacetimes  photons shell and inner and outer ergoregions in the extended plane $(a,r)$, for all the Kerr  \textbf{BHs} and  the \textbf{NSs}, at different angle $\sigma$.
Yellow curve is the photons shell boundaries, $\sigma_\pm=\sigma$, in the   extended plane.
Metric Killing bundles are the  colored  curves $\omega_\pm=\omega=$constant.  Each curve is tangent to the horizons curve $a_\pm\equiv \sqrt{r(2-r)}$ (bounding the black \textbf{BH} region) in a  point,  correspondent to an inner or an  outer \textbf{BH} horizon, whose relativistic  angular velocity,  $\omega_H^-$ or  $\omega_H^+$,   respectively,   is the  metric bundle angular velocity $\omega_\pm$.
The angular velocity defining the metric bundle is identified therefore  by  the tangent point   of the bundle curve  to the horizons curve, in the extended plane. Each point of the metric bundle curve is an horizon replica, of the horizon distinguished by  its tangent point.
Pink curves in  Figs\il(\ref{Fig:PlotturenesaAsecc1})  are the outer horizons replicas, that is all the orbits $r$, in all the spacetimes (at fixed $\sigma$), defined by the parameter $a$,  in the extended plane,  where a limiting relativistic
 photon frequency is an  outer horizon replica, for all the \textbf{BH} with spin $a\in[0,1]$, at fixed angle.

\begin{figure}
\centering
\includegraphics[width=8cm]{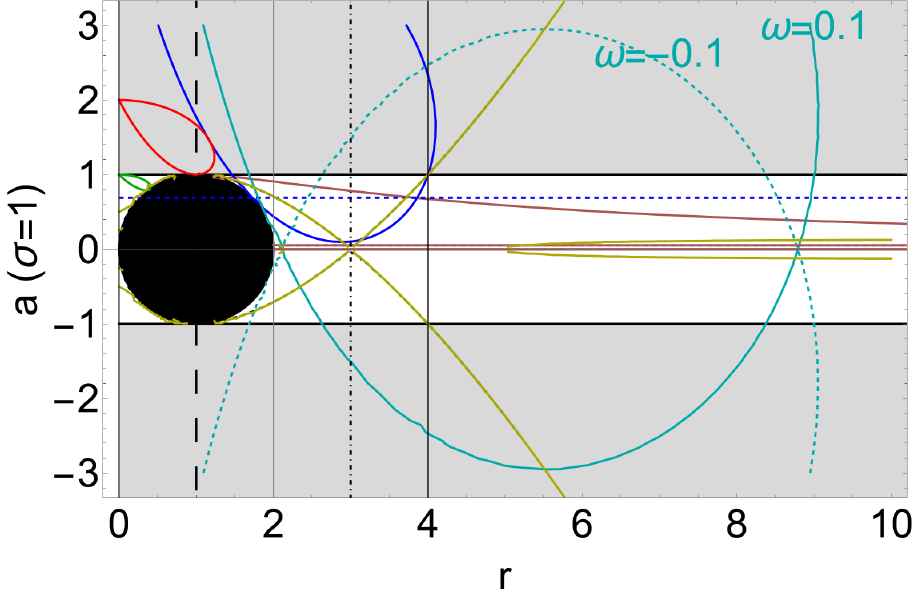}
\includegraphics[width=8cm]{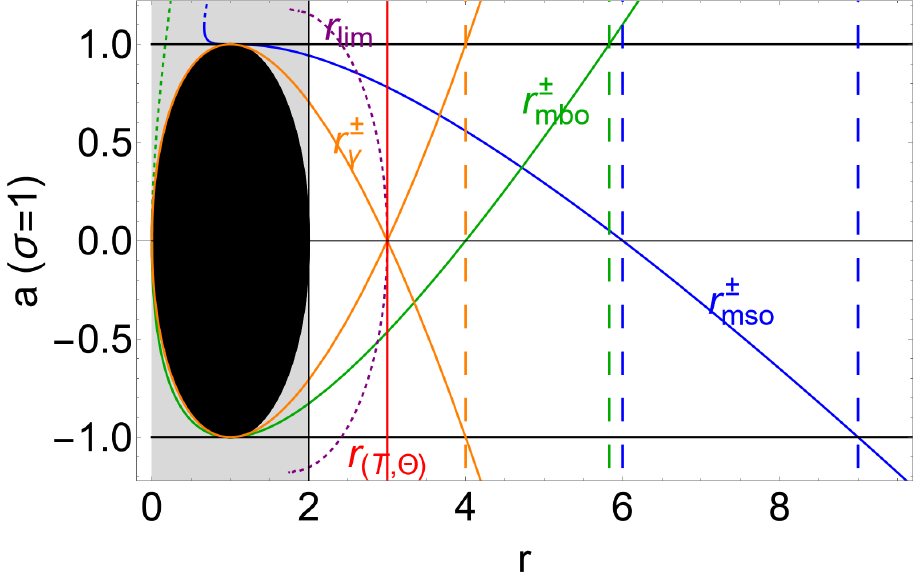}
\caption{Left panel: Counter--rotating horizon replicas in the extended plane for the angle $\sigma=1$. Bundles angular velocities $\omega_\pm=\pm 0.1$ are shown as cyan curves in the plane where $a\in[-3,3]$. For further details see also description  in Figs\il(\ref{Fig:PlotturenesaAsecc1}). In this panel the gray region in the \textbf{NS} part of the extended plane. Right panel:
in the extended plane on the equatorial plane radii $r_{\lim}$ of Eq.\il(\ref{Eq:rlimite}), the photon circular orbits $r_\gamma^\pm$, the marginally stable (bounded) orbit $r_{mso}^\pm$ ($r_{mbo}^\pm$)  and $r_{(\Ta,\Theta)}$ in Eq.\il(\ref{Eq:rthetat-definition}), for counter--rotating and co--rotating fluids are shown. Vertical lines show the  cases for the \textbf{BH} spins $a\in\{\pm1, 0\}$. (Gray region is the ergoregion)}\label{Fig:Plotturecc11}
\end{figure}

\begin{figure}
\centering
\includegraphics[width=7cm]{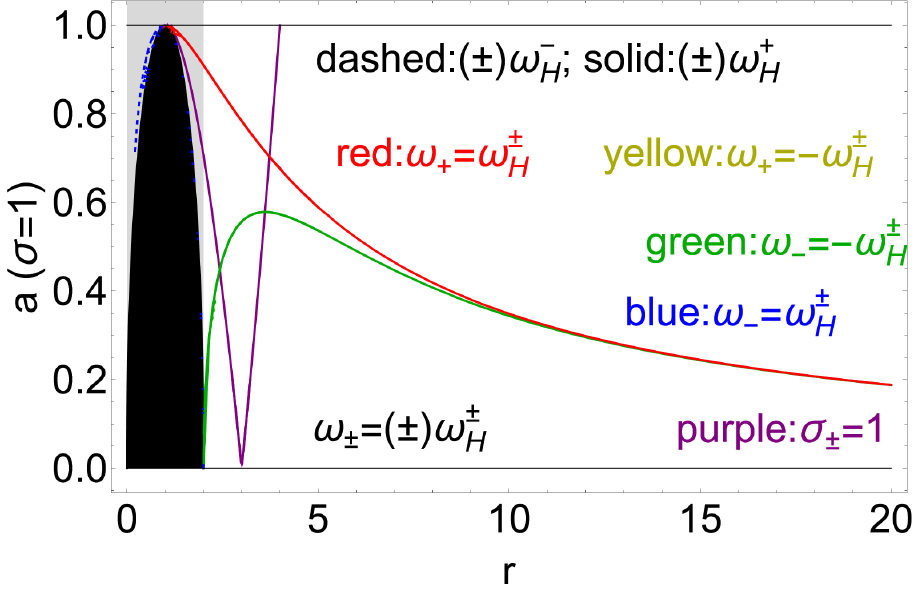}
\includegraphics[width=7cm]{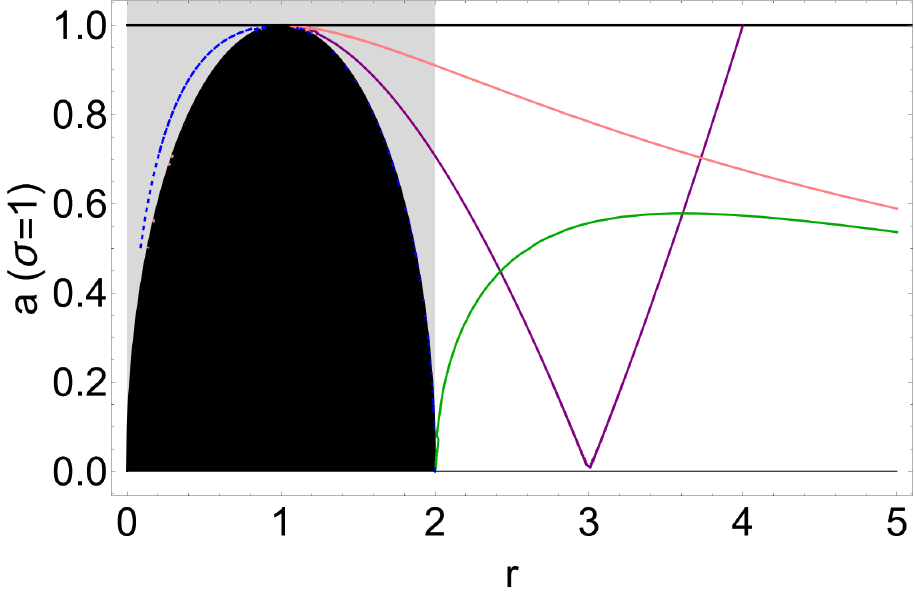}
\includegraphics[width=7cm]{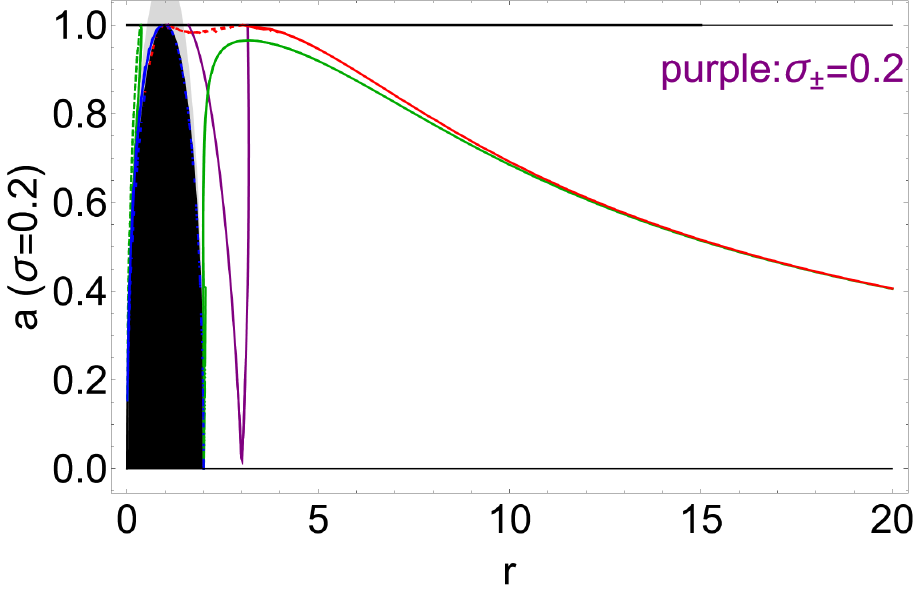}
\includegraphics[width=7cm]{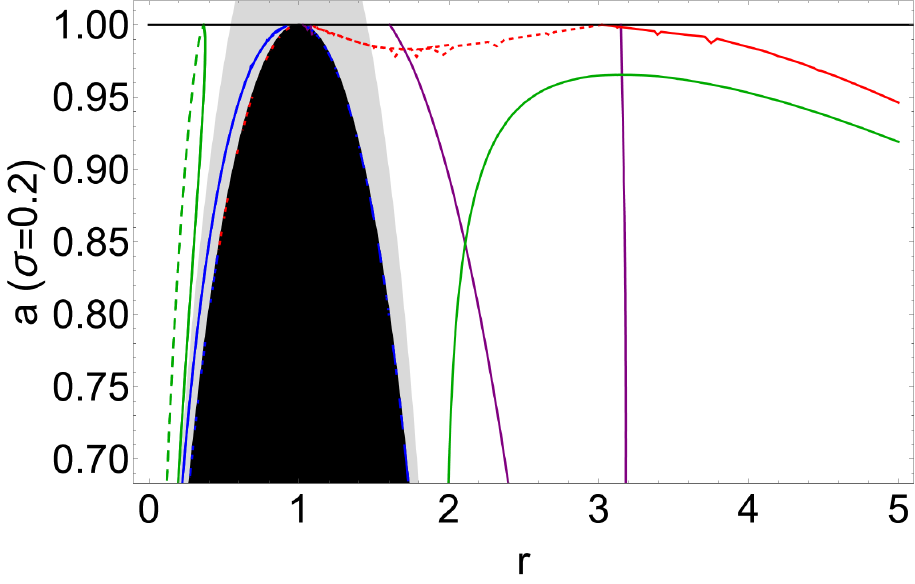}
\includegraphics[width=7cm]{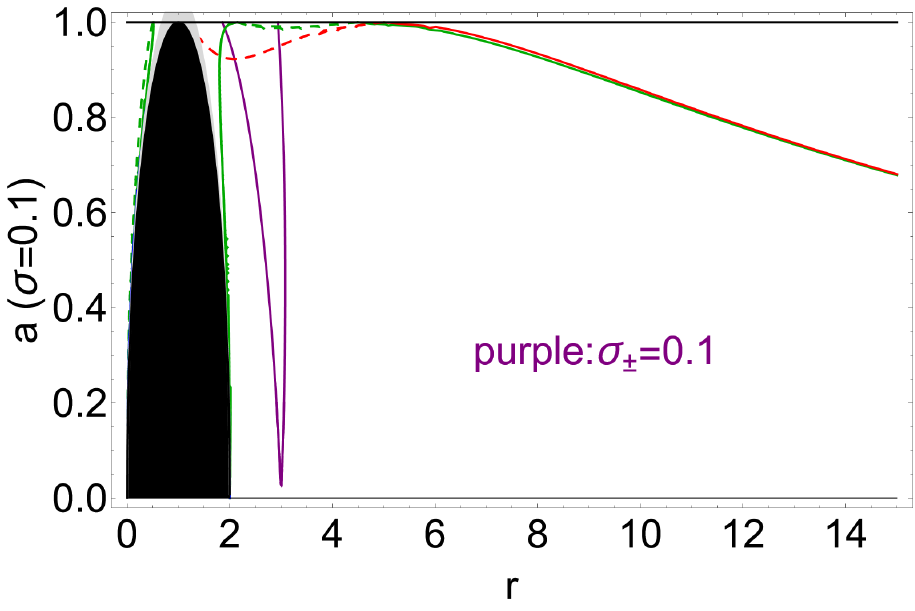}
\includegraphics[width=7cm]{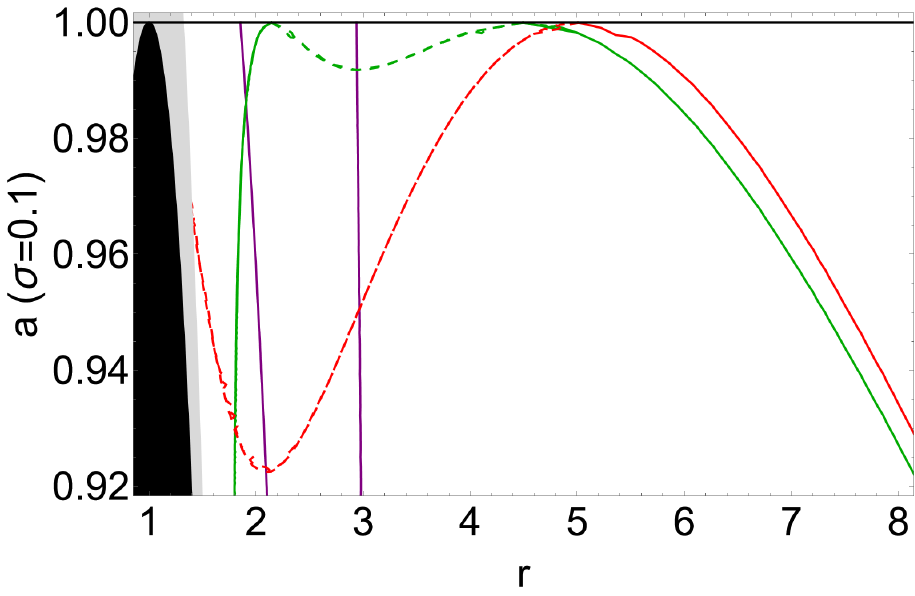}
\includegraphics[width=7cm]{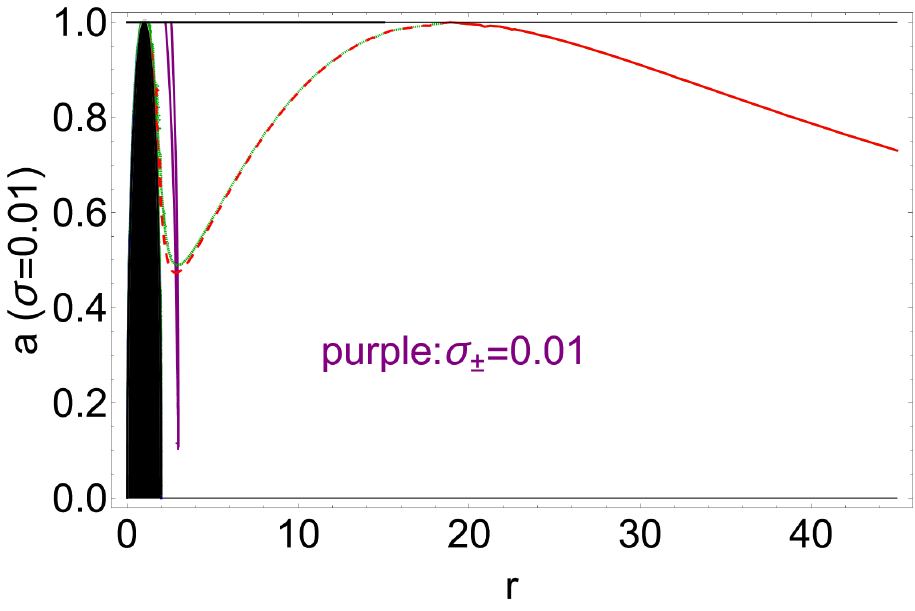}
\includegraphics[width=7cm]{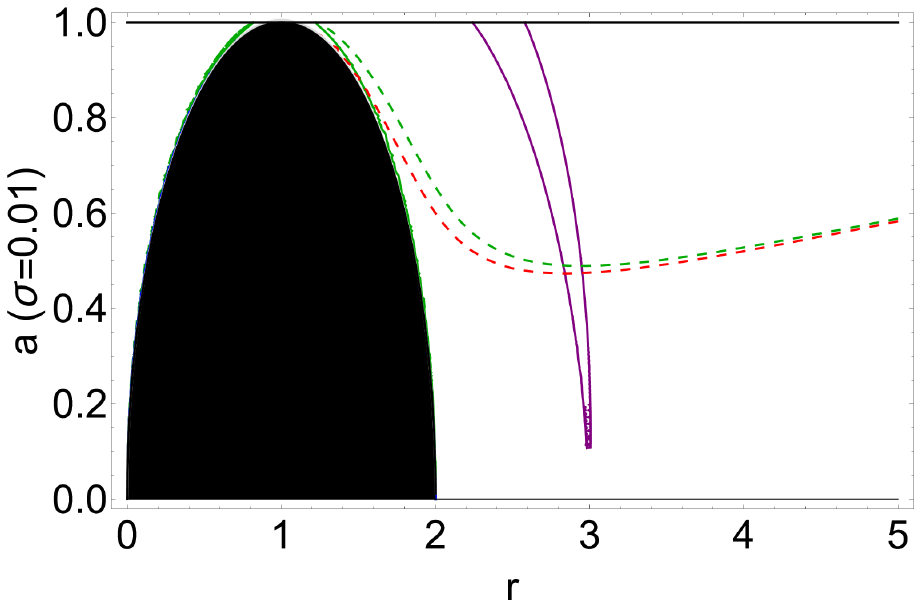}
\includegraphics[width=7cm]{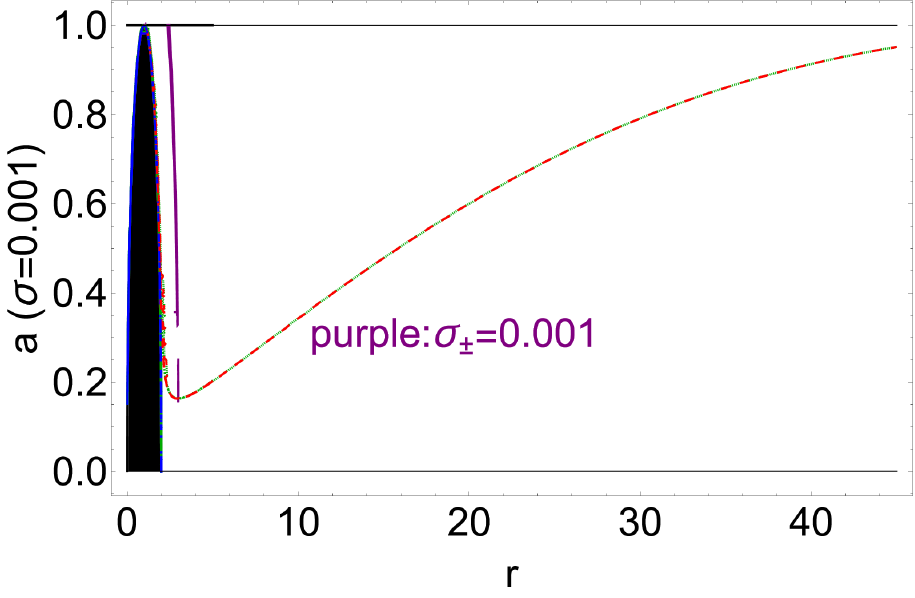}
\includegraphics[width=7cm]{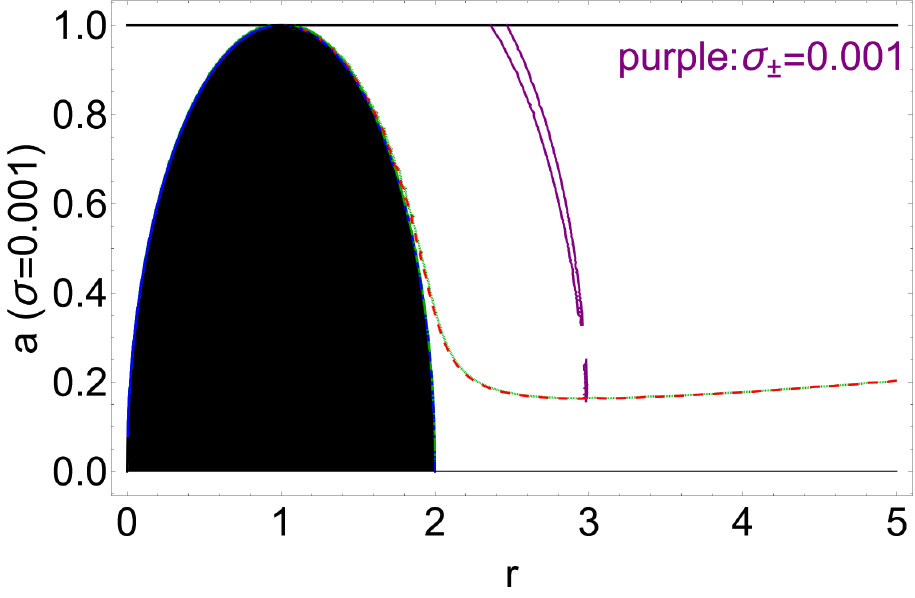}
\caption{All the \textbf{BH} co--rotating and counter--rotating, inner and outer horizons replicas in the \textbf{BH} region of the extended plane for selected angles $\sigma$ as written on the panel. For details on the extended plane features see description of Figs\il(\ref{Fig:PlotturenesaAsecc1}).  Each row is for a fixed angle $\sigma$. Right column panels are  close--up views of the left column panels. Purple curve is the photons shell boundaries  $\sigma_\pm=\sigma$ in all the \textbf{BH} spacetimes. Other colored curves are the horizons replicas $\omega_\pm=\pm\omega_H^\pm$ of all the \textbf{BHs}.}\label{Fig:PlotturenesaAu3}
\end{figure}

\begin{figure}
\centering
\includegraphics[width=8cm]{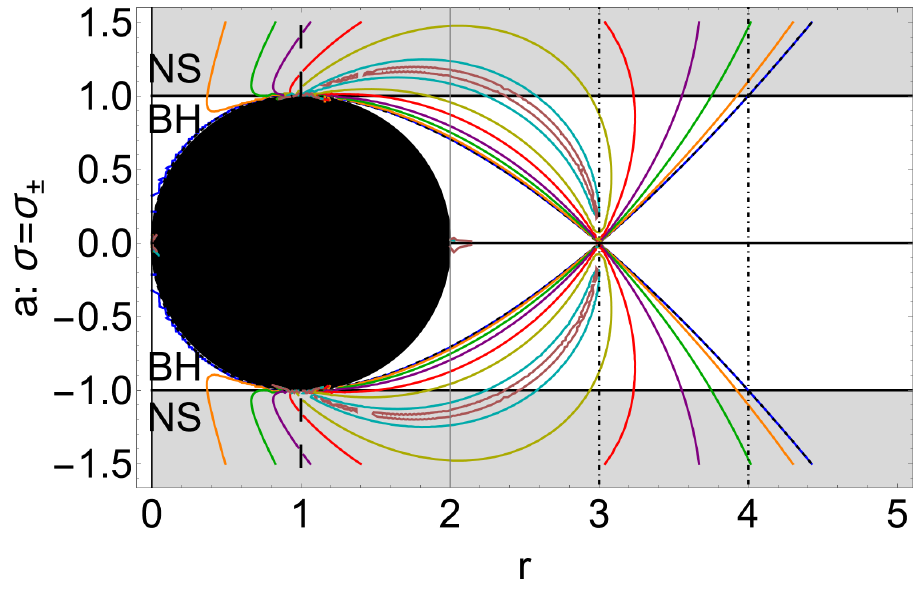}
\includegraphics[width=8cm]{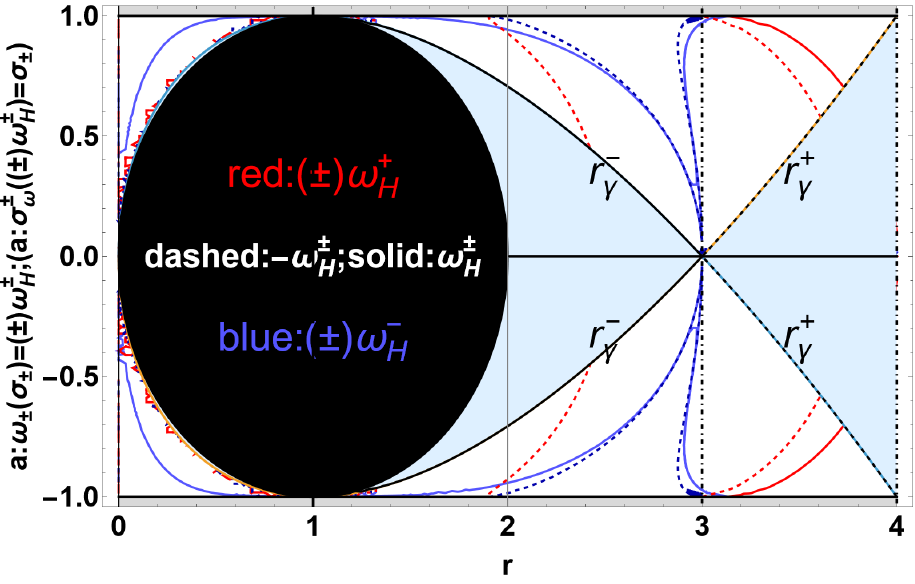}
\includegraphics[width=8cm]{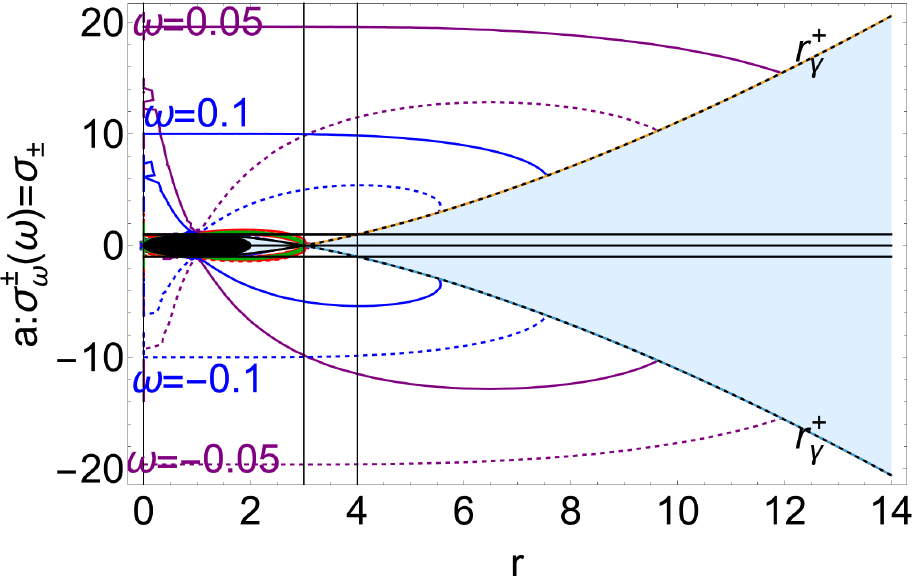}
\includegraphics[width=8cm]{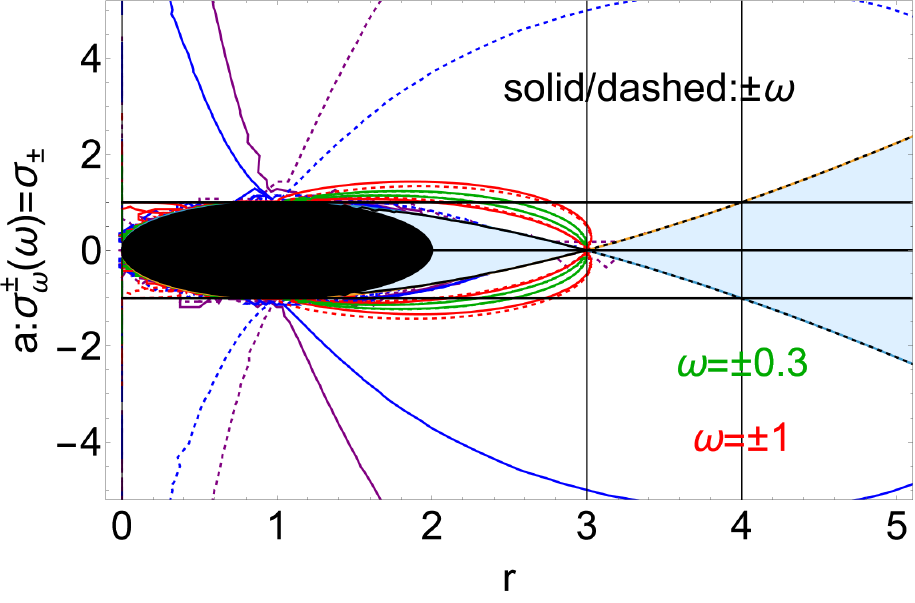}
\caption{Photons shell in the extended plane. Upper left panel: curves $\sigma_\pm=\sigma=$constant in the extended plane $(a,r)$.  $\sigma_\pm$ are the photons shell boundaries expressed in terms of the angle $\sigma\in[0,1]$. Central black region is bounded by  horizon curve in the extended panel, including  the inner $(r_-\in]0,1])$ and  outer $r_+\in[1,2]$ horizons  of all the \textbf{BHs} with spin in $a\in[-1,1]$. Black curve bounding all the $\sigma_\pm=\sigma$ colored curves are the photons shell boundaries  for $\sigma=1$ (the equatorial planes), limited by the radii $r_\gamma^\pm(a)$ which are the counter--rotating and co--rotating photon circular orbit on the equatorial plane, respectively. At $r=3$ there is, on the line $a=0$ the photon circular orbit,
at $r=4$ and $a=\pm 1$ is the counter--rotating  photon circular orbit (on the equatorial plane). Gray regions are the \textbf{NS} regions of the extended plane. Upper right panel:  the horizons replicas located on the photons shells are shown in the extended plane as   the solutions $a:\omega _{\pm }\left(\sigma _{\pm }\right)=(\pm )\omega _H^{\pm }$, or equivalently $a:\sigma _{\omega }^{\pm }\left((\pm )\omega _H^{\pm }\right)=\sigma _{\pm }$, where $\omega _H^{\pm }$ are the outer and inner horizon angular frequency respectively. The sign $(\pm)$ for $(\pm )\omega _H^{\pm }$ are for counter--rotating and co--rotating horizons angular velocity respectively. $\sigma _{\omega }^{\pm }$ is defined in Eq.\il(\ref{Eq:sigmaomega}). Note these, replicas are not defined in the shaded region included in the curves $a: r=r_\gamma^\pm$. Bottom left panel: metric Killing bundles on the photons shell boundaries for fixed angular velocities signed on the panel  in the \textbf{BH} and \textbf{NS} regions of the extended plane. Bottom right panel is a close up-view of the bottom left panel where the  \textbf{BH} region is highlighted.}\label{Fig:PlotturenesaAbrxesR}
\end{figure}

As  example, we singled--out,  in  Figs\il(\ref{Fig:PlotturenesaAsecc1}),   the outer horizon replica for the angular velocity  $\omega_H^+=0.2$,   pertaining    to the  \textbf{BH} with spin  $a=a_{\odot}\equiv 0.689655$.  The metric bundle with frequency  $\omega_H^+=0.2$  is the blue curve, tangent to the outer  horizons curve in  the point  $a=a_{\odot}$ and $r=r_+(a_{\odot})$. The outer horizon replica in the  spacetime $a=a_{\odot}$ is  given by the crossing  of the metric Killing bundle  (the blue curve) with the  horizontal line $a=a_{\odot}$ (dotted blue curve) in the plane. (There is only one outer horizon replica in this spacetime). The outer horizon replica  on the photons shell  boundary is given by the cross of the pink curve  with the yellow curve.
 The photons shell boundary curves in the extended  plane strongly depend on the angle $\sigma$, and it is clear it is not defined in all the \textbf{NSs} spacetimes.
 The curves of the outer horizons  replica, at small angle $\sigma$,  approaches the line $a=1$ (extreme Kerr \textbf{BH}), at fixed  $r$, where there are also  the inner horizons  replicas curves (dashed pink curve)  with $\omega_\pm=\omega_H^-$.
The curves of the outer horizons replica decrease with $r$ i.e. horizons replicas  exist,, for faster spinning attractors, for small  radii (in the regions close to the central  \textbf{BH}). For slowly spinning attractors, the  outer horizon replica move to farther
 radii.
On the other hand,  the inner horizon replicas are defined, for faster spinning  \textbf{BHs},   for small angles $\sigma$ and   radii $r$ (see \cite{2024NuPhB100816700P,2021EPJC...81..258P}).
In Figs\il(\ref{Fig:Plotturecc11}) we consider   the extended plane for  angle $\sigma=1$, where $a\in[-1,1]$.  In particular we show  the relation between the co--rotating and counter--rotating horizons replica (left  panel) and the geodesic radii,  with $r_{\lim}$ and  radius $r_{(\Ta,\Theta)}$ of the photons shell (right panel).
Angular velocities $\omega_\pm=\pm 0.1$ are shown  in the left panel, where  the \textbf{NS} part of the extended plane is in  gray region (note, in left panel, gray region is the ergoregion).
A focus  on the set of  co--rotating and counter---rotating inner and outer horizons replicas in the extended plane (with $a\in[0,1]$) is in Figs\il(\ref{Fig:PlotturenesaAu3}), showed for different angle $\sigma$,   in relation to the \textbf{BH} photons shell.
Notably, there is a small, bounded  class of faster spinning \textbf{BH}  attractors  where, in a fixed spacetime,   there can be  two replicas, defined close to the central attractor.
 For  large $\sigma$, this occurs  for the counter--rotating outer horizon replica, i.e, $\omega_-=-\omega_H^+$. Decreasing the angle $\sigma$,  other replicas are defined also  for slowly spinning  \textbf{BHs}.
In Figs\il(\ref{Fig:PlotturenesaAbrxesR}) we consider again the   whole range of spin in  $a\in[-1,1]$, investigating the properties of the photons shells in  extended plane.  In the upper left panel the photons shell boundaries,  $\sigma_\pm=\sigma=$constant, are shown in the extended plane $(a,r)$, for all values of $\sigma\in[0,1]$. The photons shell, for $\sigma=1$ (the equatorial plane), is  bounded by the radii $r_\gamma^\pm(a)$.
  Increasing  $\sigma\in[0,1]$, the coloured curves approach  the black curves, $r=r_\gamma^\pm(a)$. Upper right panel of Figs\il(\ref{Fig:PlotturenesaAbrxesR})  shows  the co--rotating and counter--rotating  horizons replicas located on the photons shells. Hence   the plot shows  the solutions $a:\omega _{\pm }\left(\sigma _{\pm }\right)=(\pm )\omega _H^{\pm }$ or, equivalently, $a:\sigma _{\omega }^{\pm }\left((\pm )\omega _H^{\pm }\right)=\sigma _{\pm }$  (where $\sigma _{\omega }^{\pm }$ is defined in Eq.\il(\ref{Eq:sigmaomega})).
  These are  therefore the set of replicas crossing  the photons shell boundaries.  Note, these replicas  are not defined in the  regions  bounded by the curves $a: r=r_\gamma^\pm$ (shaded blue regions in  Figs\il(\ref{Fig:PlotturenesaAbrxesR})).
   Bottom  panels of Figs\il(\ref{Fig:PlotturenesaAbrxesR}) show the   metric Killing bundles   for fixed angular velocities  $\omega$, on the photons shell boundaries, in the \textbf{BH} and \textbf{NS} regions of the extended plane.  The curves  represent the solutions  $a:\sigma _{\omega }^{\pm }\left(\omega\right)=\sigma _{\pm }$, for fixed $\omega$.
   As  $\sigma _{\omega }^{\pm }$ is a solution of $\omega=\omega_\pm$, the curves represent, for fixed $\omega=$constant, the solutions
$\omega_\pm(a,\sigma_\pm)=\omega=$constant, that is the set  points
$(a,r)$, in the extended plane, where the photon circular orbit has  relativistic angular velocity $\omega=$constant on the photons shell boundary, i.e. on the plane  provided by $\sigma=\sigma_\pm (a,r)$.
Note that there is $\omega_H^->1/2$  and $\omega_H^+<1/2$.
\subsubsection{Red-shift function in the extended plane }\label{Sec:absidered-extended-plane}
\begin{figure}
\centering
\includegraphics[width=8cm]{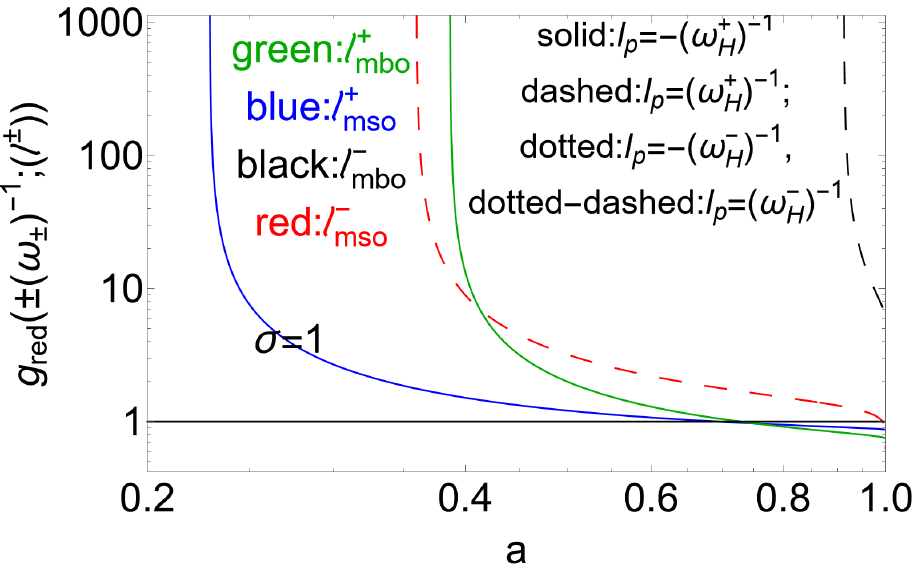}
\includegraphics[width=8cm]{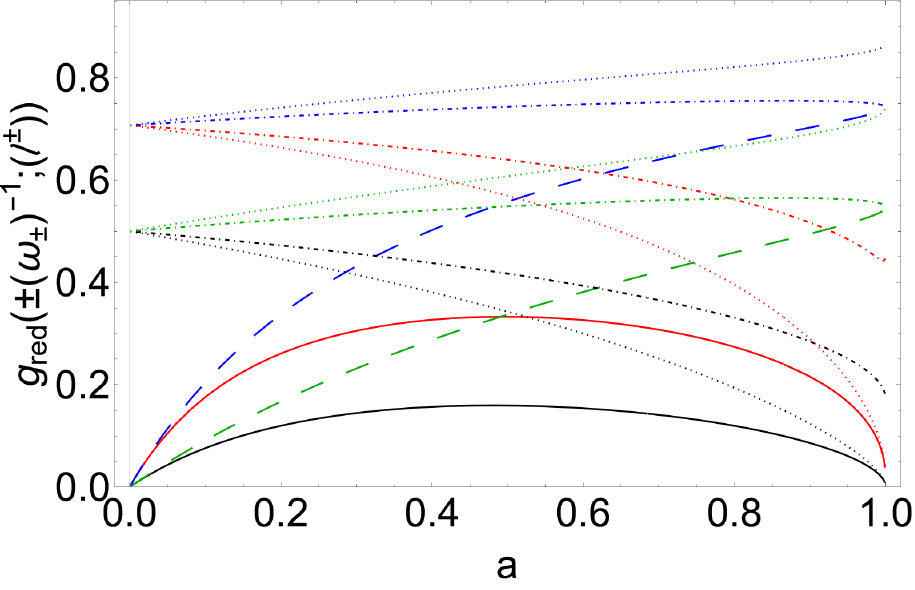}
\caption{Red-shift function $g_{red}$  is shown as function of the \textbf{BH} spin $a\in[0,1]$ and  evaluated for  the photon impact parameter $\ell_p=\pm 1/\omega_H^\pm$, where $\omega_H^\pm(a)$ are the angular velocity of the outer and inner \textbf{BH} horizons respectively. The gas specific  angular  momentum is  $\ell=\ell^\pm(r)$, for counter--rotating and co--rotating fluids respectively,  which are evaluated in this analysis on the radii $r_{mso}^\pm$ and $r_{mbo}^\pm$ of the geodesic structure, according to the color code of the left panel. (There is $\Omega=\Omega(\ell^\pm(r),r,\sigma)$ evaluated on fixed radii for $\sigma=1$ (the equatorial plane)).}\label{Fig:PlotergosclienC}
\end{figure}

In Figs\il(\ref{Fig:PlotergosclienC}) the red-shift function $g_{red}$  is evaluated for  the photon impact parameter $l_p=\pm 1/\omega_H^\pm$\footnote{It is easy to see that bundles can similarly be defined in terms of a characteristic  momentum $\ell=1/\omega$ \cite{2024NuPhB100816700P,2021EPJC...81..258P}.}.

The gas specific  angular  momentum is  $\ell=\ell^\pm(r)$, for counter--rotating and co--rotating fluids respectively,   which are  evaluated, in  Figs\il(\ref{Fig:PlotergosclienC}), on the radii $r_{mso}^\pm$ and $r_{mbo}^\pm$ of the geodesic structure. Thus,  the relativistic angular velocity of the gas is $\Omega=\Omega(\ell^\pm(r),r,\sigma)$,  evaluated on fixed radii for $\sigma=1$ (note, in this case $r$ is the cusp or a center of a torus or a cusp of a proto-jets according to the constraints discussed in Sec.\il(\ref{Appendix:INNer-Edge})).
It is interesting to note the different solutions for blue-shifted and red-shifted signals, in the spacetimes of the  faster and slowly spinning attractors.
With  $l_p=-1/\omega_H^+$ and $\ell\in(\ell_{mso}^-,\ell_{mbo}^-)$ (solid black and red lines), the emission is red-shifted. However,  the trend with spin (right panel) shows  a similar situation  for  the  faster   and slowly spinning  attractors, with a minimum of red-shift for $a\approx 0.5$.
Signals with $l_p=1/\omega_H^+$ and $\ell\in (\ell_{mso}^-,\ell_{mbo}^-)$ (dashed black and red lines)  are blue--shifted.
Signals with $l_p=-1/\omega_H^+$ and $\ell\in (\ell_{mso}^+,\ell_{mbo}^+)$ (solid  blue  and green lines)  are mostly  blue--shifted, and red-shifted for the  faster \textbf{BHs}.
All the other analyzed profiles  are red-shifted,  having different dependence from the  the  \textbf{BH} spin.  For $l_p=\pm1/(\omega_H^-)$, and part of $l_p=1/\omega_H^+$,  all the signals are red-shifted.
In general  there is $\ell(\omega_H^\pm)\neq 1/\omega_H^\pm$,  and $\Omega(1/\omega_H^\pm)\neq \omega_H^\pm$. Correspondingly, we could consider the signals red-shift,  assuming  three  different conditions on the gas particle orbits, with \emph{ i.} $\Omega=\omega_H^\pm$,  \emph{ii.} $\ell= 1/\omega_H^\pm$, or also  \emph{ iii.}   $\ell=\ell(\omega_H^\pm)$.
We focus here on the condition
 when $\Omega(\ell)=\omega_H^\pm$.
We  proceed by considering the functions $g_{red}(\omega_H^\pm)$, irrespectively by the  specifics of the emitting particle orbital model.
Alternatively,  we can discuss  the  solutions  where  $\ell^\pm:\ell^\pm=\ell(\omega_H^\pm)$ or, equivalently,   $\ell^\pm=\ell:\Omega(\ell)=\omega_H^\pm$, and considering also the correspondent  counter--rotating orbits.

\begin{figure}
\centering
\includegraphics[width=5cm]{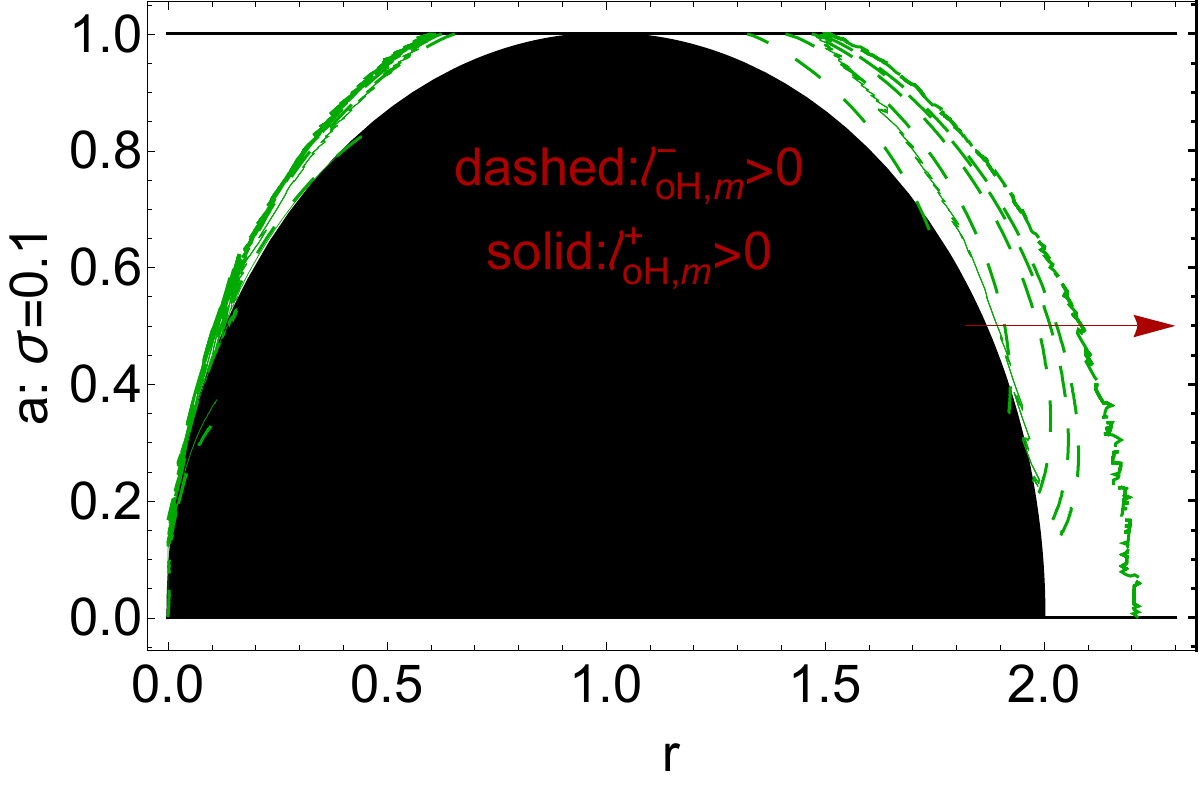}
\includegraphics[width=5cm]{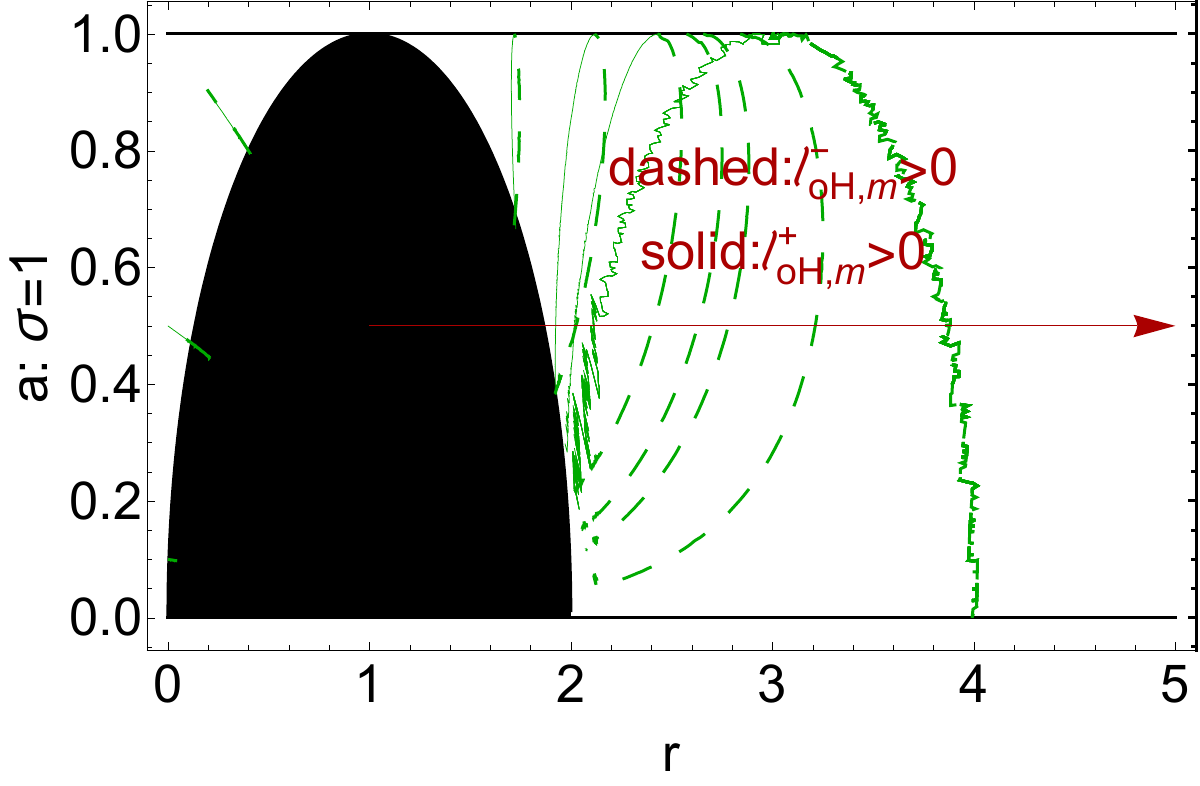}
\includegraphics[width=5cm]{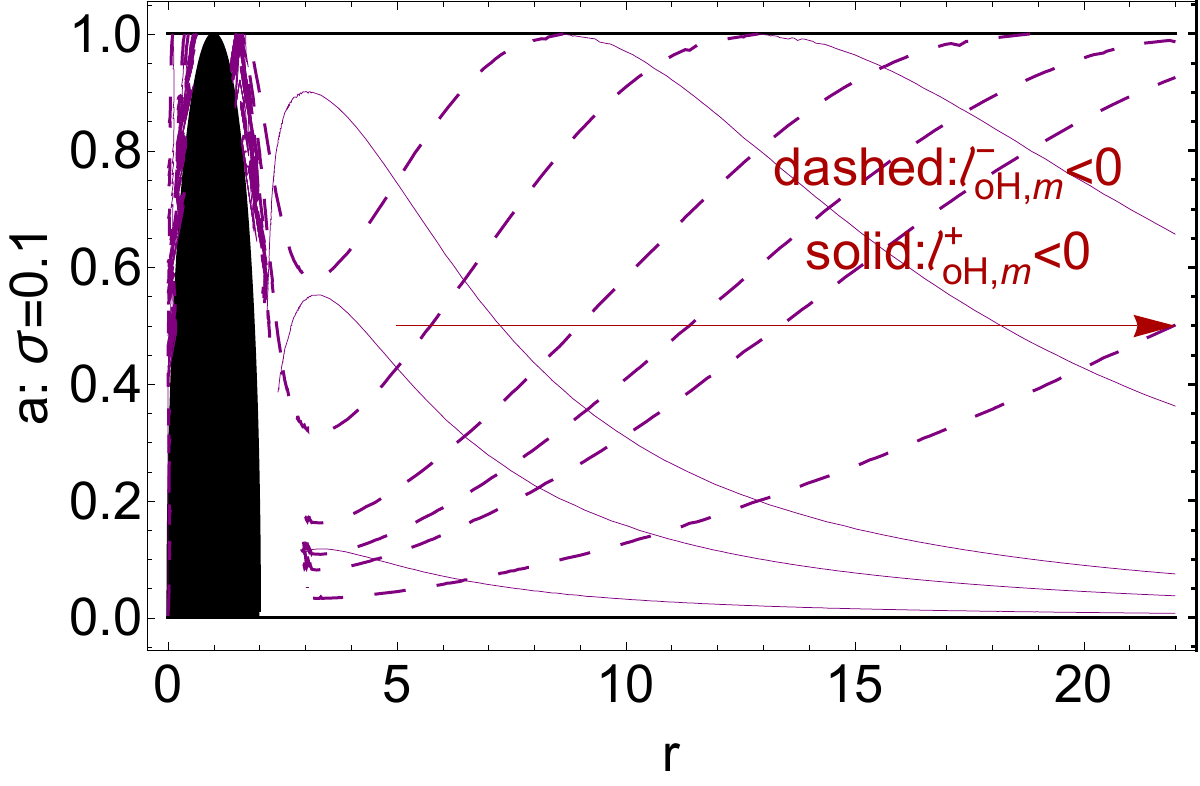}
\includegraphics[width=5cm]{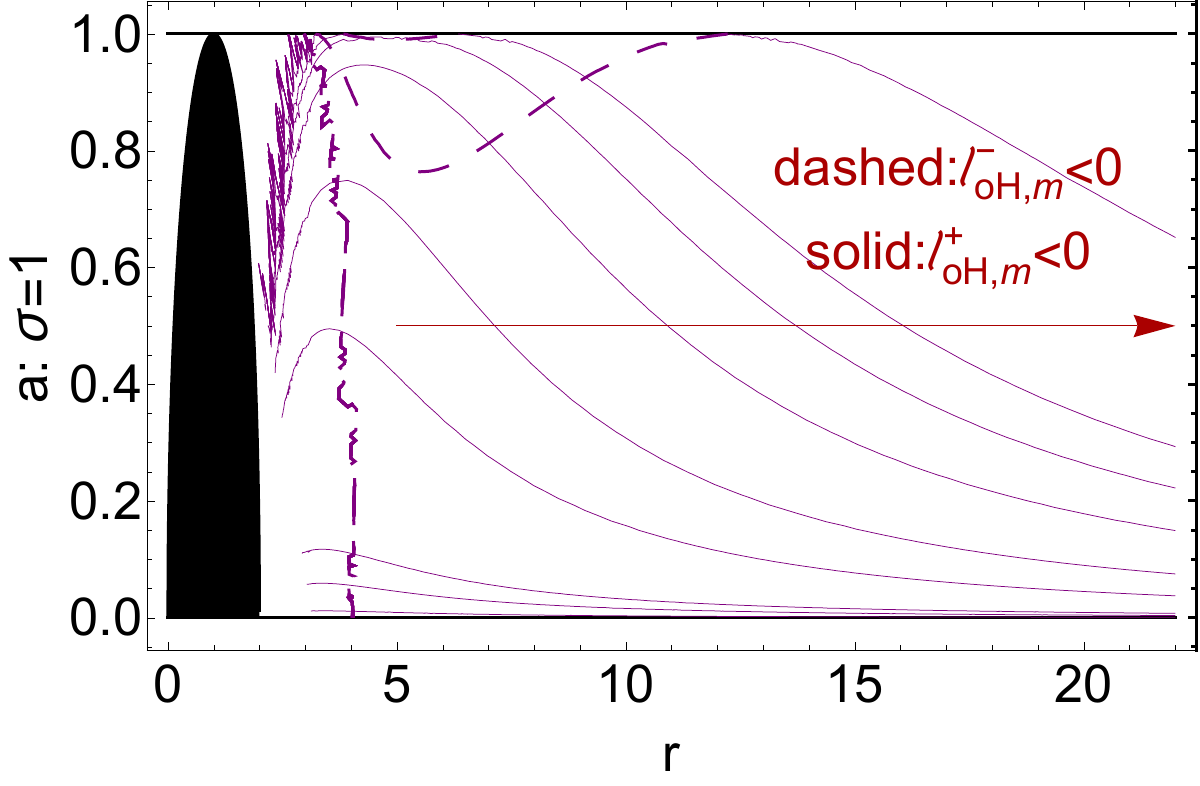}
\includegraphics[width=5cm]{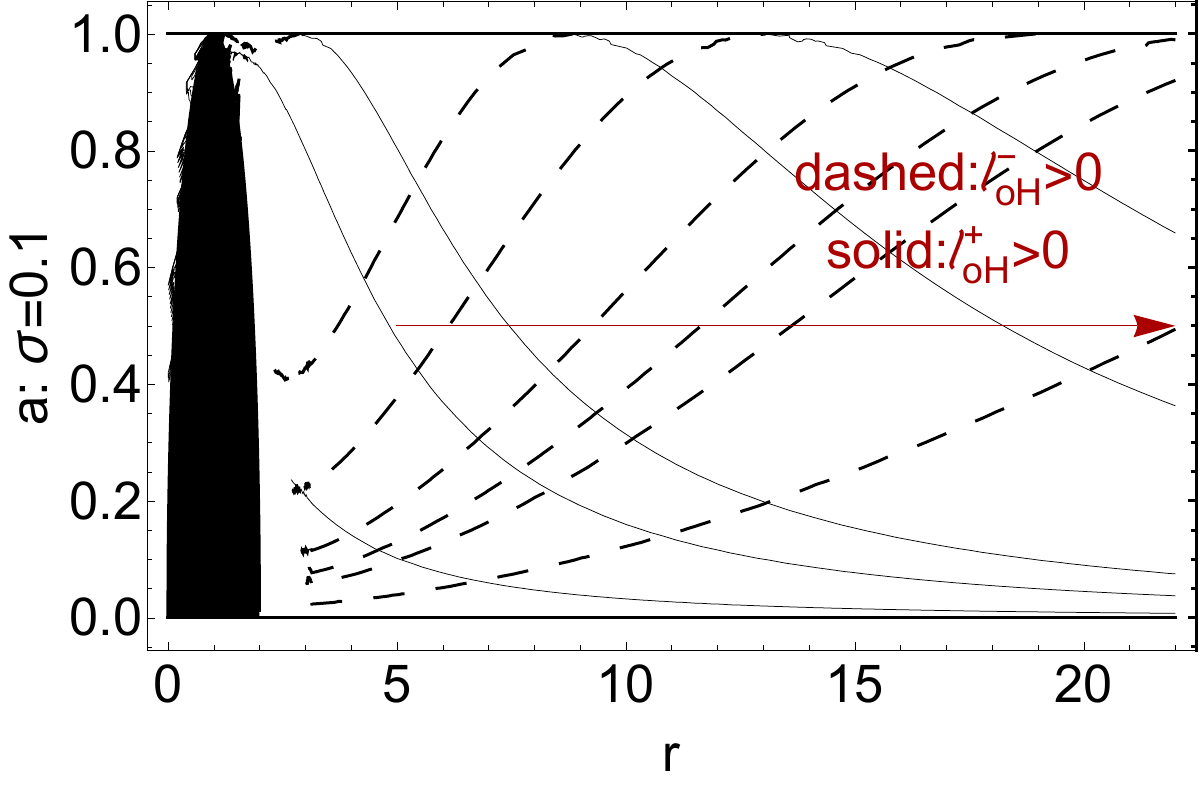}
\includegraphics[width=5cm]{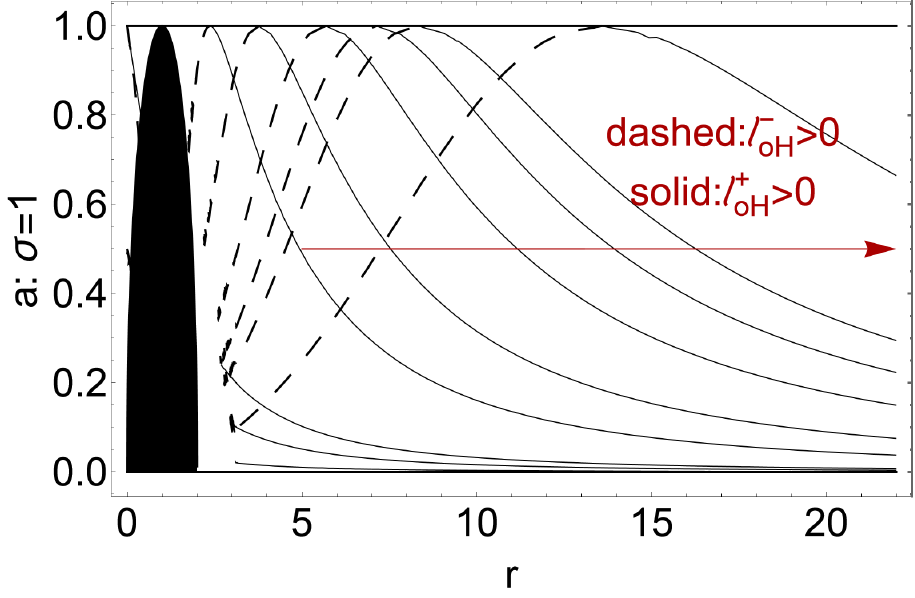}
\caption{Conditions $\ell_{oH}^\pm\lessgtr 0$ and  $\ell_{oH,m}^\pm\lessgtr 0$ in the extended plane $(a,r)$ for different angles $\sigma$. The black region is the \textbf{BHs} regions  $(r\in[r_-,r_+])$ in the extended plane. The arrow indicates the increasing  values of  $\ell_{oH}^\pm$=constant$\lessgtr 0$ and  $\ell_{oH,m}^\pm$=constant$\lessgtr 0$ in magnitude. The specific angular momenta $\ell_{oH}^\pm$ and $\ell_{oH,m}^\pm$ are defined in Eq.\il(\ref{Eq:lokmmdelf}).}\label{Fig:PlotiPLaneA}
\end{figure}

Note,  for  $\Omega=\omega_H^\pm$,  the red-shift function is well defined for $l_p<1/(\pm\omega_H^\pm)$, with further  constraints on  the coordinates  $(r,\sigma)$,  from the normalization condition on the particle.  A particular class of  solutions are the radii which are also the  points of the radial distribution $\ell^\pm(r)$, i.e., the  cusp  or center  radius of the disk. %,  o
Therefore, there is %$\Si
\bea\label{Eq:lokmmdelf}
  &&
\ell_{oH}^+:\Omega=\omega_H^+,\quad\mbox{where}\quad \ell_{oH}^+\equiv\frac{r_- \sigma  \left[a^2 \sigma  \Delta-\left(a^2+r^2\right)^2\right]+4 a^2 r \sigma }{2 a\left[a^2 (\sigma -1)- r \left(-\sqrt{1-a^2} \sigma +r+\sigma -2\right)\right]},
\\\nonumber
&& \ell_{oH,m}^+:\Omega=-\omega_H^+,\quad\mbox{where}\quad \ell_{oH,m}^+\equiv\frac{\sigma  \left(r_- \left(a^2+r^2\right) \Sigma +2 a^2 r \left[-\sqrt{1-a^2} \sigma +\sigma +2\right]\right]}{2a \left[a^2 (\sigma -1)+ r \left(-\sqrt{1-a^2} \sigma -r+\sigma +2\right)\right]},
\\\nonumber
&& \ell_{oH}^-:\Omega=\omega_H^-,\quad\mbox{where}\quad\ell_{oH}^-\equiv \frac{\sigma  \left(-r_+ \left(a^2+r^2\right)\Sigma -2 a^2 r \left[\sqrt{1-a^2} \sigma +\sigma -2\right]\right)}{2 a\left[a^2(\sigma -1)- r \left(\sqrt{1-a^2} \sigma +r+\sigma -2\right)\right]},
\\\nonumber
&& \ell_{oH,m}^-:\Omega=-\omega_H^-\quad \mbox{where}\quad\ell_{oH,m}^-\equiv\frac{\sigma  \left[r_+ \left(a^2+r^2\right) \Sigma +2 a^2 r \left(\sqrt{1-a^2} \sigma +\sigma +2\right)\right]}{2 a \left[a^2 (\sigma -1)+ r \left(\sqrt{1-a^2} \sigma -r+\sigma +2\right)\right]}.
  \eea
  In Figs\il(\ref {Fig:PlotiPLaneA}) are  the  conditions  where the specific angular momenta, $\ell_{oH}^\pm$ and $\ell_{oH,m}^\pm$, are well defined,  with $\ell_{oH}^\pm\lessgtr 0$ and  $\ell_{oH,m}^\pm\lessgtr 0$. Curves in the extended plane $(a,r)$, for different angles $\sigma$, are  $\ell_{oH}^\pm=$constant$\lessgtr 0$ and  $\ell_{oH,m}^\pm=$constant$\lessgtr 0$. The arrow indicates the increasing  values of $\ell$,  in magnitude. Solutions exist in general for a bounded range of values of  the \textbf{BH} spin.
  At fixed spin $a$,  there are several curves   with   two replicas with same specific angular momentum and horizon angular velocity.
  It is noted how  the condition  $\ell^\pm_{oH,m}>0$  occurs  close to the \textbf{BH} attractor, with   the radius $r$   decreasing with  the angle  $\sigma$.
At large angles, there are  mainly  $\ell_{oH,m}^+<0$. The values of the specific angular momenta increase, in magnitude, with the radius $r$.

\begin{figure}
\centering
\includegraphics[width=8cm]{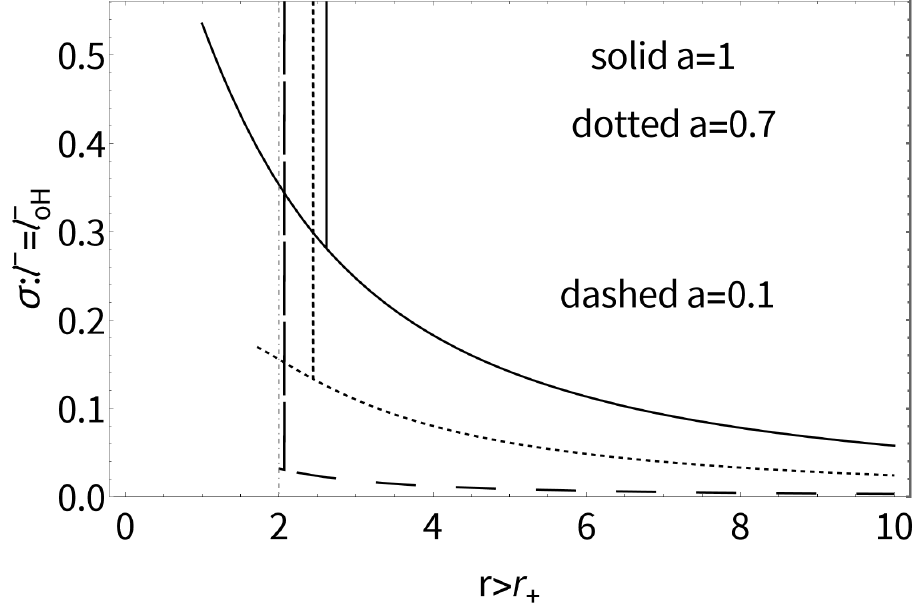}
\includegraphics[width=8cm]{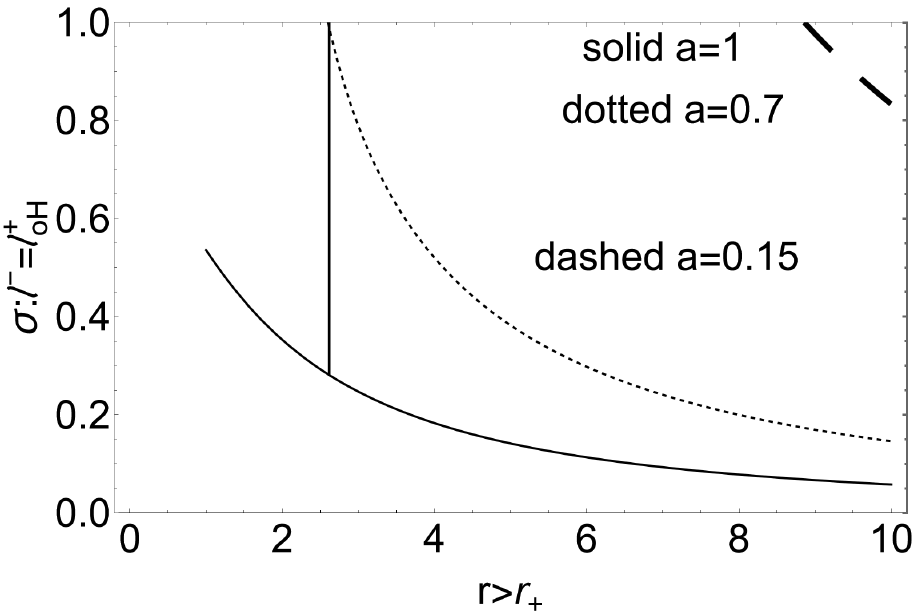}
\caption{The angle $\sigma\equiv\sin^2\theta\in[0,1]$ such that $\ell^-(r)=\ell_{oH}^-$ (left panel) and $\ell^-(r)=\ell_{oH}^+$ (right panel) as function of the radius $r$ for different \textbf{BH} spin $a$. (There have been no solutions of  $\ell_ {oH}^ += \
\ell^+(r)$ and $\ell_ {oH}^ -= \ell^+(r)$). See also Figs\il(\ref{Fig:PlotAPoc1}). The specific angular momenta $\ell_{oH}^\pm$ and $\ell_{oH,m}^\pm$are defined in Eq.\il(\ref{Eq:lokmmdelf}). $\ell^\pm(r)$ are the radial distribution of critical points of density (and pressure) in  counter--rotating and co--rotating toroids, respectively. }\label{Fig:PlotAClooA}
\end{figure}

In Figs\il(\ref{Fig:PlotAClooA}) the
 angle $\sigma:\ell^-(r)=\ell_{oH}^\pm$ is shown  as function of the radius $r$, for different \textbf{BH} spin $a$. The curves are the collections of points $(r,\sigma)$, in a fixed spacetime, where
 the specific angular momenta  $\ell=\ell_{oH}^\pm$  coincide with co-rotating radial distribution of specific angular momentum, that is the points are the  extreme points of pressure and density in the disks.
 Therefore,  these (centers or cusps)  radii  are also  replicas, and the relativistic angular velocity is $\omega_H^\pm$, respectively. There have been no solutions of  $\ell_ {oH}^ \pm= \
\ell^+(r)$.
For  $\ell^-(r)=\ell_{oH}^-$, the angle  $\sigma$ increases with the spin $a$, and decreases with the radius  $r$.
 Solutions are   close to the \textbf{BH} poles.

\begin{figure}
\centering
\includegraphics[width=8cm]{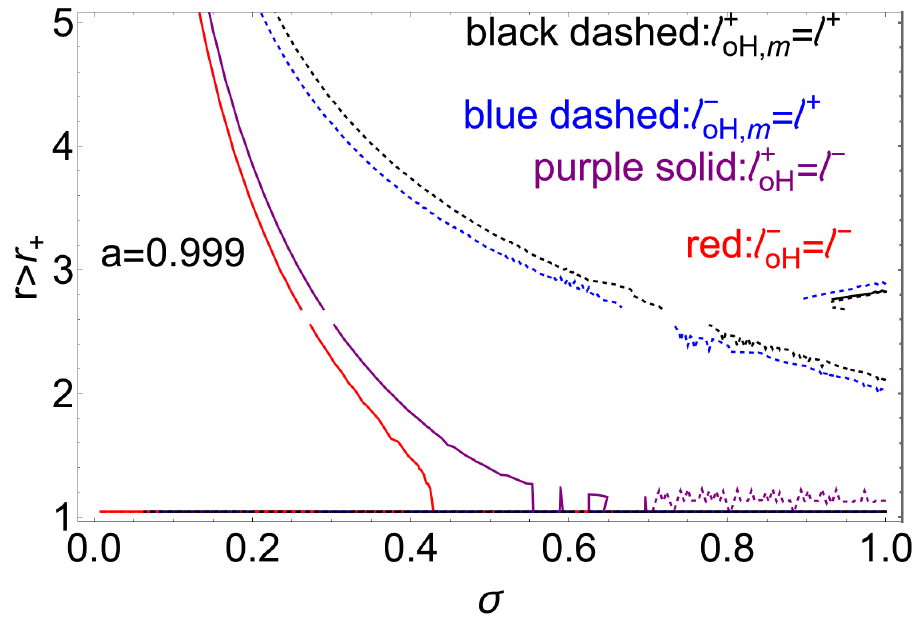}
\includegraphics[width=8cm]{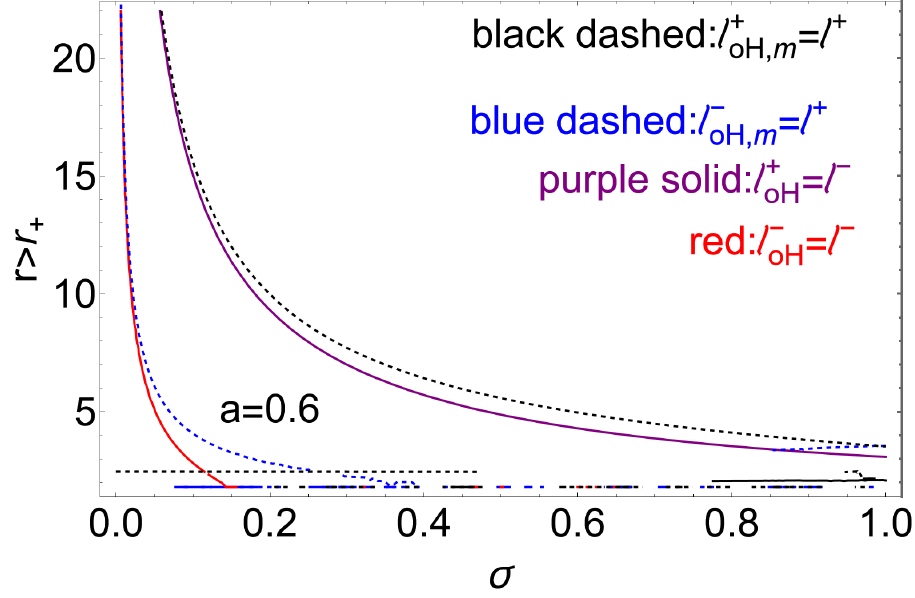}
\includegraphics[width=8cm]{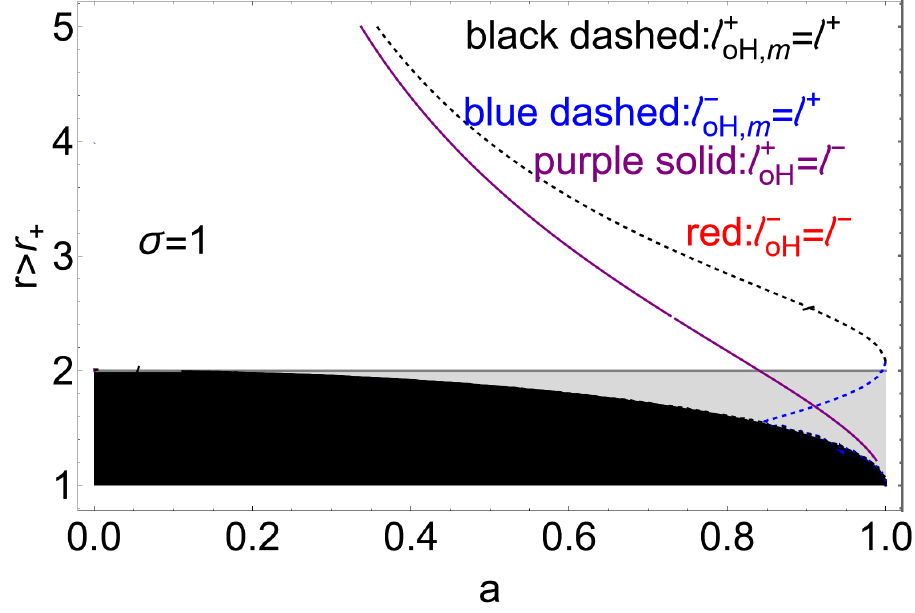}\includegraphics[width=8cm]{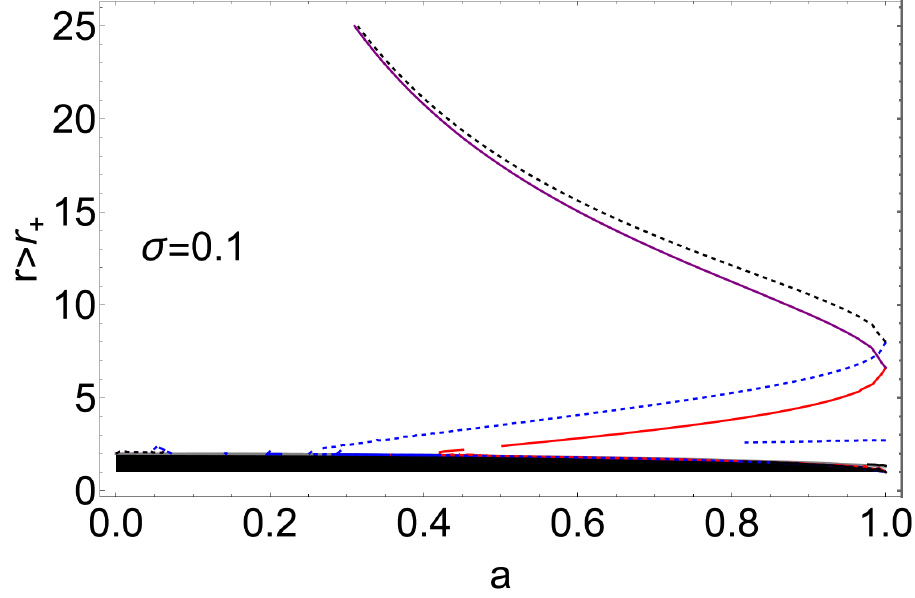}
\caption{Upper panels: The radius  $r>r_+$ such that $\ell^\pm(r)=\ell_{oH}^\pm$  and $\ell^\pm(r)=\ell_{oH,m}^\pm$ as fvunction of the angle  $\sigma\in[0,1]$ for  \textbf{BH} spin $a=0.999$ (left panel) and $a=0.6$ (right panel). (Note, there have been no solutions of  $\ell_ {oH}^ += \
\ell^+(r)$ and $\ell_ {oH}^ -= \ell^+(r)$). The specific angular momenta $\ell_{oH}^\pm$ and $\ell_{oH,m}^\pm$ are defined in Eq.\il(\ref{Eq:lokmmdelf}). $\ell^\pm(r)$ are the radial distribution of critical points of density (and pressure) in  counter--rotating and co--rotating toroids, respectively. See also Figs\il(\ref{Fig:PlotAClooA}). Bottom panels: solutions $r:\ell^\pm(r)=\ell_{oH}^\pm$  and  $r:\ell^\pm(r)=\ell_{oH,m}^\pm$ are shown as functions of the spin $a$ for angle $\sigma=0.1$ (right panel) and on the equatorial plane (left panel). Black region is the \textbf{BH} $(r<r_+)$ and gray region is the outer ergoregion.}\label{Fig:PlotAPoc1}
\end{figure}

 For  $\ell^-(r)=\ell_{oH}^+$ the angle $\sigma$ decreases with the spin $a$  and the radius $r$. Solutions could be found also at the equatorial plane, for small $r$ and spin $a$.
The radii  $r:\ell^-(r)=\ell_{oH}^\pm$  and $r:\ell^+(r)=\ell_{oH,m}^\pm$ are plotted, in Figs\il(\ref{Fig:PlotAPoc1}), as function of the angle  $\sigma$, for  \textbf{BH} spin $a=0.999$ (upper left panel), and $a=0.6$ (upper right panel), and as functions of the spin $a$, for angle $\sigma=0.1$ (bottom right panel), and on the equatorial plane (bottom left panel).
Solutions $r:\ell_{oH}^-=\ell^-$  exist for small angles $\sigma$, and increase with the \textbf{BH}  spin $a$ and decrease with $\sigma$.
Solutions  $r:\ell_{oH}^+=\ell^-$ decrease with $\sigma$ and $a$. Solutions  $r:\ell_{oH}^-=\ell^+$ decrease with $\sigma$ and increases with $a$. Solutions  $r:\ell_{oH,m}^-=\ell^+$ decrease with $\sigma$ and increases with $a$.
For large $\sigma$, solutions are defined for faster spinning attractors.
Solution  $r:\;\ell_{oH,m}^+=\ell^+$ decreases with $\sigma$ and $a$. (For fixed spin, there is one orbit $r$, for each case).

For slowly spinning attractors,  and for small angles and  radii, there is %
\bea
r(\ell_{oH,m}^+=\ell^+)>r(\ell_{oH}^+=\ell^-)>r(\ell_{oH,m}^-=\ell^+)>r(\ell_{oH}^-=\ell^-),
\eea
  for large spin and for small $\sigma$,  there is
  \bea
  r(\ell_{oH,m}^+=\ell^+)>r(\ell_{oH,m}^-=\ell^+)>r(\ell_{oH}^+=\ell^-)>r(\ell_{oH}^-=\ell^-)
  \eea
  --see Figs\il(\ref{Fig:PlotAPoc1}).

\begin{figure}
\centering
\includegraphics[width=8cm]{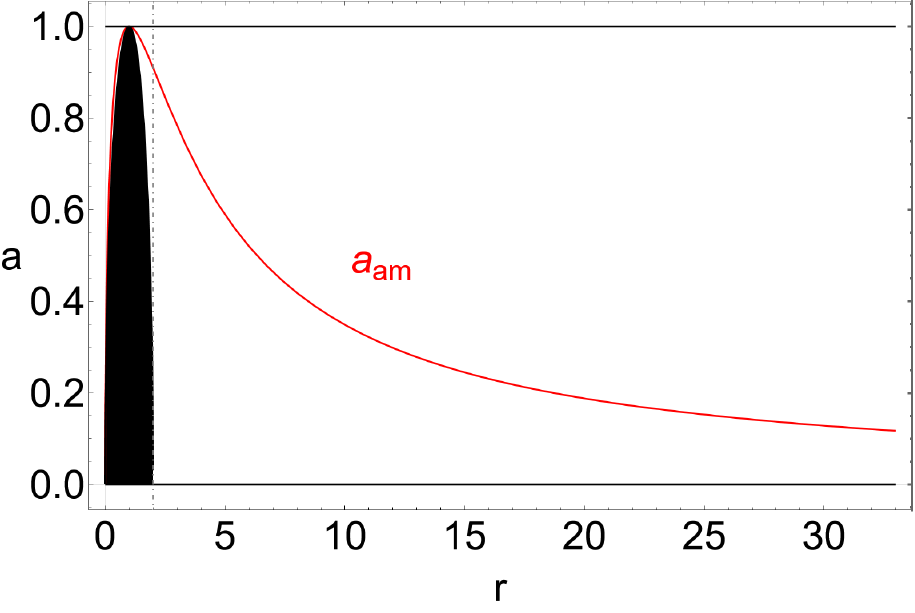}
\includegraphics[width=8cm]{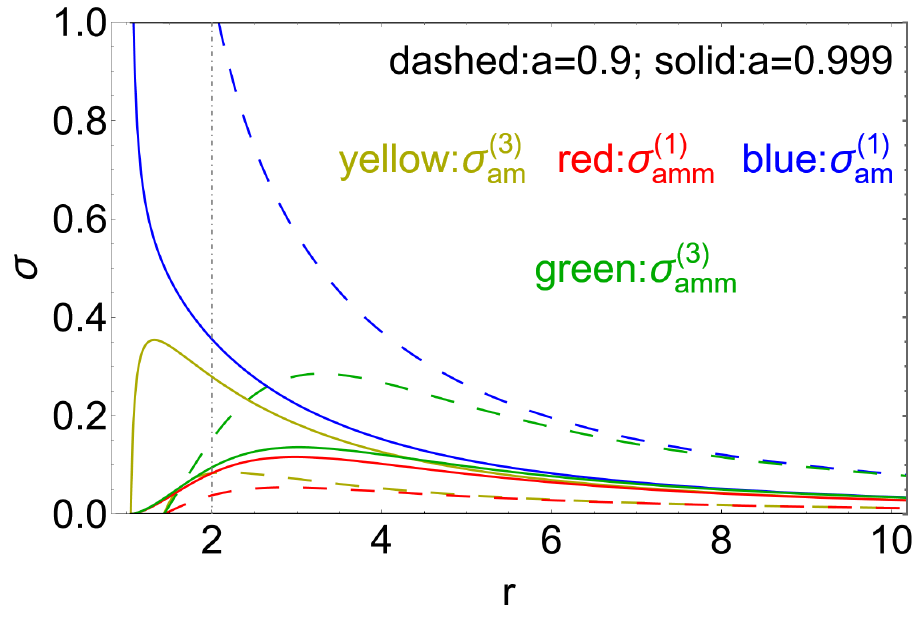}
\caption{Left panel: spin $a_{am}$  in Eq.\il(\ref{Eq:amdefnition}) is shown as function of the radius $r$ in the  extended plane $(a,r)$. Black region is the \textbf{BH} regions $[r_-,r_+]$ in the extended plane.  Right panel: limiting angles $\{\sigma_{am}^{(1)},\sigma_{am}^{(3)}\}$ of Eq.\il(\ref{Eq:aigmam13}) and  $\{\sigma_{amm}^{(1)},\sigma_{amm}^{(3)}\}$  of Eq.\il(\ref{Eq:aigmamm13}) as functions of the radius $r$ for different spins as signed on the panel. }\label{Fig:PlotAmorl}
\end{figure}

The red-shift function is $g_{red}$ is well defined when
 $\sqrt{-g_{\phi\phi} \Omega^2-2 g_{t\phi} \Omega -g_{tt}}={1}/{\dot{t}}>0$ (from Eq.\il(\ref{Eq:succeff-uo})). Here, assuming $ \Omega=\pm\omega_H^\mp$, the conditions are, in details, as follow
\bea\label{Eq:condition-100-Van}
&&
\mbox{for} \quad \Omega=-\omega_H^-:\quad\left(
r\in]1,2], a\in]a_\pm,1], \sigma\in[0,\sigma_{amm}^{(1)}[\right)
,\quad
\left(r>2, a\in]0,1], \sigma\in[0,\sigma_{amm}^{(1)}[\right);
\\\nonumber
&&
\mbox{for} \quad\Omega=-\omega_H^+:
\quad
\left(r\in]1,2], a\in]a_\pm,1], \sigma\in[0,\sigma_{amm}^{(3)}[\right),
\\
&&\nonumber \qquad\qquad \qquad\qquad \qquad\qquad
\left[r>2, \left(a\in]0,a_{am}],  \sigma\in[0,1]\right),
\left(a\in]a_{am},1],
\sigma\in[0,\sigma_{amm}^{(3)}[\right)\right];
\\\nonumber
&&
\mbox{for} \quad \Omega=\omega_H^-:
\quad
(r\in]1,2],a\in]a_\pm,1], \sigma\in[0,\sigma_{am}^{(3)}[),\quad
(r>2, a\in]0,1], \sigma\in[0,\sigma_{am}^{(3)}[);
\\\nonumber
&&
\mbox{for} \quad\Omega=\omega_H^+:
\quad
\left[r\in]1,2], \left(a\in]a_\pm,a_{am}], \sigma\in[0,1[\right);\quad  \left(a\in]a_{am},1], \sigma\in[0,\sigma_{am}^{(1)}[\right)\right],\\
&& \qquad \qquad\qquad \qquad \nonumber
\left[r>2, (a\in]0,a_{am}], \sigma\in[0,1]),
(a\in]a_{am},1],
\sigma\in[0,\sigma_{am}^{(1)}[)\right].
\eea
The spin $a_{am}$ is
\bea\label{Eq:amdefnition}
a_{am}\equiv \sqrt{\frac{r \left[4\mathcal{ S_D}-r \left(r \left[r-\mathcal{S_D}\right]+12\right)\right]}{2(r+2)^2}},\quad\mbox{where}\quad  \mathcal{S_D}\equiv  2+\sqrt{r (r+4)+20}.
\eea
Function $a_{am}$ decreases with the radius $r$--see  Figs\il(\ref{Fig:PlotAmorl})--left panel.
The angles $\sigma_{am}^{(1)}$ and  $\sigma_{am}^{(3)}$, in Eqs\il(\ref{Eq:condition-100-Van}),
are solutions of  the  equation
\bea&&
\{\sigma_{am}^{(1)},\sigma_{am}^{(3)}\}: \sum_{i=0}^4\sigma^i \mathcal{J}_i=0,\quad\mbox{where}
\\\nonumber
&&
\mathcal{ J}_0\equiv 16 a^2\Delta,\quad \mathcal{J}_1\equiv  8 \left[a^6+2 a^4 \left(r^2-3\right)+a^2 r \left(r^3-4 r+8\right)-2 r^4\right],
\\&&\nonumber
\mathcal{J}_2\equiv  a^8+a^6 (r-2) (3 r+8)+a^4 [(r-2) r (r+2) (3 r+4)+48]+a^2 r [r (r [r (r [r+2]+4)-8]+16)-32],
\\&&\nonumber
\mathcal{ J}_3\equiv -2 a^4  \left[a^4+2 a^2 \left(r^2-2\right)+r^4-8 r+8\right],\quad \mathcal{J}_4\equiv  a^6\Delta,
\eea
or, explicitly, there is
\bea\label{Eq:aigmam13}
&&
\sigma_{am}^{(1)}\equiv\frac{\mathbf{\chi_{III}}-\sqrt{4 \mathbf{\chi_{S}}-\frac{\mathbf{\chi_{I}}}{\sqrt{\mathbf{\chi_{V}}}}}-2 \sqrt{\mathbf{\chi_{V}}}}{4} \quad \mbox{and}\quad
\sigma_{am}^{(3)}\equiv\frac{\mathbf{\chi_{III}}-\sqrt{4 \mathbf{\chi_{S}}+\frac{\mathbf{\chi_{I}}}{\sqrt{\mathbf{\chi_{V}}}}}+2 \sqrt{\mathbf{\chi_{V}}}}{4},
\eea
(see Figs\il(\ref{Fig:PlotAmorl})), where
\bea&&\nonumber
\mathbf{\chi_{V}}\equiv-\frac{64 \left(a^2-1\right) (r-1)^2}{a^4  \Delta^2},\quad \mathbf{\chi_{III}}\equiv\frac{2 \left[a^4+2 a^2 \left(r^2-2\right)+r^4-8 r+8\right]}{a^2 \Delta},\quad\mbox{and}
\\\nonumber
&& \mathbf{\chi_{I}}\equiv-\frac{512 \left(a^2-1\right) (r-1) \left[a^4 (r+1)+2 a^2 \left[r \left(r^2+r-6\right)+2\right]+r^3 \left(r^2+r-8\right)+16 r-8\right]}{a^6  \Delta^3},
\eea
with
 {\small
  \bea\nonumber
&&
\mathbf{\chi_{S}}\equiv\frac{a^8+4 a^6 \left(r^2+2\right)+2 a^4 r \left(3 r^3+8 r-40\right)+4 a^2 \left[r \left(r \left[r \left(r^3+2 r-24\right)-8\right]+80\right)-32\right]+(r-2) r (r+2) \left[\left(r^3+4 r-16\right) r^2+64\right]+128}{a^4 \Delta^2}.
\eea}
The angles $\sigma_{amm}^{(1)}$ and  $\sigma_{amm}^{(3)}$, in Eqs\il(\ref{Eq:condition-100-Van}),
are solutions of  the following equation
\bea&&
\nonumber
\{\sigma_{amm}^{(1)},\sigma_{amm}^{(3)}\}:\sum_{i=0}^4 \mathcal{K}_i \sigma^i =0,\quad\mbox{where}
%\sigm
%\sigma ^2 J_2
\\\nonumber
&&\mathcal{K}_0\equiv 16 a^2 \Delta^2,\quad\mathcal{  K}_1\equiv   8\Delta  \left[a^6+2 a^4 \left(r^2-3\right)+a^2 r \left(r^3-4 r-8\right)-2 r^4\right],
\\\nonumber
&&\mathcal{ K}_2\equiv a^2\left[a^8+4 a^6 \left(r^2-4\right)+a^4 \left(6 r^4-32 r^2+48 r+48\right)+\right.
\\\nonumber
&&\qquad \qquad \left.
4 a^2 r^2 \left[r \left(r^3-4 r+16\right)+24\right]+r^2 \left(r \left[r \left(r^4+16 r+32\right)-64\right]+64\right)\right],
\\\nonumber
&&
\mathcal{K}_3\equiv -2 a^4\Delta  \left[a^4+2 a^2 \left(r^2-2\right)+r^4+8 r+8\right],\quad
\mathcal{K}_4\equiv a^6\Delta^2,
%\\&
\eea
and, explicitly,
there is
\bea\label{Eq:aigmamm13}
&&
\sigma_{amm}^{(1)}\equiv \mathbf{\mathbf{\chi_{II}}}-\frac{1}{2} \sqrt{\mathbf{\mathbf{\chi_{P}}}-\frac{\mathbf{\mathbf{\chi_{O}}}}{4 \sqrt{\mathbf{\chi_{N}}}}}-\frac{\sqrt{\mathbf{\chi_{N}}}}{2},\quad\mbox{and}\quad \sigma_{amm}^{(3)}\equiv \mathbf{\chi_{II}}-\frac{1}{2} \sqrt{\mathbf{\chi_{P}}+\frac{\mathbf{\chi_{O}}}{4 \sqrt{\mathbf{\chi_{N}}}}}+\frac{\sqrt{\mathbf{\chi_{N}}}}{2};
\eea
(see  Figs\il(\ref{Fig:PlotAmorl})), where
\bea\nonumber
&& \mathbf{\chi_{N}}\equiv -\frac{64 \left(a^2-1\right) (r+1)^2}{a^4 \Delta^2},\quad \mathbf{\chi_{II}}\equiv\frac{a^4+2 a^2 \left(r^2-2\right)+r^4+8 r+8}{2 a^2 \Delta},\quad\mbox{with}
\\\nonumber
&&
 \mathbf{\chi_{O}}\equiv-\frac{512 \left(a^2-1\right) (r+1) \left[a^4 (r-1)+2 a^2 (r-1) \left(r^2+2\right)+r^3 [(r-1) r+8]+16 r+8\right]}{a^6 \Delta^3};
  \eea
and
{\small
\bea\nonumber
&&\mathbf{\chi_{P}}\equiv\frac{a^8+4 a^6 \left(r^2+2\right)+2 a^4 r \left(3 r^3+8 r-24\right)+4 a^2 \left[r \left(r \left[r \left(r^3+2 r-8\right)-8\right]-16\right)-32\right]+r \left(r^2+4\right) \left[\left(r^3-4 r+16\right) r^2+64\right]+128}{a^4 \Delta^2}.
\eea
}

In Figs\il(\ref{Fig:PlotAmorl})--right panel, the  limiting angles, $\{\sigma_{am}^{(1)},\sigma_{am}^{(3)}\}$ and  $\{\sigma_{amm}^{(1)},\sigma_{amm}^{(3)}\}$,  are plotted as functions of the radius $r$, for different spins. The   limiting angles decrease with the spin, and  increase with the radius $r$. Angles $\sigma_{am}^{(3)}$ and  $\{\sigma_{amm}^{(1)},\sigma_{amm}^{(3)}\}$   have a maximum at small radii $r$.
There is $\sigma_{am}^{(1)}\in[0,1]$, while  $\{\sigma_{amm}^{(1)},\sigma_{amm}^{(3)}\}$ and  $\sigma_{am}^{(3)}$ are defined for small angles $\sigma$.

\begin{figure}
\centering
\includegraphics[width=8cm]{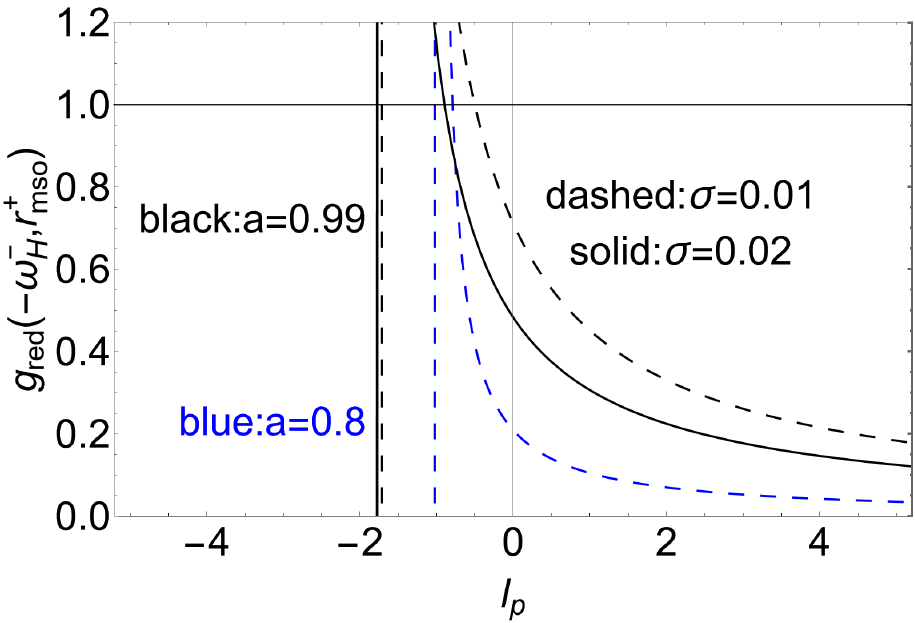}
\includegraphics[width=8cm]{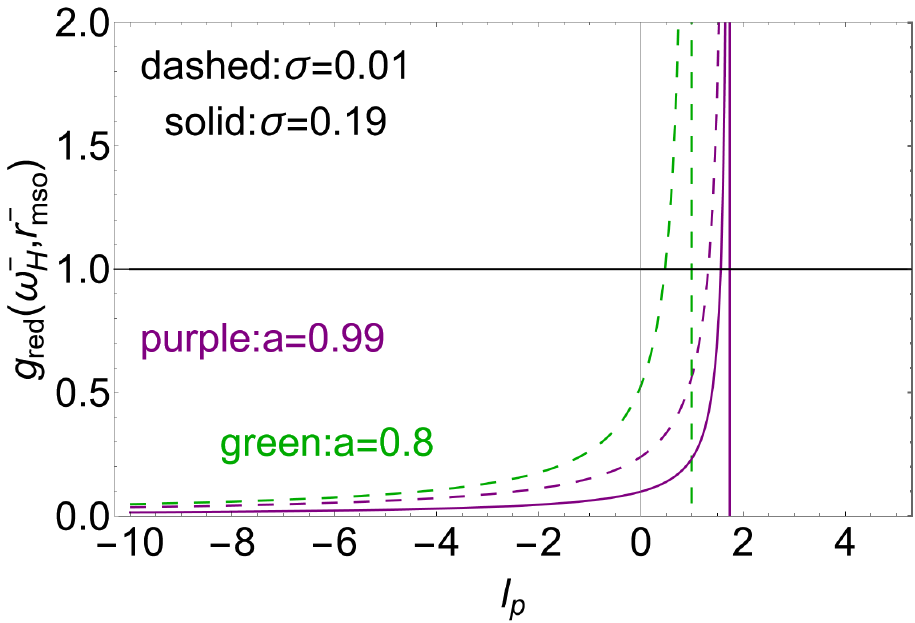}
\includegraphics[width=8cm]{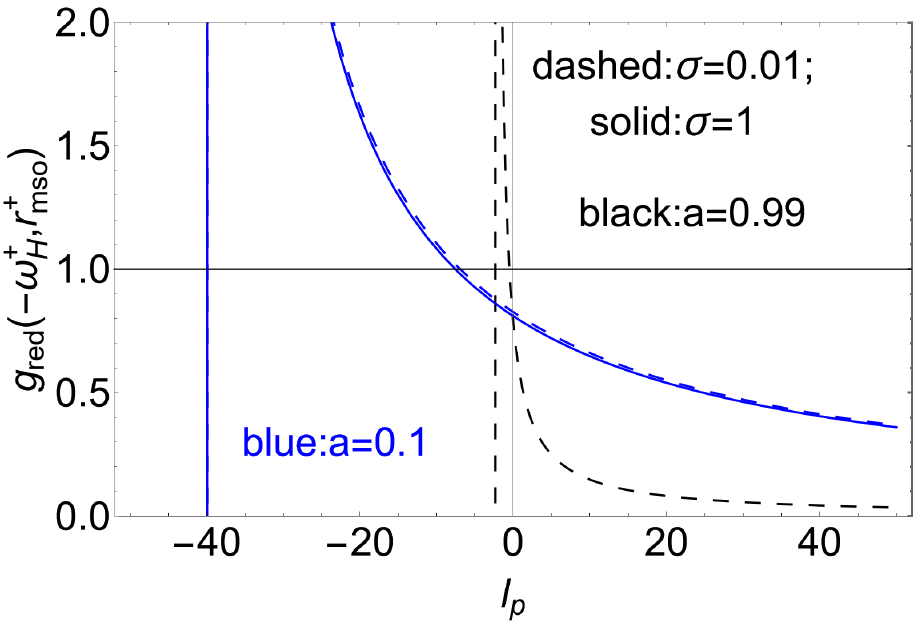}
\includegraphics[width=8cm]{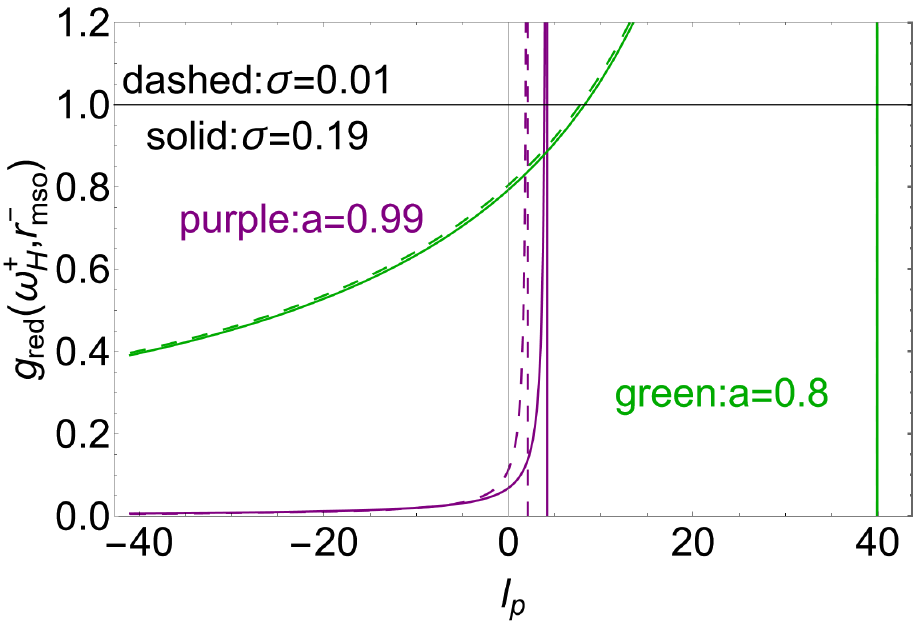}
\caption{The red-shift function $g_{red}$ evaluated on the relativistic angular velocity  $\Omega=\pm \omega_\pm$ where $\omega_H^\pm$ are the outer and inner \textbf{BH} horizon angular velocities and on the  radii  $r_{mso}^\pm$, as function of the photon impact parameter $l_p$, for different \textbf{BH} spin $a$  and angles $\sigma$ as signed on the panels.}\label{Fig:PlotAmorlB}
\end{figure}

In Figs\il(\ref{Fig:PlotAmorlB}) the red-shift function $g_{red}$ is evaluated on the relativistic angular velocity  $\Omega=\pm \omega_\pm$ and on the radii  $r_{mso}^\pm$,  and it is shown as  function of the photon impact parameter $l_p$, for different \textbf{BH} spin $a$ and angles $\sigma$.
 For faster spinning attractors,  solutions exist, in general, for  smaller values of $\sigma$.  The signal is red-shifted for larger part of the $l_p$ values.

\section{Conclusion}\label{Sec:Conclusions}
We
   studied the red--shift function $g_{red}$ in relation to the Kerr \textbf{BH} horizon replicas.  Function $g_{red}$ expresses the frequency shift of a signal emitted from a  circularly orbiting particle.
In some cases the horizons  replicas    are  close to \textbf{BH}  poles. This is a property of the replica occurring  particularly for the inner horizons replicas.
The collection of all the  horizons replicas in \emph{one spacetime} is provided by   the stationary observers light surfaces (Figs\il(\ref{Fig:Plotlocalwesit})) in that spacetime, and all the  characteristic frequencies of all the   horizons replicas in a fixed spacetime are the limiting   angular velocities  $\omega_\pm$ of the stationary observers  this  spacetime. Therefore,   in   Sec.\il(\ref{Sec:limitng-velocitues})  we  constrained  $g_{red}$  using the properties  of  these limiting light--surfaces.
In particular,  in Sec.\il(\ref{Sec:photon-shells})     the extreme points of the limiting frequencies $\omega_\pm$  are set    in  relation with the \textbf{BH} photons shell boundaries.
While in  Sec.\il(\ref{Sec:redffhidft-shellsreplicas}) we related directly   the
 horizons replica red-shift function    to the spacetime  photons shell and, consequently, to  the \textbf{BH} shadow boundaries.
Horizons  replicas can be co--rotating and counter--rotating  with respect to the central attractor,  and  the differences among the orbits of  these two classes replicas, are  examined in  more details in Sec.\il(\ref{Sec:co--rotating-counter-rpliAs}).  In Sec.\il(\ref{Sec:absidered-extended-plane}) we specified  the properties of the red-shift function read in the extended plane (where the horizons replicas can be defined as points of metric Killing bundles--Figs\il(\ref{Fig:PlotturenesaAsecc1})) in terms of the photons impact parameter.
According to this  analysis, the orbits could be identified  through  the  red--shifting  or  blue--shifting of the emitted  signals. Some of these signals have been also connected to  accretion disks,  studying emission from  the surface of the disk or accretion flow and  proto--jet-emission. We  examined, in particular, the  frequency--shift   from the horizons  replica  related to  the disks surface, inner edge and accretion flows,
for an equatorial, general relativistic, axially symmetric accretion disk, orbiting around the central Kerr \textbf{BH}  attractor.
However, from methodological view-point we re--phrased   many  quantities  significant on the accretion disks physics    in the extended plane, where the replica, collected in metric Killing bundles, and their properties  are defined   across  the set of all  Kerr  geometries.

{Definition of replica is  tied to  definition of the  bundle  characteristic frequency.  A (co--rotating) horizon replica by definition has  same frequency (angular velocity) $\omega$ as the inner \textbf{BH} horizon, if $\omega=\omega_H^->1/2$, or the  outer \textbf{BH} horizon, if $\omega=\omega_H^+\in [0,1/2[$, of the (fixed) \textbf{BH}  in the  spacetime of (the observer and) the replica.
(The    counter--rotating horizons replicas, having $a\omega<0$, were also examined.  While the properties of the counter-rotating replicas are detailed  in our analysis,  general  considerations outlined below  hold for co--rotating and counter--rotating replicas.)
Frequencies  $\omega_H^\pm$ are functions of the \textbf{BH} (dimensionless) spin $a$ only (see Eq.\il(\ref{Eq:omegahmo})), and the measure of these quantities  uniquely identifies the central  \textbf{BH}. The bundle characteristic  frequency is constant along the bundle, for each point  $(r,\theta)$  in the  (fixed) spacetime  where it is defined. At fixed angle, for $\sigma=$constant,  the characteristic frequency of the bundle  depends  on the spin $a$ only,  this frequency is also  a frequency  of the stationary observers   light surfaces  in that spacetime (all points of a bundle are points of stationary observers  light--surfaces)--see Eq.\il(\ref{Eq:limiting-stationary}) and Figs\il(\ref{Fig:Plotlocalwesit}). (We remark that  these light--surfaces constrain several aspects of \textbf{BH} astrophysics for example, the magnetosphere of the central object \cite{Komissarov,Uz05,BZ77,Z77,Punsly,Mahlmann18,KMCK}).  Identifying $\omega$ at different points would allow the  replica  and, therefore, the central \textbf{BH} identification. That is,  identifying  the replica surfaces  (see Figs\il(\ref{Fig:PlotBri0999mm}) and Figs\il(\ref{Fig:Plotlocalwesit}))  implies  to  recover direct information of  the central attractor spin--mass ratio.  Replicas have interesting implications also in the analysis of the spacetime regions  near the poles of the Kerr \textbf{BH} and its rotational axis, as these spacetimes regions  are filled with replicas of the outer and, particularly, inner horizons replicas  also at  larger angles, therefore  this analysis  could constitute also  a framework to explore  further  this  special region of  a spinning attractor spacetime--see Figs\il(\ref{Fig:PlotBri0999mm}) and Figs\il(\ref{Fig:Plotlocalwesit}).  The possible identification of a spacetime   replica could  be grounded on the consideration that replicas are special points on the spacetime light surfaces. As seen in this analysis,  replicas are also connected  to the  \textbf{BH} shadow boundary. The boundaries of the photon shells are, at any vew angle,  on the line $\beta=0$ of the celestial plane  $(\alpha,\beta)$,   and they are the extremes of the light--surfaces frequencies $\omega_\pm$ located, respectively, in  regions  of the  celestial plane defined by the line  $\alpha=0$, corresponding to the radius $r_{\lim}(a)$ of the spacetime, which is a function  of the spin $a$ only--- see Eqs\il(\ref{Eq:estremeoemgapojnts}) and \ref{Eq:estremeoemgapojnts1}). As  investigated in Sec.\il(\ref{Sec:redffhidft-shellsreplicas}) redshift emission can provide  another way of identifying space-time replicas.   Replicas can be on  \textbf{BH} shadow boundaries   within the conditions considered in Sec.\il(\ref{Sec:absidered-extended-plane}),  providing  a further way of mapping light surfaces with replica points--see also \cite{Pugliese:2021ivl,2024NuPhB100816700P,Pugliese:2021aeb}). }

Observation of the \textbf{BH}  shadow boundary could be a
  second phenomenological   aspect possibly hosting replica emission. It has been proved the presence of photons from the  horizons  replicas, at certain view--angles for certain spins, on the \textbf{BH}  shadow boundaries. Within these conditions,  we proceeded  constraining  the  replicas  in relation to the  \textbf{BH}  photons shell.
The analysis  eventually resulted into a map,  dependent  on the   \textbf{BH} spin, of the  red-shifting and blue-shifting  regions parametrized by the   photons impact parameter, view-angle, and  emitted  angular velocity.
This results   provides  a scenario  where, in particular,  it could be possible to distinguish co--rotating  and  counter-rotating inner and horizons   replicas
 depending   on the view angles $\sigma$,  in different classes of   \textbf{BHs} spacetimes,  distinguished by the \textbf{BH} dimensionless spin.

\section*{Data Availability Statement }
This manuscript has no associated data.
[Authors' comment: Data sharing not applicable to this article as no
datasets were generated or analysed during the current study.]

\end{document}